\documentclass[a4paper]{article}
\usepackage[margin=25mm]{geometry}
\usepackage{moreverb,url, dsfont, soul, dsfont,amsmath, amsthm}
\usepackage[round]{natbib}
\usepackage[colorlinks, bookmarksopen, bookmarksnumbered, citecolor=blue, urlcolor=red]{hyperref}
\usepackage{moreverb, url, soul, subcaption, threeparttable, multicol, multirow, lscape, makecell, bm, bbm, soul}
\usepackage{enumitem}
\usepackage[title]{appendix}
\usepackage{mathtools, color, xcolor}
\usepackage{tabularx, ltablex, floatrow, inputenc}
\usepackage[]{graphicx}
\usepackage{color, soul, threeparttable, multirow, multicol, setspace}
\usepackage{amsmath, amssymb, amsfonts, caption, subcaption, tikz, booktabs, colortbl, dcolumn}
\usepackage{verbatim}
\usepackage{xr}
\usepackage{xr-hyper}
\usepackage{hyperref}

\newcommand{\bigCI}{\mathrel{\text{\scalebox{1}{$\perp\mkern-10mu\perp$}}}}

\providecommand{\keywords}[1]
{   \small	
	\textbf{{Keywords}} #1}

\title{\bf Overlap, matching, or entropy  weights: \\ what are we weighting for?}
\author{Roland A. Matsouaka$^{1,2,*}$, Yi Liu$^{1}$, Yunji Zhou$^{1}$
	\\
	\small $^{1}$Department of Biostatistics and Bioinformatics, Duke University, Durham, NC, USA \\
	\small $^{2}$Program for Comparative Effectiveness Methodology, Duke Clinical Research Institute, Durham, NC, USA\footnote{Corresponding author: Roland A. Matsouaka; \tt{roland.matsouaka@duke.edu}}
}
 
\date{} % Comment this line to show today's date

\begin{document}
	\maketitle
	\begin{abstract}

    There has been a recent surge in statistical methods for handling the lack of adequate positivity when using inverse probability weights (IPW). However, these nascent developments have raised a number of questions. Thus, we demonstrate the ability of equipoise estimators (overlap, matching, and entropy weights) to handle the lack of positivity.  Compared to IPW, the equipoise estimators have been shown to be flexible and easy to interpret. However, promoting their wide use requires  that researchers know clearly why, when to apply them and what to expect.  

    In this paper, we provide the rationale to use these estimators to achieve robust results. We specifically look into the impact imbalances in treatment allocation can have on the positivity and, ultimately, on the estimates of the treatment effect. We zero into the typical pitfalls of the IPW estimator and its relationship with the estimators of the average treatment effect on the treated (ATT) and on the controls (ATC). Furthermore, we also compare IPW trimming  to the equipoise estimators. We focus particularly on two  key points: What fundamentally distinguishes their estimands? When should we expect  similar results? Our findings are illustrated through Monte-Carlo simulation studies and a data example on healthcare expenditure.
	\end{abstract} \hspace{10pt}
	\keywords: {Positivity; propensity scores; equipoise; overlap weights; matching weights, entropy weights.} 
       
\maketitle
	
\newpage
	%%%%%%%%%%%%%%%%%%%%%%%%%%%%%%%%%%%%%%%%%%%%%%%%%%%%%%%%%%%%%%%%%%%%%%%%%%%%%%%%%%%%%%%%%%%%%%%%%%%%%%%%%%%%%%%%%%%%%%%%%%%%%%%%
	%%%%%%%%%%%%%%%%%%%%%%%%%%%%%%%%%%%%%%%%%%%%%%%%%%%%%%%%%%%%%%%%%%%%%%%%%%%%%%%%%%%%%%%%%%%%%%%%%%%%%%%%%%%%%%%%%%%%%%%%%%%%%%%%
	\section{Introduction}\label{Sect-Intro}

	To assess the effect of a new treatment regimen ($Z=1$) over a standard (or control) treatment ($Z=0$) based on  data from an observational study, using causal identification a number of assumptions must be made, including the positivity assumption. For instance, to estimate the average treatment effect (ATE), this assumption requires $0<e(x)<1$, where $e(x)=P(Z=1|X=x)$ is the propensity score (PS), i.e., the probability of treatment assignment, given the vector of baseline covariates $X$ \citep{rosenbaum1983central,rubin1997estimating}. 
	The positivity assumption ensures that the distributions of the related baseline covariates have a good overlap and hence a good common support \citep{petersen2012diagnosing,li2018addressing}. 
 
     The inverse probability weighting (IPW) estimator for ATE assigns to study participants weights that are inversely proportional to their respective PSs. Thus, IPW creates a pseudo-population of participants, corrects for observed covariates distributions imbalances between the treatment groups, and adjusts for (measured) confounding bias inherent to most non-randomized studies. Nevertheless, when PSs are equal to (or near) 0 or 1, there is  violation (or near violation) of the positivity assumption, which we often refer to as lack of adequate positivity \citep{petersen2012diagnosing}. Violations (or near violations) of the positivity assumption occur either at random (or stochastically), i.e., by chance due the data (or underlying model) characteristics or when some subgroups of participants can never (or barely) receive one of the treatment options under study. This can lead to moderate or even poor overlap of the distributions of the PSs and may result in large IPW weights, especially when the ratio $[e(x)(1-e(x))]^{-1}$ is highly variable  \citep{li2013weighting, zhou2020propensity}.  As such, IPW may put a large amount of weights on a small number of observation, which can  unduly  influence the estimation of the treatment effect.  
	
	While violations of the positivity assumption can be remedied by either PS trimming or truncation, recent advancements have introduced methods that aim to overcome the limitations of these {\it ad hoc} solutions. Some of these novel methods propose bias-corrected estimators  \citep{chaudhuri2014heavy, ma2020robust, sasaki2022estimation}, while other  reparametrize the PS estimation, by adding {\it a priori} covariate balancing constraints to modify  the PS model  \citep{graham2012inverse,imai2014covariate}.  Some consider direct optimization techniques to derive sample weights 
	under  covariate constraints  \citep{hainmueller2012entropy,zubizarreta2015stable,wong2017kernel,hirshberg2017two} or redefine the target population altogether and bypass the need to account for the lack of positivity  \citep{li2018balancing,matsouaka2020framework,zhou2020propensity}.  %to locate a suitable region of common support to point identify the ATE for a subpopulation (that may or may not be the population of interest) defined by the common support and achieve internal validity of the ATE estimator.
   
	  \subsection{The positivity assumption and propensity score weighting methods}\label{sec:invPSmethods}
   
	  The literature defines two  specific violations of the positivity assumption: random (i.e., by chance) and structural violations \citep{westreich2019epidemiology, petersen2012diagnosing}. % Hence, depending on the type of violation, there may be different methods available to infer causality. Unfortunately, this distinction is not always made explicit in many applications.     
    Random (or stochastic) violation of the positivity assumptions  arise by happenstance, e.g.,  when the sample size is small or  the PS model is misspecified.  In such cases, increased sample size, bias-corrected IPW trimming,  PS reparameterization or direct optimization  offer better alternatives to estimate ATE \citep{chaudhuri2014heavy,ma2020robust,sasaki2022estimation}.  Alternatively, methods for equipoise treatment effect, i.e., the overlap weight (OW), matching weight (MW), and Shannon's entropy weight (EW) estimators \citep{matsouaka2020framework,li2018addressing}, can also be considered. These estimators target treatment effects defined within the subgroup of participants for whom  treatment equipoise exists. 
	  
  % Structural violation of positivity is common The lack of positivity is inherent to the characteristics of the target population and may not be easily be mitigated by study eligibility criteria. 
	  
	As noted by \cite{petersen2012diagnosing}, violations of the positivity assumption can lead to substantial bias and sometimes  an increased variance of the causal effect estimator. While checking the PS distributions (or the PS weights) between treatment groups can help assess such violations, it is important to recognize that well-behaved weights alone may not guarantee the satisfaction of the positivity assumption \citep{ma2020robust, petersen2012diagnosing}. \textcolor{black}{A better investigation into violations of the positivity assumption must always be preceded by an expert-knowledge elicitation of the data at hand, the scientific questions  as well as the source and the nature of the data at hand.}

	 \subsection{The positivity assumption and imbalance in treatment allocations}\label{sec:invPSmethods}
	  Correct estimation of the treatment effect is challenging when treatment (or exposure) allocation is rare   \citep{pirracchio2012evaluation,rudolph2022estimation,hajage2016use}. 
		Nevertheless, assessment of treatment effects with small proportion of treated participants is a common occurrence, particularly  in pharmacoepidemiologic observational studies of drugs  \citep{hajage2016use,platt2012positivity}. The evaluation of the risk-benefit profile of a newly released drug is often conducted using observational studies where data on the effectiveness and safety of the drug are collected during routine care  \citep{schneeweiss2007developments,rassen2012newly}.  \citet{schneeweiss2011assessing} provide an example in the comparative effectiveness of newly marketed medications, which presents additional challenges. These challenges include potential bias due to patient channeling toward the newly marketed medication (due to patient, provider, and system related factors), shifts in the user population (due to varying background characteristics and comorbidities), timely data availability issues, and a smaller number of users in the initial months of marketing. As \citet{schneeweiss2011assessing}  indicated, ``Of these challenges, channeling is often the biggest threat to the validity of nonrandomized studies\dots''   Therefore, there is a pressing need for the use and development of sound statistical methods that also aim at consistent and robust estimation of the treatment effects when the lack of positivity is expected or unavoidable. 
	
	While some authors have investigated the use of PS methods when  the proportion of participant is small, their focus has primarily been on traditional PS methods, overlooking alternative methods that are well suited for lack of positivity. These alternatives go beyond the traditional use of PS matching, truncation, or trimming   \citep{hajage2016use,franklin2017comparing,austin2011tutorial}. 
 
 Causal inference, being inherently a missing data problem \citep{holland1986statistics}, we often overlook  the fundamental task of any causal estimator: to use the available data to adequately input unknown potential outcome values. For IPW, this means weighting participants to create a pseudo-population where causal inference can be drawn. When the treatment allocation is imbalanced and extreme weights emerge, estimationn and inference of the treatment effects relies heavily on a few participants with extremely large inverse probability weights, which can introduce severe bias due to data disparity. Therefore, regardless of whether violations of the positivity assumption are structural or not, it is crucial to ensure that the estimated treatment effects are disproportionally driven  by a small number of outlying participants, especially  if there is a substantial treatment allocation imbalance. For instance, in a tutorial for PS analysis, Austin uses  a sample of current smokers  discharged alive from a hospital following an acute myocardial infarction  \citep{austin2011tutorial}. What is remarkable in this paper are the small proportion (32.20\%) of patients who did not benefit from in-patient smoking cessation counseling,  the wide range of estimated PSs, the presence of a few extreme  weights, and the results on the 3-year survival  outcomes (binary and time-to-event). Thus, reading this paper, we can't help but raise some questions. Were the different conclusions drawn were solely due to methodological differences?  In fact, some of these methods showed a significant reduced risk of mortality, while others just indicated that the treatment effect was not different from the null.  Could the discrepancy in the study conclusion be solely driven by the differences in the selected methods and their underlying estimands? Did  the imbalance in the number of participants between the two treatment groups play also a role? As we will demonstrate in this paper, we believe both the choice of specific methods and the imbalance in treatment allocation play a preeminent role.
	
	\subsection{How about trimming or truncating extreme weights?}\label{sec:trimming_methods}
 
	\textcolor{black}{The IPW estimator of the ATE targets  $E[B/A]$, with $B =  (Z-e(X))Y$ and  $A=  e(X)(1-e(X)).$ When there is a violation of the positivity assumption, some observations have $A\approx 0$, which can have unduly influences on the na\"ive sample mean of $B/A$.} The primary objective of trimming (i.e., dropping participants) and truncation (i.e., capping weights)  is to curtail such undue influences and provide a stable estimator. %The expectation, rationale, and heuristic of trimming and truncation are that they can help achieve a good bias-variance trade-off, enhance the internal validity of the study, and improve the efficiency of the resulting treatment effects. 
 Trimming (or truncating) participants with $A\approx 0$ (above given thresholds) is a common practice, as it effectively constraints the weights within reasonable bounds. However, the resulting estimate is often highly sensitive to the choice of on the threshold(s) \citep{chaudhuri2014heavy, ma2020robust, sasaki2022estimation}. Unfortunately, the choice of a threshold is often ad hoc and subjective as they rest solely on the user's discretion  \citep{crump2006moving}. 

 In many applications, a user-selected threshold can  drastically change the number of participants we discard or for whom we curtail the weights (see, for instance, \citep{zhou2020propensity}). This  tremendously affects  the finite-sample performance of the estimator, influencing both its bias and  efficiency. Moreover, the corresponding estimator may not target the ATE based on the original population since, for instance, under structural violation of the positivity assumption they  both can alter their target  estimands and the underlying populations of interest, depending on the threshold considered \citep{chaudhuri2014heavy, zhou2020propensity}. \textcolor{black} {For example, ATE trimming by a threshold $\alpha\in(0,0.5)$ shifts its target population to the population of participants whose PS (of receiving either treatment or control) is inside the interval $(\alpha, 1-\alpha)$.}  \textcolor{black}{
Often, standard trimming and truncation method result to a non-negligible bias (even asymptotically)  when estimating ATE, which may have some inference implications. 
%If we want to target the overall population---via trimming or truncation (to circumvent poor overlap or violations of the positivity assumption)---we must instead consider some bias-correction strategies.
} 

\textcolor{black}{Landmark bias-correction strategies for trimming have been proposed. \cite{chaudhuri2014heavy} proposed a bias-corrected, tailed-trimmed IPW estimator of the ATE, based on the tail behavior of $|B/A|$. Their estimator is robust and asymptotically valid, even under substantial limited over of the PS distributions.  Rather than trimming on the ratio $B/A$, \cite{ma2020robust} and \cite{sasaki2022estimation} considered trimming observations with $A\approx 0$ to build flexible, biased-corrected estimators that allow for larger trimming and hence smaller variances or faster rates of convergence.
 Robustness of the estimator by \cite{ma2020robust} is achieved by combining resampling with a local polynomial-based bias-correction technique, where a data-driven threshold  is selected by minimizing the mean squared error. \cite{sasaki2022estimation}, on the other hand, leverage the smoothness
of the conditional moment function $a \mapsto E[B|A = a]$ to achieve more robust inference and faster convergence rate.  }

\textcolor{black}{The above strategies (implicitly) assume random violation of the positivity assumption, under which true ATE exist and can be point estimated; their respective trimming and bias correction solutions aim at improving inference. Unlike these papers, our proposed estimators do not even rely on the positivity assumption; thus, they are applicable to either random or structural violation of the positivity assumption. In addition, they automatically focus on the area of common support,  identify a specific subgroup of participants where the estimate has a strong internal validity, without involving the outcome $Y$.  %They estimate treatment effects using an estimand defined on the region of common support for the subpopulation for which there is equipoise.
}%{\color{red} \underline{Comment}: More detailed reviews? We need to think about how to make a connection of reviewing on these methods to this paragraph and next. Possible aspects: practical usage?  }

     Furthermore, when there is an important imbalance in treatment allocation (i.e., the proportion of participants in one of the treatment groups is small), if often exacerbates the lack of adequate positivity, which can lead to more trimming or truncation in one group instead of both. Such practices not only reduce  the number of participants  in the final sample (after trimming), but also influence the contribution of those from whom  extreme weights are capped (by truncation).  \textcolor{black}{This even further complicates point estimation and inference when structural violations of the positivity assumption are expected.} Therefore, there is a growing interest for new practical methods that do not leave room to manually and subjectively pick a threshold; methods that can leverage inherent data-driven mechanisms to control the impact of extreme weights and provide robust assessments of  treatment effects.	
	
	\subsection{Are there better alternatives?}\label{sec:trimming_methods} 
	The overlap weight (OW), matching weight (MW), and Shannon's entropy weight (EW) estimators (hereafter referred to as equipoise treatment effect estimators  \citep{matsouaka2020framework})  effectively circumvent the lack of positivity without specifying any user-driven threshold. Besides, they provide both better causal estimations and  higher effective sample sizes  \citep{li2013weighting,li2018balancing, zhou2020propensity, li2021propensity}.  However, it remains to see whether   imbalances in treatment allocation can directly affect  their estimations  (compared to  IPW estimation of  ATE) and to what extent. 
 
	Therefore, the main objective of this paper is to provide a formal assessment of the  impact of equipoise estimators (i.e., OW, MW, EW) on the treatment effect estimation in studies where there is a disproportionate distribution of treated or control participants in the population and how it relates to the lack of positivity. %This helps clarify our motives and their usefulness: indicate clearly whether weighting is warranted, ensure that balance is always reached, and dispel any confusion we might have on the pertinence and  appropriateness of the underlying statistical inference. 
	\textcolor{black}{The title of our paper: ``Overlap, matching and entropy weights: what are we weighting for?'' is thus a call to action, i.e., to  delve into the unique characteristics of these equipoise estimators, which grant them the flexibility to address the lack of positivity effectively  \citep{zhou2020propensity}.   In the process, we showcase how OW, MW, and EW methods can be  strategically used to estimate treatment effects and make asymptotically correct inferences of the corresponding estimators, when there is a violation of the positivity assumption.}
	
	The rest of the paper is organized as follows. We  start in the next section with key questions that help define the purpose of our study and what we intend to accomplish. Then, in Section \ref{sec:bal.wgts}, we introduce  notations and present the family of balancing weights. We specify their related estimands and define the corresponding estimators. Of particular interest are questions related to what is being estimated and what the target populations are when using these balancing weights? %These  shed lights on the purpose IPW when we estimate ATE. 
    Next, we explore how the estimators are impacted by the proportion of treated participants and provide proper interpretations of their estimates. %In light of what we have uncovered in this paper, we discuss how the ATE may estimate a measure that is counter-intuitive to our expectations or lacks  clinical (or epidemiological) importance when the proportion $p$ of treatment participants is small ($<30\%$) or large ($>70\%$). 
    The illustrative example in Section \ref{sec:key_questions} sets the scene for the main idea of this paper and informs the simulations in Section \ref{sec:simulations}. %Finally,  we make the case for the use of OW, MW, or EW when the goal is to estimate a treatment  effect from a total population perspective. 
	
%	In practice, while it is essential to clearly formulate our  scientific questions of interest, we should also expect to  provide appropriate responses with accurate,  efficient, and robust inference. Hence, 
 We evaluate the performance of the estimators  using Monte-Carlo simulation studies in Section  \ref{sec:simulations}, covering three different treatment allocations, under various treatment effects and model specifications. The methods are illustrated, in Section \ref{sec:data},  to evaluate the impact of racial disparities on healthcare expenditure. %In this data set, there is imbalance in the probability $p$ of group representation and it is a well known problem that some participants have outlying PSs  \citep{li2018balancing}. 
	Technical proofs, additional simulation details and data application results are presented in the Supplemental Material. 
	
	%%%%%%%%%%%%%%%%%%%%%%%%%%%%%%%%%%%%%%%%%%%%%%%%%%%%%%%%%%%%%%%%%%%%%%%%%%%%%%%%%%%%%%%%%%%%%%%%%%%%%%%%%%%%%%%%%%%%%%%%%%%%%%%%
	\section{We do not always get what we expect}
	\subsection{What are we trying to estimate?}\label{sec:key_questions}
	%The goal of any PS weighting method is to re-weight the study participants. This create a pseudo-population in which the treatment groups are both balanced and comparable, with respect to their measured covariate distributions \citep{hernan2016using, hernan2018causal,cole2008constructing}. Comparing treatment groups from this pseudo-population would then lead to consistent estimates of the treatment effects. The standard IPW to estimate the average treatment effect (ATE) up- or down-weight up participants to represent the pseudo-population, depending on their respective PSs. It over-represents treated participants with smaller propensity and control participants with a higher propensity of being treated. When the number of these outlying participants is critical, they can drastically influence the estimation of the average treatment effect, leading to what is called the ``tyranny of the minorities''  \citep{lin2013agnostic}. Therefore, 
	Estimating ATE using IPW is problematic in the presence of observations with $e(x)(1-e(x))\approx 0$ as the method purposefully over-represents their contributions and can distort treatment effect estimation \citep{zhou2020propensity}. In medical research, patients with with $e(x)(1-e(x))\approx 0$ are often those who exhibit distinct baseline characteristics, commorbidities, specific counterindications, restrictions, or outlying outcome measures.

	% \citet{hirano2003efficient} recommend that we may ``... restrict attention to a subpopulation for which there is sufficiently large probability of observing both treated and untreated units'' by removing extreme, outlying participants while still aiming for a fair assessment of treatment effects. However, the question of how to achieve this balance remains fundamental. 
 
    On the other hand, imbalance in the treated allocation can also lead to $e(x)(1-e(x))\approx 0$. %the proportion of treated participants also has an impact on the quality of the estimand we can determine. 	
    When the proportion of treated participant is small, many studies focus on the average treatment effect on the treated (ATT), i.e., the average effect in participants who ultimately received the active treatment.  Nevertheless, the assessment of the average treatment effect (ATE) is de rigueur in some contexts, such as pharmacoepidemiologic observational studies of drugs  \citep{hajage2016use,platt2012positivity, rassen2012newly, schneeweiss2007developments}.  Unfortunately, a smaller proportion of treatment participants can lead to  a lack of positivity and the presence of extreme PS weights, complicating the assessment of ATE.  Furthermore, because ATE provides the treatment effect that we would have obtained if everyone (treated and controls alike) is assigned to one treatment option versus being everyone is assigned to the alternative treatment option,  IPW plays an intriguing balancing act when the treatment groups are inadequately  allocated, as we  illustrate below.  
	 
	%%%%%%%%%%%%%%%%%%%%%%%%%%%%%%%%%%%%%%%%%%%%%%%%%%%%%%%%%%%%%%%%%%%%%%%%%%%%%%%%%%%%%%%%%%%%%%%%%%%%%%%%%%%%%%%%%%%%%%%%%%%%%%%%
	\subsection{Are we actually getting what we hope for?}\label{sec:illustration}

	\subsubsection{An illustrative example} 
 
 To illustrate the influence of $p= P(Z=1)$ in estimating different causal estimands, we conducted a simulation study as follows. 
	For $N=10^4$, we generated  $X=(1,X_1,X_2)$ such that $X_1\sim \mathcal{N}(6,9), X_2\sim \text{Bern}(0.75)$ and a treatment assignment $Z = \text{Bern}(\text{expit}(X\beta))$, with $\beta \sim (\beta_0, 0.2, 0.8)'$. %Through different choices of $\beta_0$, we can obtain different proportions of participants in the treatment group ($p=P(Z=1)$) and assess different mean values of the PSs. Hence, 
	We chose $\beta_0=-3.5$ and 0 to obtain $p = 17.14\%$ (small) and $83.27\%$ (large), respectively.  %The corresponding histograms of the PSs (denoted $e(X)$) are shown in Figure \ref{fig:ps-toy}. 
	Next, we generated the outcome variable $Y = ZY(1)+(1-Z)Y(0)$, where $Y(0) = -X_1+2X_2+\varepsilon$ and $Y(1) = Y(0)+(X_1+X_2)^2$, with $\varepsilon\sim \mathcal{N}(0,1)$.
	
% 	\begin{figure}[ht]
% 		\begin{center}
% 			{\includegraphics[trim=0 5 0 10, clip, width=0.85\textwidth]{ps_toy.png}}
% 		\end{center}
% 		\caption{PS distrib(X_1+X_2)^2utions for small (left) and large (right) proportion of treated participants.}\label{fig:ps-toy}
% 	\end{figure}

	Using the potential outcomes,  we calculated  ATE = $E[Y(1)-Y(0)]$  and estimated the PSs via a logistic regression model. Then, we determined ATT, ATC, ATO, ATM, and ATEN using formula \eqref{eq:estimand_gen} and those in Table \ref{tab:wgts_summary}. The results, presented in Table \ref{tab:toy_effects}, indicate the   the variances of PSs of the treated to the control groups are similar ($r\in [0.5,2]$),  per the rule of thumb of  \cite{rubin2001using}. Furthermore, when $p$ is small ($17.14\%$), the ATE is closer to ATC whereas the equipoise estimands (ATO, ATM and ATEN) are closer to ATT. On the other hand, when $p$ is large ($83.27\%$), ATE is closer to ATT where equipoise estimators (ATO, ATM and ATEN) are closer to ATC. 
	
	\begin{table}[htbp]
		\begin{center}
		\caption{Causal effects under two different treatment allocations $p$}
		\label{tab:toy_effects}
		\begin{threeparttable}
			\begin{tabular}{ccccccccccccccccccccccc}
				\toprule
				&&&	&  \multicolumn{3}{c}{IPW estimands} &  & \multicolumn{3}{c}{Equipoise estimands}  \\\cmidrule(lr){5-7}\cmidrule(lr){9-11}
				Scenario & $p$ &		$r$	&	& ATE & ATT & ATC & &  ATO & ATM & ATEN  \\ \midrule

				1 & $17.14\%$& $1.29$ 	&& 54.75  & 76.54 & 49.62 & & 69.87 & 75.88 & 66.08  \\
				2 & $83.27\%$& $0.63$ 	& & 54.75 & 58.01 & 35.22  &  & 38.79 & 35.33 & 42.05  \\
				\bottomrule
			\end{tabular}
			\begin{tablenotes}
			\setlength\labelsep{0pt}
				\footnotesize
				\item $r$: ratio of variances of the propensity scores  (treatment vs. control)
			\end{tablenotes}
		\end{threeparttable}
		\end{center}
	\end{table}
	
	\subsubsection{Is it really a surprise?}
	In part, the above results on the ATE are  not surprising. The ATE depends on how the treatment effects (ATT and ATC) vary given the treatment groups  and, more importantly, it weighs more one or the other depending on which group has the higher proportion of participants.  The relationship $ATE = pATT+ (1-p)ATC$,  indicates what  matters more when there is imbalance in treatment allocations. The influence of $p$ in estimating  ATE is far from trivial; it  can indeed have a substantial impact. When the treatment effect is heterogeneous and the proportion of treated participants $p$ is small (resp. large), ATE puts more  emphasis on the subpopulation of control (resp. treated) participants. %For instance, for small $p$, the ATE is closer to the treatment effect on this subpopulation, the average treatment effect on the controls (ATC). 
 However, it is questionable that  whether the ATC is the estimand that matters the most when dealing with a smaller number of treated participants. Usually, in this situation, most study designs and analytical methods target instead the ATT, which has a completely different interpretation than the ATC  \citep{austin2021applying,greifer2021choosing}. In addition, in finite sample (as we will demonstrate in the simulations) and when $p$ is small, the estimate of the ATC is less precise, as it relies on outcomes from a small number of treated participants (comparators) to serve as counterfactuals for the larger percentage of controls participant. Similar observations can be made when $p$ is large, where ATE is now closer to ATT, even though the estimation of ATT relies on outcomes from a smaller fraction of control participants as conterfactuals for the treated participants. 
 
    This behavior of ATE is contrary to what we usually aim for, if we make an analogy with a rare-exposure study, where we use a large pool of controls to find better matches for the limited number of exposed cases. This is also counter-intuitive to our commom practice when using matching (either on the PS or on some covariates) and our target is the ATT or an ATT-like estimator  \citep{stuart2010matching}. We match efficiently when we have at least as many controls as treated participants.  Therefore, relying heavily on  estimating the (population)  ATE when we have a smaller percentage of treated participants may systematically fails to reach our target of inference. Fortunately, as indicated in Table \ref{tab:toy_effects}, that goal can be achieved by purposefully targeting a specific subpopulation of patients using a different set of weights.
	
	\subsubsection{What if we target a different population of participants?}
	The results obtained from ATO, ATM, and ATEN are not really surprising either.   \cite{li2018balancing} alluded to  similar results for  overlap weights when explaining how flexible and data-adaptive they are vis-\`a-vis values of the PSs $e(X)$. They noted that  the overlap weights $(w_1(x), w_0(x))\propto (1-e(x), e(x)) \approx \left(\frac{0.25}{e(x)}, ~\frac{0.25}{1-e(x)} \right)$ when $e(X)\approx 0.5,$ in which case they are proportional to IPW weights for the ATE. For smaller $e(X)$, $(w_1(x), w_0(x))\approx \left(1, ~\frac{e(x)}{1-e(x)} \right)$ in which case $w_1(x)$ and $w_0(x)$ resemble ATT weights. Finally, for  larger $e(X)$, the overlap weights $(w_1(x), w_0(x))\approx \left(\frac{1-e(x)}{e(x)}, ~1 \right) $ in which case they resemble ATC weights. 	
	We naturally conjecture that, under some regularity conditions,  $p=P(Z=1) = E\left[ E(Z|X)\right] =E(e(X))$ (i.e., the first moment of the PSs $e(X)$)  might be sufficient to reflect how overlap weights weigh ATT and ATC, thus extending the above observations from  \citet{li2018balancing}. 
	
	 \cite{matsouaka2020framework} formalized and proved that asymptotically this phenomenon generalizes not only to ATO, but also to ATM and ATEN. To do so, they showed that ATO, ATM, and ATEN estimators are expected to yield similar estimated results. Besides, they demonstrated the direct links between the density distributions of PSs within each treatment group. Finally, they established the connections between the proportion of treated participants $p=P(Z=1)$, the ratio of variances of the PSs $r$, and whether one might expect to obtain, in theory, a result  close to ATT or ATC when using ATO, ATM, or ATEN. For instance, in the case of overlap weights (as well as all other equipoise estimators), Table \ref{tab:toy_effects} confirms that, when $p$ is small,  the emphasis is on the subpopulation of treated participants  and the estimand is indeed close to ATT. 
 
    While the results from Table \ref{tab:toy_effects} align with the theoretical results in the literature (see, for instance, \cite{matsouaka2020framework}), we still have a limited understanding of what ATO, ATM, ATEN truly identify and estimate in the presence of treatment allocation imbalances and under finite sample sizes. The ratio $r$  also appears to play a role  to determine whether these estimands  lean heavily toward ATT or not. In addition, with finite sample sizes, estimation of ATE using IPW weights when the proportion of participants $p$ is small (or large) is challenging by the issues of limited overlap and the influence of extreme weights. Hence, we also need to explore finite-sample statistical and numerical properties of estimating ATE via IPW weights as well as ATO, ATM, and ATEN. Therefore,  we will need to run simulation studies to find out, whether ATE and equipoise weight estimators lead towards two different alternative estimands (ATT vs. ATC) in finite sample size when the proportion of treated participants $p$ is small or large.
	%%%%%%%%%%%%%%%%%%%%%%%%%%%%%%%%%%%%%%%%%%%%%%%%%%%%%%%%%%%%%%%%%%%%%%%%%%%%%%%%%%%%%%%%%%%%%

		%%%%%%%%%%%%%%%%%%%%%%%%%%%%%%%%%%%%%%%%%%%%%%%%%%%%%%%%%%%%%%%%%%%%%%%%%%%%%%%%%%%%%%%%%%%%%%%%%%%%%%%%%%%%%%%%%%%%%%%%%%%%%%
	%%%%%%%%%%%%%%%%%%%%%%%%%%%%%%%%%%%%%%%%%%%%%%%%%%%%%%%%%%%%%%%%%%%%%%%%%%%%%%%%%%%%%%%%%%%%%%%%%%%%%%%%%%%%%%%%%%%%%%%%%%%%%%%%%%%%%%%%%%%%%%%%%%%%%%%%%%%%%%%%%%%%%%%%%%%%%%%%%%%%%%%%%%%%%%%%%%%%%%%%%%%%%%%%%%%%%%%%%%%%%
	\section{Balancing weights}\label{sec:bal.wgts}
	\subsection{Theoretical background, notation and assumptions} \label{subsec:notation}
	Let us denote $Z=z$  the treatment indicator ($z=1$ for treated and  $z=0$ for control), $Y$ a continuous outcome, and ${X}=(X_0, X_1, \ldots, X_p)$ the matrix of baseline covariates, where $X_0=(1,\ldots, 1)'$.  The observed data $\mathcal{O}=\{ (Z_i, X_i, Y_i): i=1\dots, N \}$ are a sample of $N$ participants drawn independently from a large population of interest. 
	We adopt the potential outcome framework \citep{neyman1923applications, imbens2015causal}, and assume that  for any randomly chosen participant in the population, there exists a pair of potential outcomes  $(Y(0), Y(1))$,  where $Y(z)$ is the   outcome that would been observed if, possibly contrary to fact, the individual were to receive treatment $Z=z$. 	In addition, we make the following assumptions: 
 \begin{enumerate}
     \item Consistency: $Y=ZY(1)+(1-Z)Y(0),$ i.e., for each individual,  the observed outcome $Y$  matches the potential outcome  $Y(z)$ for the treatment  $Z=z$ they received.
     \item Stable-unit treatment value assumption (SUTVA): there is only one version of the treatment and the potential outcome $Y(z)$ of an individual does not depend on another individual's received treatment, as it is the case when participants' outcomes interfere with one another  \citep{rosenbaum1983central}.
     \item Unconfoundness: $E[Y(z)|X]=E[Y(z)|X,Z=z],$ $z=0,1$.%, i.e., the potential outcomes are independent of $Z$ given the vector of covariates ${X}$.
 \end{enumerate}	
 The PS, $e({x})=P(Z=1|{X=x})$, is the conditional probability of treatment assignment  given the observed covariates. Under the unconfoundness assumption, the PS is a balancing score since $X\bigCI Z|e({X})$. This implies that participants with the same PS have similar distributions of their  observed baseline covariates ${X}$ regardless of their treatment assignment  \citep{rosenbaum1983central,rosenbaum1984reducing,rubin1997estimating}. Therefore, instead of controlling for the whole vector of multiple covariates $X$ to estimate treatment effects, one can leverage this property of the PS $e({X})$ to derive unbiased estimators of the treatment effect. 	
	Since the PS $e(X)$ is usually unknown  in non-randomized studies, we estimate it by postulating a model
	$e({X};{\beta})=P(Z=1|{X}; {\beta})$, for some parameter vector $\beta$. 
	
	More often, the goal is to estimate the average treatment effect (ATE) $\tau=\displaystyle E[\tau(X)]$ from the data, where  $\tau(x)=E[Y(1)-Y(0)|X=x]$ is the conditional average treatment effect (CATE). %, conditional on covariate values $X=x$. %Note that when the CATE is constant, then ATE =  CATE and the treatment is said to be homogeneous. Otherwise, it is considered heterogeneous. In the latter case, we may focus our interest on the effect of treatment on  specific subgroups of the covariate distributions. Therefore, 
	In this paper, we consider the weighted average treatment effect (WATE) 
	\begin{align}\label{eq:estimand_gen}
		\tau_{g}&=\displaystyle \frac{\displaystyle E[g(X)\tau(X)]}{\displaystyle E(g(X))},
		%=C^{-1}\!\displaystyle {\displaystyle\int \tau(x)f(x)g(x)\mu(dx)},~~\text{with}~~ C={\displaystyle\int g(x)f(x)\mu(dx)}
		%&=\displaystyle \frac{\displaystyle\int \mathbb{E}[(Y(1)-Y(0))|{X}={x}]f(x)g(x)\mu(dx)}{\displaystyle\int g(x)\mu(dx)}
	\end{align} 
	where $g(x)$ is a known selection (or tilting) function and $\tau$ corresponds to the identity function $g(x)  = 1$. %of the covariates with respect to a base measure $\mu,$ which  we have equated to the Lebesgue measure, without loss of generality.
 
    The estimand $\tau_{g}$ encompasses a large class of causal estimands, depending of the function g  \citep{li2018balancing,hirano2003efficient, crump2006moving}. 
    The selection function  $g(x)$ delimits and specifies the target subpopulation defined in terms of the covariates $X$.   Higher values of $g(x)$ indicates the regions of the covariates space with higher weights in the target subpopulation.  In addition, the function $g$ helps with  the interpretation of the corresponding treatment effect and characterizes the related weights (see Table \ref{tab:wgts_summary}).

\begin{figure}[ht]
    	\begin{center}
    	\includegraphics[trim=0 5 0 5, clip, width=0.6\textwidth]{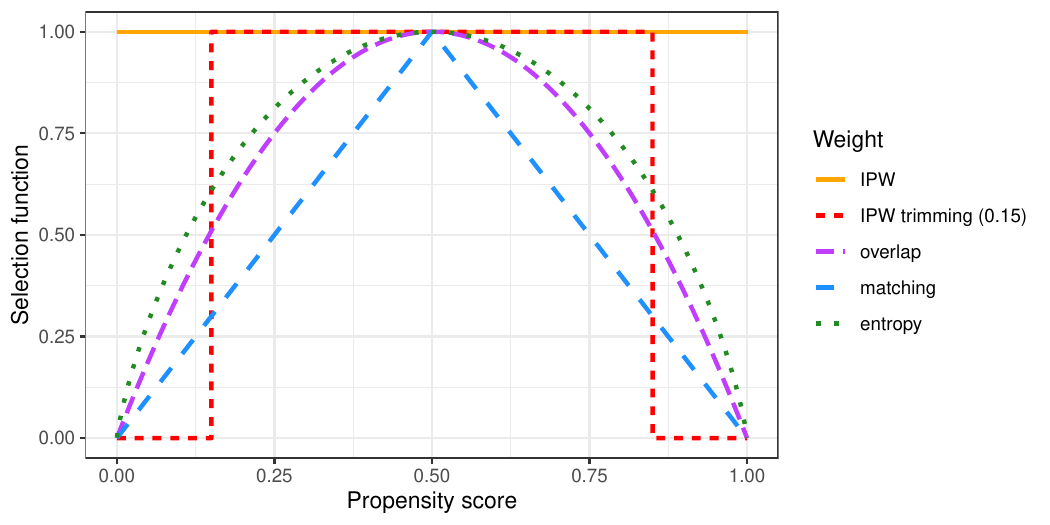}  
    	\caption{Selection functions for IPW (with and without trimming) and  equipoise estimands.}\label{fig:etewts}
       	\end{center}
    \end{figure}

The equipoise selection functions assign higher weights to observations around the middle of the PS spectrum, at $e(x)= 0.5$, and gradually and smoothly downweight the contributions of those at the tails of PS spectrum (as shown in  Figure \ref{fig:etewts}), which bypass the need to enforce positivity assumption. As demonstrated by \cite{matsouaka2020framework}, the  selection functions for clinical equipoise are closely related. This  explains why, in practice, one can expect to obtain statistically similar results with ATO, ATM, or ATEN, as they all target similar populations of participants, i.e., those for whom there is clinical equipoise. % The use of the selection function $g(x)=e(x)\left(1-e(x)\right)$ date back from the work of \cite{crump2006moving} on the optimally weighted average treatment effect (OWATE), which we refer to as the average treatment effect on the overlap (ATO), in this paper, following the nomenclature of  \citet{li2018balancing}. 
%Finally, the equipoise estimators target estimands that are different from the typical ATE. They control the unduly influence of outliers while improving efficiency, increase the balance in the distributions of the observed covariates betweem treatment groups, and stress on the internal validity of the study at hand like with randomized clinical trials  
Additional insights on the equipoise estimators can be find in the references hereafter \citep{li2018addressing, matsouaka2020framework, thomas2020overlap}. %Their corresponding subpopulations can be described in terms of the covariates values as in shown in  \cite{thomas2020overlap}. 

	\begin{table}[htbp] \small
		\caption{Examples of selection function, target population, causal estimand, and weights}\label{tab:wgts_summary}
		\begin{center} 
			\begin{threeparttable}
				\begin{tabular}{rccccccccccccccccccccc}
					\toprule
					Target population  & Selection function $g(x)$ & Estimand  &   Method\\ \cmidrule(lr){1-4} 
					overall &   $1$  &  ATE  &   IPW          \\%\addlinespace
					treated &  $e(x)$   &  ATT  & IPW Treated \\%\addlinespace 
					control &  $1-e(x)$  &       ATC  &       IPW Control \\ %\addlinespace
					\multirow{1}{*}{\makecell{trimmed}} &   $I_{(\alpha_1, \alpha_2)}(x)=\mathds{1}({\{\alpha_1\leq e(x)\leq 1-\alpha_2\}})$     &  OSATE  & IPW Trimming\\ %\addlinespace
					\multirow{1}{*}{\makecell{truncated}} &   $I_{(\alpha_1, \alpha_2)}(x) + J_{\alpha_1}(e(x))^zJ_{1-\alpha_2}(1-e(x))^{1-z}$     &     &       IPW Truncation\\ 
                equipoise &  $e(x)\left(1-e(x)\right)$  &       OWATE$\big/$ATO   &       OW\\ %\addlinespace
				equipoise &  $\min\{e(x), 1-e(x)\} $  &     ATM     &      MW\\
				equipoise &  $-[e(x)\ln(e(x))+(1-e(x))\ln(1-e(x))] $  &     ATEN     &      EW\\
					\bottomrule
				\end{tabular} 
				\begin{tablenotes}
					\footnotesize
					
					%	\item $I_{\alpha}(x)=\mathds{1}({\{\alpha\leq e(x)\leq 1-\alpha\}})$; $\mathds{1}(\boldsymbol{\cdot})$ is the standard indicator function; 
					\item %$g$ is the selection function;  
					$\mathds{1}(\boldsymbol{\mathcal{.}})$ is the  indicator function and %,i.e., $\mathds{1}(\boldsymbol{\mathcal{C}})=1$ when $\boldsymbol{\mathcal{C}}$ is satisfied and 0 otherwise; %$T_{\alpha}^z(x)=I_{\alpha}(x) + zJ_{\alpha}(e(x))+ (1-z)J_{\alpha}(1-e(x))$ 
					$J_{\alpha}(e(x))={\alpha}^{-1} e(x){\mathds{1}({\{ e(x)<\alpha\}})}$, with $\alpha, \alpha_1, 1-\alpha_2 \in (0, 0.5),$  $z\in \{0,1\}$. %for $\alpha \in (0, 0.5),$ where
					%	\item  $T_{\alpha}^z(x)=I_{\alpha}(x) + zJ_{\alpha}(e(x))+ (1-z)J_{\alpha}(1-e(x))$.
					% The related weights are $ w_z(x)=\widehat g(x) \widehat e(x)^{-z}(1-\widehat e(x))^{z-1}$.	
					%   
					%			\item   $h(x)=e(x)\left(1-e(x)\right)$;  $u(x)=\min\{e(x), 1-e(x)\};$  	$v_K(x)=\min\left\{1, Ku(x)\right\},$  $K\geq 1.$
				\end{tablenotes}
			\end{threeparttable}
		\end{center} 
	\end{table} 
	 		 
	The weighted treatment effect $\tau_{g}$ can be estimated by	the weighted, H\'ajek-type  \citep{hajek1971comment} estimator
	\begin{equation}\label{eq:li_est}
		\widehat\tau_{g}= \displaystyle\sum_{i=1}^{N}\left[ \frac{Z_i\widehat  w_1(x_i)}{N_{\widehat w_1}}-\frac{	(1-Z_i)\widehat w_{0}(x_i)}{N_{\widehat w_0}}\right] Y_i,~~~\text{with}~N_{\widehat w_k}=\displaystyle\sum_{i=1}^{N} Z_i^k(1-Z_i)^{1-k}\widehat w_k(x_i),~k=0, 1
	\end{equation} 
	where  $\widehat w_k(x) =\widehat g(x) \widehat e(x)^{-k}(1-\widehat e(x))^{k-1},$ with  $\widehat g(x)=g(x; \widehat \beta)$, $~\widehat e(x) =e(x;\widehat \beta),$ and  $\widehat \beta$ an estimator  of  $\beta$  \citep{li2013weighting, li2018balancing}. 
	The weights $w_k(x)$  balance the weighted distributions of the covariates between the two treatment groups  \citep{li2018balancing}. Thus, the name {\it balancing weights}.

	We also consider augmentations of the estimators $\widehat\tau_g$ based on  outcome regression models. For ATE, ATT and ATC, they correspond to the following augmented  estimators, with $\widehat w_k(x) =\widehat g(x) \widehat e(x)^{-k}(1-\widehat e(x))^{k-1},$ $k=0, 1$: 
	     \allowdisplaybreaks\begin{align}\label{eq:ate_dr}
		\widehat \tau_{\text{ATE}}^{\text{dr}}&=\sum_{i=1}^{N}\left[\frac{Z_i\widehat  w_1(x_i)\left\lbrace Y_i- \widehat m_1(x_i)\right\rbrace}{{N}_{\widehat w_1(x_i)}} - \frac{(1-Z_i)\widehat w_0(x_i)\left\lbrace Y_i- \widehat m_0(x_i)\right\rbrace}{N_{\widehat w_0(x_i)}} %\nonumber\\&
		+\frac{\left\lbrace  \widehat m_1(x_i) -  \widehat m_0(x_i)  \right\rbrace}{N} \right], ~~g(x)=1; \nonumber\\
		%\label{eq:ate_dr}
		\widehat \tau_{\text{ATT}}^{\text{dr}}&=\sum_{i=1}^{N}\left[\frac{Z_i\left\lbrace Y_i- \widehat m_0(x_i)\right\rbrace}{\displaystyle{N}_{\widehat w_1(x_i)}} - \frac{(1-Z_i)\widehat  w_0(x_i)\left\lbrace Y_i- \widehat m_0(x_i)\right\rbrace}{\displaystyle N_{\widehat  w_0(x_i)}}\right], ~\text{with}~ g(x)=\widehat e(x_i); \\
		\label{eq:att_dr}
		\widehat \tau_{\text{ATC}}^{\text{dr}}&=\sum_{i=1}^{N}\left[\frac{Z_i\widehat  w_1(x_i)\left\lbrace Y_i- \widehat m_1(x_i)\right\rbrace}{\displaystyle N_{\widehat w_1(x_i)}} - \frac{(1-Z_i)\left\lbrace Y_i- \widehat m_1(x_i)\right\rbrace}{\displaystyle N_{\widehat  w_0(x_i)}}\right], ~\text{where}~ g(x)=(1-\widehat e(x_i)) \nonumber%\label{eq:atc_dr}
	\end{align}
        where we postulate parametric models $\widehat m_z(X) = m_z(X;\widehat \alpha_z)$, i.e., the outcome regression models, with parameter $\widehat \alpha_z$ to estimate $m_z(X) = E[Y(z)|X],~z=0,1$ \citep{matsouaka2020framework, matsouaka2023variance}. 
 
	For equipoise estimands (ATO, ATM and ATEN), we consider the augmented estimators  \begin{align*}%\label{eq:tau_aug}
		\widehat\tau_{g}^{\text{aug}} = & \displaystyle
		\sum_{i=1}^{N}\left[ \frac{Z_i\widehat  w_1(x_i)\{Y_i-\widehat m_1(x_i)\}}{{N}_{\widehat w_1(x_i)}}-
		\frac{(1-Z_i)\widehat w_{0}(x_i) \{Y_i-\widehat m_0(x_i)\}}{\displaystyle{N}_{\widehat w_0(x_i)}}\right]+\displaystyle
		\sum_{i=1}^{N}g(x_i)\{\widehat m_1(x_i) - \widehat m_0(x_i)\}\big/\displaystyle\sum_{i=1}^{N} g(x_i). 
	\end{align*}  
While the estimators \eqref{eq:ate_dr} are doubly-robust (i.e., they are consistent when at least one of the PS or the outcome regression models is correctly specified), the equipoise estimators $\widehat\tau_{g}^{\text{aug}}$ are not. The latter are consistent only if the  PS model is correctly specified \citep{matsouaka2020framework,mao2018propensity}. %This is why we distinguish them to the DR estimators for ATE, ATT and ATC in (\ref{eq:ate_dr})--(\ref{eq:atc_dr}) above.

For all the estimators, we derived their close-form sandwich variance estimators based on the asymptotic normal approximation and M-theory (see Supplemental Material \ref{apx:proofs}). 
	The performance of the weights $\widehat w_k(x),~k=0,1,$ can be assessed through the effective sample size (ESS)  
	\begin{align*} %\label{eq:effssize}
		\widehat{ESS}= \left( \displaystyle\sum_{i=1}^{N} \widehat w(x_i)^2\right) ^{-1} \left( \displaystyle\sum_{i=1}^{N} {\widehat w(x_i)}\right)^{2},~~ \text{where} ~~\widehat w(x_i)=z_i\widehat w_1(x_i)+(1-z_i)\widehat w_0(x_i), ~~ z_i=0,1.
	\end{align*}  
	The ESS approximates the number of independent observations, drawn from a simple random sample from the source population, that are needed to obtain an estimate with a sampling variation similar to that of the weighting observations in the target sample \citep{mccaffrey2004propensity}.	 It helps characterize the variance inflation or precision loss due to weighting \citep{zhou2020propensity}.   	 
			% Alternatively, one can also estimate directly the variance inflation based on  a “design effect” approximation  \citep{kish1985survey}
			% \begin{align*}
			% 	\widehat{\text{VI}}=(1/N_1+1/N_0)^{-1}  \sum_{z = 0}^1 \left[ \left( \displaystyle\sum_{i=1}^{N_z} {\widehat w_z(x)}\right)^{-2}\displaystyle\sum_{i=1}^{N_z} \widehat w_z(x_i)^2\right], 
			% \end{align*} 
   % where $N_z$ is the number of participants in group $Z=z \in \{0,1\}.$
	
	%%%%%%%%%%%%%%%%%%%%%%%%%%%%%%%%%%%%%%%%%%%%%%%%%%%%%%%%%%%%%%%%%%%%%%%%%%%%%%%%%%%%%%%%%%%%%%%%%%%%%%%%%%%%%%%%%%%%%%%%%%%%%%%%%%%%%%%%%%%%%%%%%%%%%%%%%%%%%%%%%%%%%%%%%%%%%%%%%%%%%%%%%%%%%%%%%%%%%%%%%%%%%%%%%%%%%%%%%%%%%%%%%%%%%%%%%%%%%%%%%%%%%%%%%%%%%%%%%%%%%%%%%%%%%%%%%%%%%%%%%%%%%%%%%%%%%%%%%%%%%%%%%%%%%%%%%%
	%%%%%%%%%%%%%%%%%%%%%%%%%%%%%%%%%%%%%%%%%%%%%%%%%%%%%%%%%%%%%%%%%%%%%%%%%%%%%%%%%%%%%%%%%%%%%%%%%%%%%%%%%%%%%%%%%%%%%%%%%%%%%%%%
	\section{Simulation}\label{sec:simulations}
	%In finite sample, in addition to the relationship $ATE= pATT +(1-p) ATC$, where $p$ is the proportion of treated participants, the IPW estimator of ATE often assigns extreme inverse probability weights to some treated (resp. control)  participants when $p$ is small (resp. large), which complicates the theoretical justification we provided in Section \eqref{sec:illustration}. Therefore, 
	% We conducted a series of Monte Carlo simulations to assess the finite-sample performance of the aforementioned estimators and evaluate the role that $p$ and $r$ (ratio of variances of PSs of the treated to the control group) play, while accounting for the impact of extreme inverse probability weights either through IPW trimming or the equipoise treatment effect estimators (OW, MW, EW).
	
	%We conduct our  expectation the  to different causal estimands. 
	
	Our data generating process (DGP) is similar to that of  \citet{li2021propensity}. %With this chosen DGP, the presence of extreme IPW weights will be related to the proportion of participants in the treatment group as the PSs are defined and calculated via logistic regression models, i.e., functions of inverse logit of the linear combination of the covariates. 
		First,  we simulated a superpopulation of $10^6$ individuals under different scenarios for desired proportions $p$ to determine the true values of the estimands under heterogeneous treatment effects. Then, to assess the finite-sample performance of the different estimators, we simulated $M=2000$ independent data sets of size $N=1000$ and allowed the $p$ to vary, as specified below. Within each data set, we estimated ATE (with and without trimming) via IPW, ATT, ATC, and the treatment effects via the equipoise estimators (OW, MW, and EW). For ATE with trimming, we trimmed observations with the PSs that fall outside of the interval $[\alpha, 1-\alpha]$, with $\alpha = 0.05, 0.1$, and $0.15$, respectively. 
  
        We summarized  and interpret the results based on the criteria laid out in Section \ref{sec:performance}. For H\'ajek-type estimators, we only report the results under correctly specified (PS) model. The results under misspecified PS models were similar to those in  \citet{zhou2020propensity} and were not reported here.
	
	\subsection{Data generating process}
	We first generated the covariates $X=(X_0, X_1,\dots, X_7)$ and the treatment assignment $Z\sim \text{Bern}([1+\exp(-X'\beta)]^{-1})$. $X_0=(1,1,\dots,1)$ and $X_4\sim \text{Bern}(0.5), X_3\sim \text{Bern}(0.4+0.2X_4)$, $(X_1, X_2)'\sim \mathcal{N}(\boldsymbol{\mu, \Sigma})$, $X_5 = X_1^2,$ $ X_6 = X_1X_2,$ and  $ X_7 = X_2^2$, where $
	\boldsymbol \mu = (X_4-X_3  + 0.5X_3X_4, X_3-X_4 + X_3X_4)'
	$, $
	\boldsymbol \Sigma = 
	X_3\begin{pmatrix}
		1 & 0.5 \\
		0.5 & 1
	\end{pmatrix} + $  
	$X_4\begin{pmatrix}
		2 & 0.25 \\
		0.25 & 2
	\end{pmatrix}
	$.

	 We selected different values of $\beta=(\beta_0, \beta_1, \ldots, \beta_7)'$ to have a range of  proportions of treated participants $p$ and variances of
PSs (treated vs. control), as shown in Table \ref{tab:coef} (see Supplemental Material \ref{subapx:sim-ps-analysis}).  The distributions of the estimated PSs are shown in Figure \ref{fig:apx-ps-sim}.	Then, we generated the outcomes $Y=ZY(1)+(1-Z)Y(0),$ with $Y(0) = 0.5+X_1+0.6X_2+2.2X_3-1.2X_4+(X_1+X_2)^2 + \varepsilon$ and  $Y(1) = Y(0) + \delta(X)$, with $\varepsilon\sim \mathcal{N}(0, 4)$. 
    We considered both a constant ($\delta(X)=4$) and a heterogeneous ($\delta(X)=4+3(X_1+X_2)^2+X_1 X_3$) treatment effect.  The true average heterogeneous treatment effects are reported in Table \ref{tab:truth-all} (Supplemental Material \ref{subapx:sim-ps-analysis}). %For instance, we calculated the true ATE (resp. ATT) by taking the average of the differences $Y(1)-Y(0)$ for each participant in the entire superpopulation (resp. the subpopulation of those with $Z=1$) out of the $10^6$ observations. The true constant treatment effect for all weighted average treatment effects (WATE) is equal to 4. 
	
	To assess the performance of the augmented estimators, we also analyzed the generated data under misspecified PS and outcome regression (OR) models. We considered 4 specific scenarios of model specification: (i) all the models are true, (ii) the PS model is true, (iii) the OR models are true, and (iv) all the models are misspecified. To misspecify the models, we removed $X_1^2, X_2^2$ and $X_1X_2$ from $Z$ and $Y$ to only use $(X_0,X_1,\dots,X_4)$ instead.
	
	\subsection{Measures and performance criteria}\label{sec:performance}
	%We estimated the ATE, trimmed ATE (with the propensity score trimmed at $\alpha = 0.05, 0.1$, and $0.15$, respectively), the ATT, and ATC all based on IPW as well as the equipoise treatment effects ATO, ATM and ATEN. %We included IPW trimming to compare its performance to that of equipoise estimators under different parametric model specifications.  
	%\citet{zhou2020propensity} have actually shown that, compared to H\'ajek-type estimators of ATE or their trimming versions, equipoise estimators are more robust to model misspecifications. Thus, we contribute to the literature by examining the whether we have the same results despite the possible sensitivity of the augmented estimators to model specifications. This is important since the augmented estimators for ATE, ATT, and ATC defined by \eqref{eq:ate_dr}--\eqref{eq:atc_dr} are actually doubly robust (DR) estimators, while those for the equipoise estimands, defined by equation \eqref{eq:tau_aug}, are not  \citep{mao2018propensity, matsouaka2020framework}. We considered 4 specific scenarios of model specification to evaluate augmented estimators: both the PS and OR models are correctly specified, only the PS model is correctly specified, only the OR model is correctly specified, and both the PS and OR models are misspecified. 
    %For H\'ajek-type estimators, we only provide results under correctly specified (PS) model to evaluate the impact of important imbalances in treatment allocation. Corresponding results under misspecified PS models were similar to those in  \citet{zhou2020propensity} and were not reported here.
	
	To evaluate the performance of different estimators as well as their sensitivity to model misspecifications, we considered the following measures: the absolute relative percent bias, ARBias$ =100\%(\widehat\tau_g-\tau_g)/\tau_g$; the relative root mean square error  RRMSE = $[(\widehat\tau_g-\tau_g)/\tau_g]^2$; and the coverage probability (CP) for 95\% Wald confidence intervals, i.e., the proportion of times $\tau_g$ was inside of its estimated confidence interval. For both ARBias and RRMSE, the smaller the measure, the better while the CP is considered significantly different from the nominal 95\% coverage level if it is outside of the interval $[0.94, 0.96].$  
 
    We also reported, in Tables \ref{tab:md1-all}--\ref{tab:md6-all} (Supplemental Material \ref{subapx:sim-allres}), other measures such as the root mean square error RMSE = $(\widehat\tau_g-\tau_g)^2$  and the relative efficiency  RE = SD/SE,  where SD is the empirical standard deviation of $\widehat\tau_g$ and SE is the estimated standard error of a $\widehat\tau_g$ from its estimated sandwich variance. %We also present RMSE in tables since it is also informative to measure precision within each method.   
	%We used the same criteria to investigate the corresponding estimates based on the augmented estimators $\tau_g^{\text{aug}}$. 
    Finally, we considered how close estimates of  ATE  and equipoise estimates were from those of ATT and ATC, which allows us to answer our main question of interest: what are we weighting for?  
	%%%%%%%%%%%%%%%%%%%%%%%%%%%%%%%%%%%%%%%%%%%%%%%%%%%%%%%%%%%%%%%%%%%%%%%%%%%%%%%%%%%%%%%%%%%%%%%%%%%%%%%%%%%%%%%%%%%%%%%%%%%%%%%%
	%%%%%%%%%%%%%%%%%%%%%%%%%%%%%%%%%%%%%%%%%%%%%%%%%%%%%%%%%%%%%%%%%%%%%%%%%%%%%%%%%%%%%%%%%%%%%%%%%%%%%%%%%%%%%%%%%%%%%%%%%%%%%%%%
	\subsection{Results}
	
	For brevity, we only present the results from a few models (Models 1, 2, 3, and 6) in this section. The complete and exhaustive simulation results for all the scenarios considered are provided in  Supplemental Material \ref{apx:simulation}, including the average ESS (by treatment groups) we obtained from the different estimators  (see \ref{tab:ess}). 
 
    The results from Model 1 (constant treatment effect) are reported to showcase the variability in estimating both ATE and ATC under small $p$ (10.05\%).  We also show the results under heterogeneous  treatment effects from Models 2, 3 and 6, where $0.5\leq r\leq 2$ on average, but $p$ ranges from 20.77\% to 79.59\%.
	\begin{figure}[htbp]
    \begin{center} {\includegraphics[trim=5 5 9 50, clip, width=0.55\textwidth]{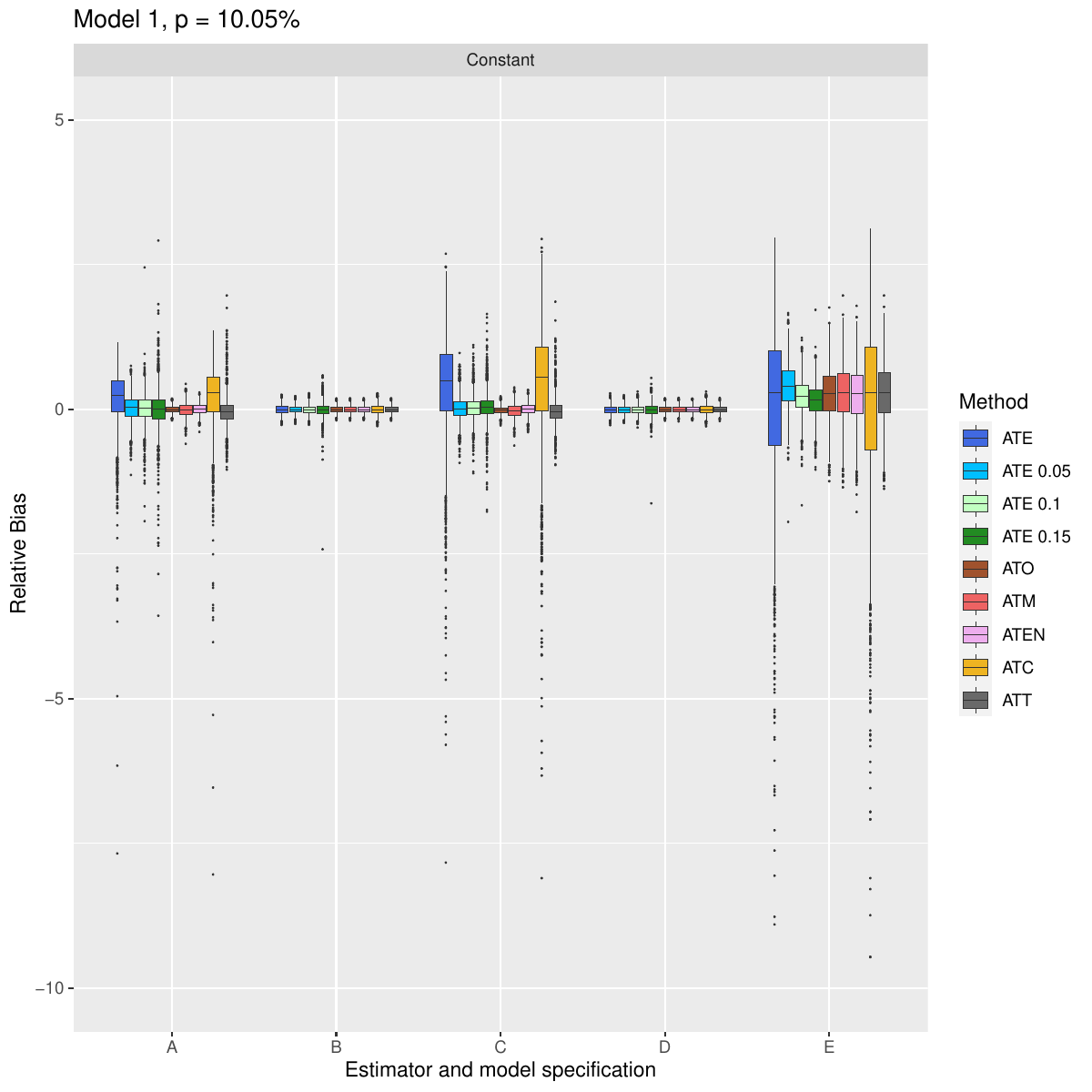}}
    \caption{Relative bias under constant treatment effect (Model 1: proportion of treated = 10.05\%)}\label{fig:RBias-main_Model1}
    \end{center}
    {\tiny A: H\'ajek-type (weighted) estimator; B (resp. C, D, and E): augmented estimator, with both the PS and OR models correctly specified (resp. only the PS model correctly specified, only the OR model correctly specified, both the PS and OR models misspecified).} 
  \end{figure}

  \begin{figure}%[htbp]
		\begin{center}
			{\includegraphics[trim=21 10 20 10, clip,  width=0.60\textwidth]{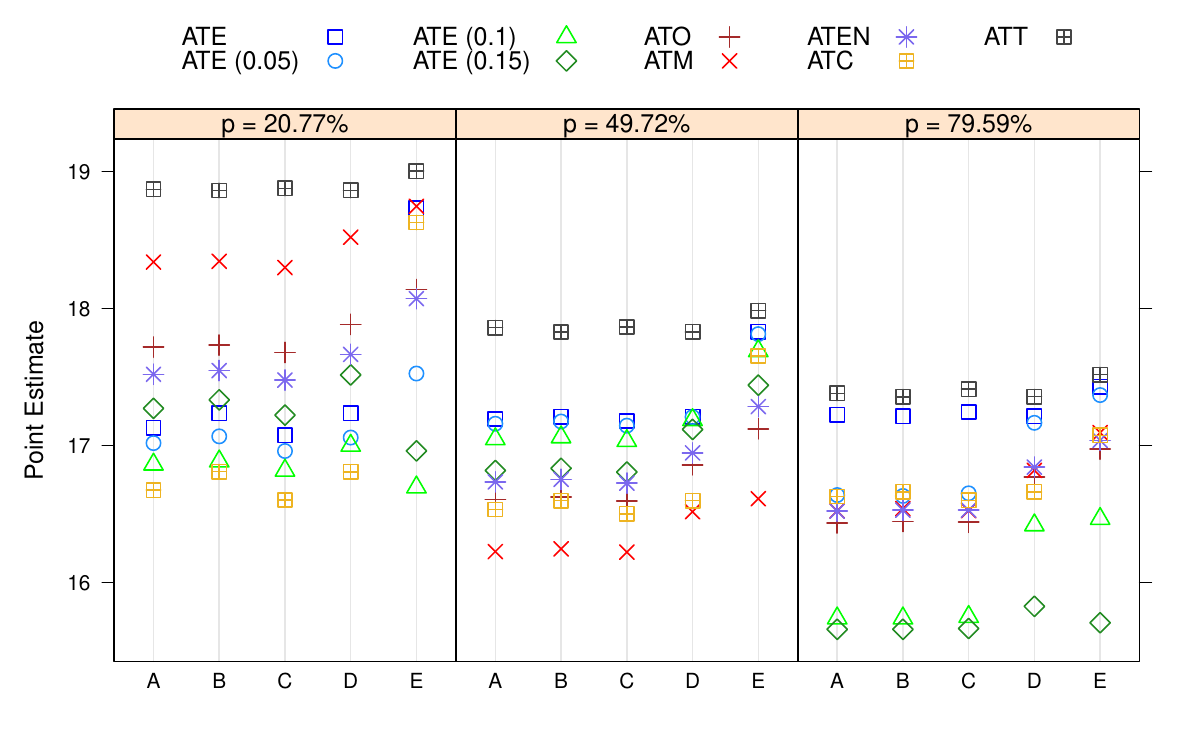}}
		\end{center}\vspace{-0.9cm}
		{\tiny  A: H\'ajek-type (weighted) estimator; B (resp. C, D, and E): augmented estimator, with both the PS and OR models correctly specified (resp. only the PS model correctly specified, only the OR model correctly specified, both the PS and OR models misspecified).%; ATE: average treatment effect; ATE ($\alpha$): ATE by trimming PS $>1-\alpha$ or PS $<\alpha$, with $\alpha=0.05, 0.1, 0.15$; ATO (resp. ATM, ATEN, ATC, and ATT): average treatment effect on the overlap (resp. matching, entropy, controls, and treated) population.
        } 
		\caption{Point estimates  under heterogeneous treatment effects (Models 2, 3, and 6).}\label{fig:xy-hetePE}
	\end{figure}	
 %As mentioned in Section \ref{sec:illustration}, the condition  $0.5\leq r\leq 2$ corresponds empirically  to equal variances of propensity scores between the two groups---following Rubin's rule of thumb  \citep{rubin2001using}. 
	
   Figure \ref{fig:RBias-main_Model1} provides the ARBias for Moddel 1 (constant treatment effect), under different specifications of the PS and outcome models. Throughout the scenarios, the bias were lower for OW, MW and EW---except when both models were misspecified. The sheer number of outlying ARBias values for IPW without trimming in estimating the ATE and ATC clearly indicates how unstable and unreliable these estimators are whenever $p$ is small.
	
	While the biases are smaller overall with true outcome regression models (scenarios B and D), they are higher with the H\'ajek-type estimator (scenario A) and whenever the PS model is misspecified (scenarios C and E). The estimates of ATE and ATC via IPW without trimming are more biased and the magnitude of their related biases are highly variable. This is even more pronounced when both models are misspecified. The estimates under IPW trimming have also similar behavior, although to lesser extent.

    Figure \ref{fig:xy-hetePE} gives the point estimates of all the WATEs under heterogeneous treatment effect of Models 2, 3 and 6.  It can be seen that, more than often, when $p$ is small and equal to 20.30\% (resp. high and equal to 79.59\%) and the variances of the PSs from the two treatment groups are roughly equal, the estimated ATEs via IPW,  without trimming, are closer to ATC (resp. ATT) while the estimated equipoise estimands (ATO, ATM and ATEN) are closer to ATT (resp. ATC)---with both H\'ajek-type (scenario A) and augmented estimators (scenarios B, C, D, and E). When $p$ is about 0.5, these estimands have similar results. This trend remains the same whether the PS and OR models are correctly specified or not. 
 
    The results from IPW trimming were intriguing and did not show any specific trend when we went from the 0.05 threshold to 0.15. When $p=20.77\%$, the point estimate from IPW trimming with the threshold of 0.05, i.e., ATE (0.05)  is always between ATE (0.10) and ATE (0.15), except in the case where both PS and OR models are misspecified. For $p=49.72\%$, ATE (0.10) is between ATE (0.05) and ATE (0.15) as one would expect. Finally, when  $p=79.59\%$, ATE (0.10) and ATE (0.15) are almost the same, while ATE (0.05) coincides with estimates from equipoise estimators in scenarios A, B, and C. In scenarios, D and E, ATE and ATE (0.05) are very close, whereas ATE (0.10) and ATE (0.15) are  far away from the ATT estimate they were expected to be close to. 
	
\begin{figure}%[htp]
		\begin{center}
			{\includegraphics[trim=10 10 20 10, clip,  width=0.65\textwidth]{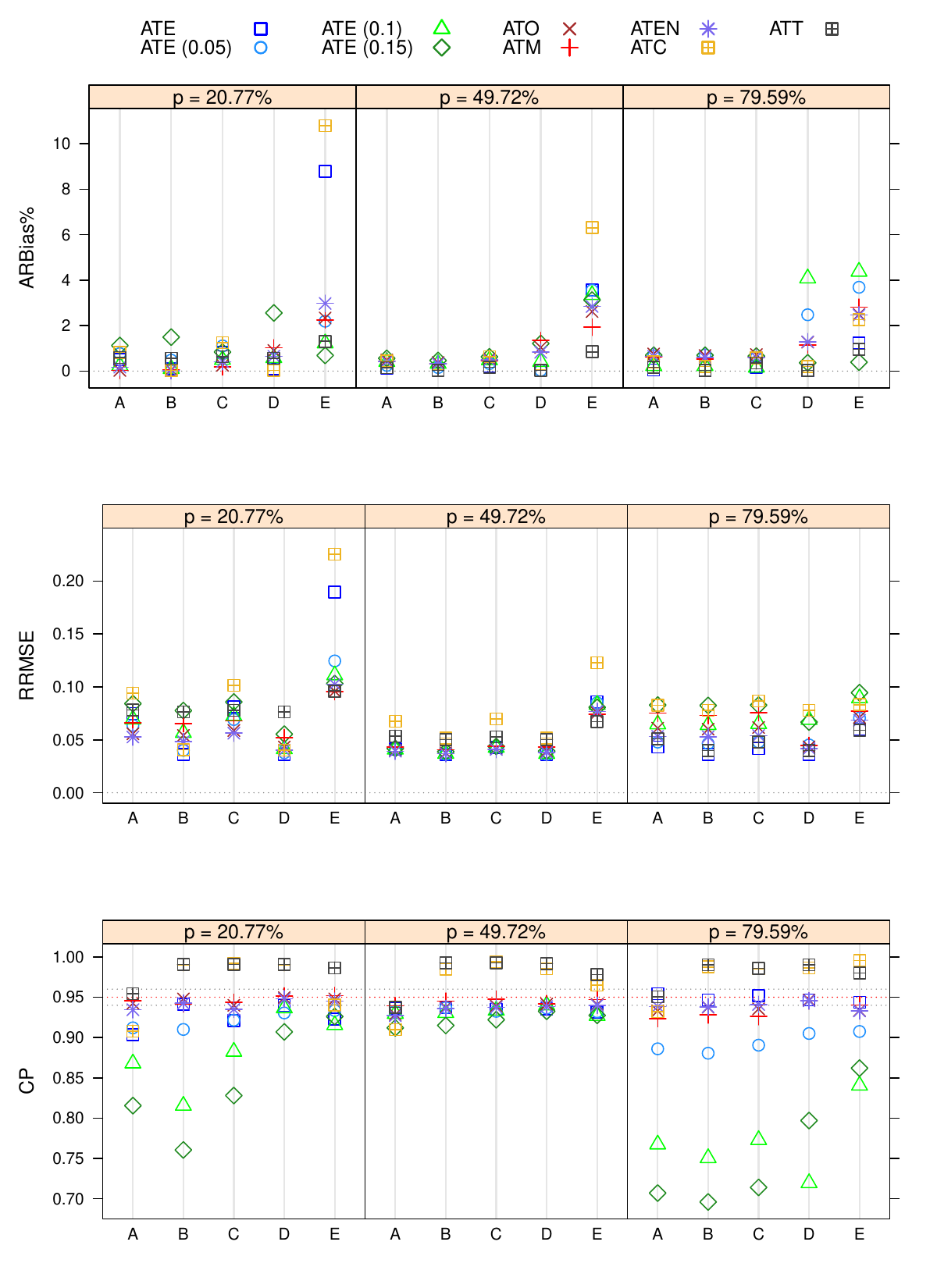}}
		\end{center} \vspace{-0.9cm}
		{\tiny %ARBias\%: absolute relative percent bias; RRMSE: relative root mean square error; CP: coverage probability;
		A: H\'ajek-type (weighted) estimator; B (resp. C, D, and E): augmented estimator, with both the PS and OR models correctly specified (resp. only the PS model correctly specified, only the OR model correctly specified, both the PS and OR models misspecified).} 
		\caption{Bias, mean-squared error, and coverage probability (under Models 2, 3, and 6).}\label{fig:xy-results}
	\end{figure}

	In Figure \ref{fig:xy-results}, while trimming seems to have a good bias-variance trade-off when we look at their RRMSEs, this contrasts with their poor CPs. Trimming appears to systematically underestimate the targeted nominal 0.95 level, with some average CPs being as low as 0.76 for ATE (0.15), when both PS and OR models are correctly specified (scenario B) and $p=20.77\%$, or below 0.92 for ATE (0.05), ATE (0.10), and ATE (0.15) in all scenarios when $p = 79.59\%$. In fact, ATE (0.10) and ATE (0.5) had the worst CPs throughout, with CPs close to 0.70 for ATE (0.15) under scenarios A, B, and C as well as for ATE (0.10) under scenario D---leaving a huge gap between the CPs  and the targeted nominal 0.95 coverage level.
	
     There are some nuances, depending on whether the values of $p$ (small, large or in the vicinity of 50\%). Nevertheless, regardless of $p$, H\'ajek-type estimators as well as  doubly robust ATE and augmented equipoise  estimators (ATO, ATM and ATEN) tend to have small and similar ARBiases and RRMSEs when at least one model is correctly specified, i.e., under scenarios A, B, C, and D. The only exceptions are ATE (0.15), which have a higher ARBias under scenarios A, B, and D when $p=20.77\%$; ATE (0.10), where  ARBias is higher for $p=79.59\%$; followed by ATE (0.05). When the two models are misspecified (scenario E), some estimators of ATE (with and without) have larger ARbiases and RRMSEs. Figure \ref{fig:xy-results} also indicates that the estimates for ATC often have larger ARbiases (in scenarios E when $p<50\%$) and RRMSEs (in most scenarios) than other estimates. 	 
     
	 Furthermore, Figure \ref{fig:xy-results} (bottom panel) show that CPs for equipoise estimators are closer to nominal 0.95 coverage level (i.e., within $[0.94,0.96]$) than ATT, ATC and trimmed ATE estimates.   
  This means that when constructing a 95\% confidence interval, the close-form sandwich variance estimators lead to more efficient equipoise and ATE (without trimming) estimators  than ATC, ATT and trimmed ATE, as expected  \citep{li2018balancing}. The confidence intervals for ATT and ATC are too conservative, while for trimmed ATE are %\st{too narrow} %{\color{blue} This can be said only if we calculated the confidence interval widths.} 
	 extremely overoptimistic. As we alluded to in the previous paragraph, most of their average coverages are below 0.90--far away from the targeted nominal 0.95 coverage level. 
	 
	 Overall, these results suggest that the equipoise estimators (OW, ME, and EW) outperformed the IPW (with and without trimming). When there is no or few outlying observations, the former either weight more ATT or ATC proportional to $c_1(1-p)ATT + c_2pATC$. The weight depends on which of the treatment groups has the smaller proportions of participants---and factors in  attributes $(c_1, c_2)$ of the ratio $r$ of variances of the PSs between the treatment groups. Such a weight assignment goes contrary to what IPW without trimming does. Our simulations also confirm the subjective results one can obtain with trimming: the more we trim the less sure we are as to whether we are improving the bias-variane trade-off or we are making it worse. Finally, the simulations also demonstrate that the augmented equipoise  estimators $\widehat \tau_g^{aug}$ are more robust to model misspecifications, whenever $p$ is extremely large or small and ratio of variances $r\approx 1$. 
	 % %%%% These below are parts of the conclusion %%%%
	 % Thus, to the question ``what are we weighting for?'', we advise that if ATT and ATC are of interest, but there is an important imbalances in the treatment allocations users can consider reporting equipoise and ATE estimates than those based on IPW trimming or ATT or ATC, especially  from the perspective that they have better ARBiases, RMSEs, and confidence intervals---constructed through the close-form sandwich variance estimators.  	
	We reached similar conclusions %with respect to the robustness and performance of variance estimations with 
	with the full set of our simulation results provided in Supplemental Material \ref{subapx:sim-allres}. %which includes all the models 1--6. We also provided the results from the constant treatment effect $\delta(X)=4$. 
 
 In addition, the effective sample sizes in Table \ref{tab:ess} concur also with the overall conclusion of our simulations. Equipoise weights yield better effective sample sizes (to make up for the treatment allocation imbalances),  when estimating treatment effects. For instance, in Model 1 (with about $10\%$ treated participants in the original sample), the ESS was 52.35\%, 69.12\%, 60.32\%,  45.73\% with, respectively, IPW and IPW trimming at 0.05, 0.10, and 0.15, while the ESS was equal to 99.25\%, 101.14\%, and 94.15\% with OW, MW and EW, respectively.

	\section{Data Application}\label{sec:data}
	
	We evaluate racial disparities in the health care expenditure using data from the Medical Expenditure Panel Survey (MEPS) (\url{https://www.meps.ahrq.gov/mepsweb/}). %; the data file name is ``HC-129: 2009 Full Year Consolidated Data File''. A similar data set has been used elsewhere to assess disparities on health care expenditure across racial or ethnic groups  \citep{cook2009adjusting}. 
	% We consider 4 race or ethnic groups, with $9830$ non-Hispanic Whites, $4020$ Blacks or African-Americans, $1446$ Asians and $5280$ Hispanics. %The latter three racial or ethic groups are considered as minorities.  
	We focus on three specific 2-by-2 comparisons: White vs. Hispanic, White vs. Black, and White vs. Asian, with White as the reference group ($Z = 1$) and the minority racial or ethnic group as control ($Z = 0$). %The objective is to evaluate the impact of the racial disparities in health care expenditure.
The proportion $p$ of White participants was, respectively, $65.06\%$ (vs. Hispanic), $70.97\%$ (vs. Black), and $87.18\%$ (vs. Asian). We run separate logistic regression  models to estimate PSs and linear regression models for  the outcome models, with 31 covariates (4 continuous and 27 categorical variables). Details on the data analysis are provided in Supplemental Material \ref{apx:data}.
\begin{figure}[h]
		\begin{center}
			{\includegraphics[trim=5 10 0 5, clip, width=0.7\textwidth]{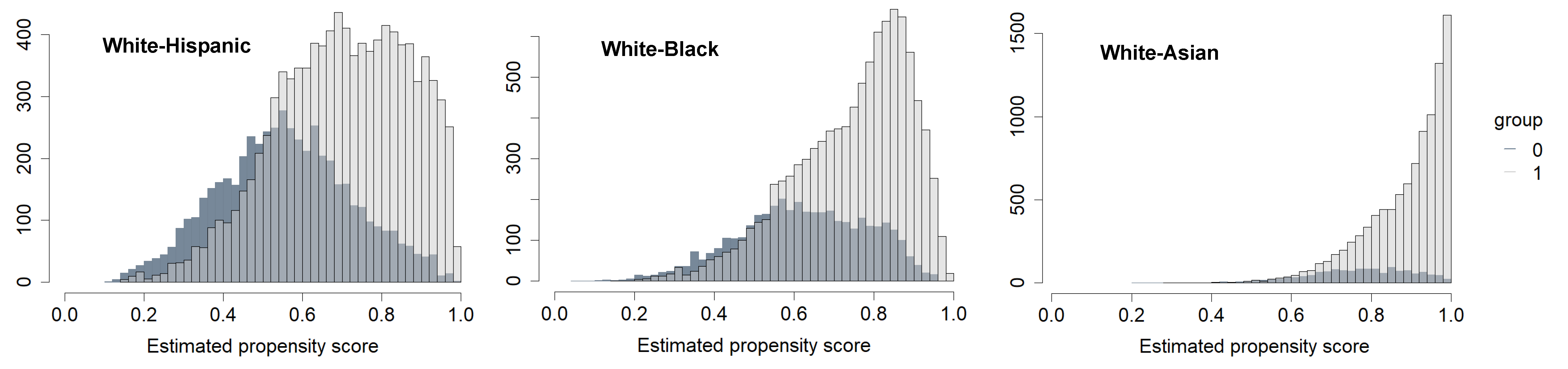}}
			
			\tiny
			\begin{tabular}{rcccccccccccccccccccc}
            \toprule
            Comparison & $N$ & Minimum & 25-th Quantile & Median & Mean & 75-th Quantile & Maximum\\ 
            
            \cmidrule(lr){1-8}
            White & 9830 & 0.14 & 0.58 & 0.71 & 0.70 & 0.84 & 1.00  \\ 
            Hispanic & 5280 & 0.11 & 0.44 & 0.55 & 0.56 & 0.67 & 0.99 \\ 
            
            \cmidrule(lr){1-8}
            White & 9830 & 0.16 & 0.65 & 0.78 & 0.75 & 0.86 & 0.99 \\ 
            Black & 4020 & 0.04 & 0.51 & 0.62 & 0.62 & 0.75 & 0.99 \\ 
            
            \cmidrule(lr){1-8}
            White & 9830 & 0.30 & 0.83 & 0.92 & 0.89 & 0.97 & 1.00 \\ 
            Asian & 1446 & 0.22 & 0.68 & 0.78 & 0.77 & 0.87 & 1.00 \\
            \bottomrule
            \end{tabular}
		\end{center}
		\caption{MEPS Data: Propensity score estimates distributions and summary statistics.}\label{fig:ps-meps}
	\end{figure}
	
	 Figure \ref{fig:ps-meps} shows the estimated PS distributions for White ($Z=1$) vs. a minority group ($Z=0$), along with the summary statistics. Overall, the histograms of the PSs indicate good overlap. Nevertheless,
	a number of control  participants have PSs near 1 in all three comparisons, especially for the White-Asian tandem. Most Asian participants have a PS greater than 0.6 and a substantial number of them have their PSs near 1. This indicates that some of participants in the minority groups (Hispanic, Black, and Asian) will have large weights in each of comparisons of interest. 
    \begin{figure}[htbp]
        \begin{center}
            {\includegraphics[trim=5 5 15 5, clip,  width=0.65\textwidth]{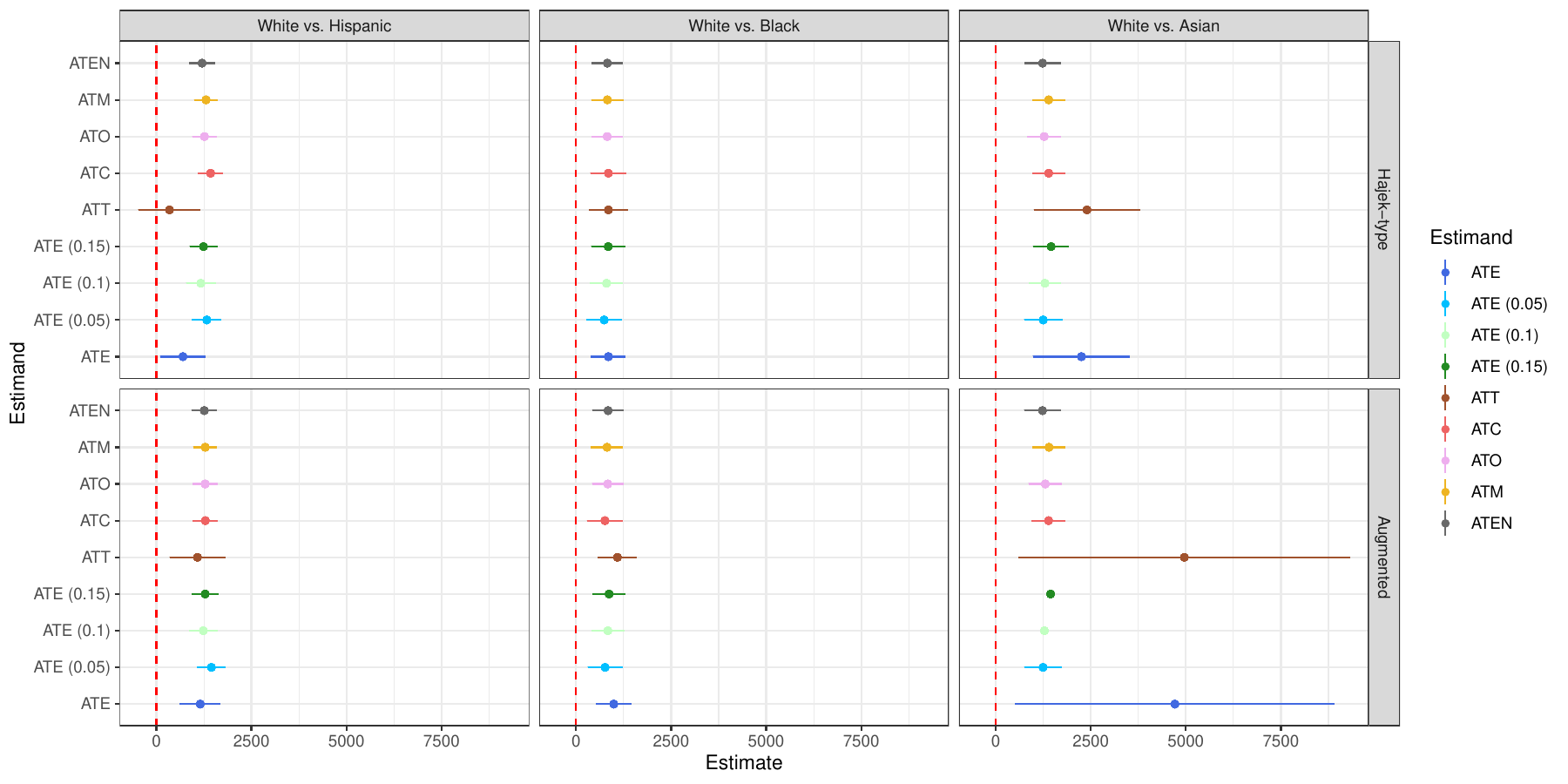}}
        \end{center}
        \caption{Racial disparities in the health care expenditure 
        %\textbf{\color{red}(Version with trimming)}
        }\label{fig:meps-forest}
    \end{figure}

	 Figure \ref{fig:meps-forest} shows the H\'ajek-type and augmented estimates of ATE, ATT, ATC, ATO, ATM and ATEN from the three comparisons, with the point estimate and the corresponding 95\% confidence interval. %The proportions of White (``treatment'', $Z=1$) in all three sub-population are greater than $0.6$, which we consider as  cases where $p$ is large. 
	 The results indicate significant health expenditure differences across racial and ethnic groups compare to White participants. The only exception is the ATT estimates in White vs. Hispanic: on averege, White participants pay  \$699.12 (see Table \ref{tab:meps-WATE}) more in health care expenditure than Hispanic participants. 
	 
	 As expected, the ATE estimates align with the ATT estimates  while equipoise estimates (ATO, ATM and ATEN) are closer to ATC estimates.  What is remarkable is the fact that the estimates of ATO, ATM, and ATM remained on the same ballpark (augmented or not) within each of the 2-by-2 comparisons. The estimates are in the range of \$1202.48--\$1285.72 for White-Hispanic, \$814.78--\$841.77 for White-Black, and   \$1229.33--\$1400.46 for White-Black (see Table \ref{tab:meps-WATE}). In contrast, the estimates of ATE (IPW without trimming) was \$699.12 using H\'ajek-type estimator and \$1154.44 using the double-robust estimator when we compare White vs. Hispanic, \$850.82 and \$ 992.82 for White vs. Black, and finally \$2253.00 and \$4712.69 for White vs. Asian.
	
	%%%%%%%%%%%%%%%%%%%%%%%%%%%%%%%%%%%%%%%%%%%%%%%%%%%%%%%%%%%%%%%%%%%%%%%%%%%%%%%%%%%%%%%%%%%%%%%%%%%%%%%%%%%%%%%%%%%%%%%%%%%%%%%%
	%%%%%%%%%%%%%%%%%%%%%%%%%%%%%%%%%%%%%%%%%%%%%%%%%%%%%%%%%%%%%%%%%%%%%%%%%%%%%%%%%%%%%%%%%%%%%%%%%%%%%%%%%%%%%%%%%%%%%%%%%%%%%%%%
	\section{Discussion and conclusion}\label{sec:discussion}

   {\color{black}\subsection{Summary}}

    In studies where the lack of positivity is highly likely (or not), or there is an imbalance in the treatment allocation (i.e., proportion of treated participants is too small or too large), the standard inverse probability weighting (IPW) method can assign disproportionately extreme weights to a handful of participants and yield a  less reliable treatment effect estimate with large variance. Practitioners of statistics often face a dilemma when deciding which path to follow: to weight or not to weight? And if they opt for weighting, which weights to use?  Indeed, there is a confusion about how to weight, which weights to choose, and whether should this be done along with trimming. While weighting can be used to target different estimands, IPW is commonly used to target the average treatment effect (ATE), the average treatment effect on the treated (ATT), or the average treatment effect on the control (ATC). Both ATT or ATC are often estimated for policy evaluation, where the  IPW  is a specific type of weights  to go after (sub)population treatment effects  \citep{greifer2021choosing}.
	   
    In this paper, we provided a coherent assessment of the different estimands of the PS weighting methods to dispel  confusion we may have about their use, when there is a violation of the positivity assumption. We examined the analytic solutions that can be explore when there is a violation of the positivity assumption. Then, we demonstrated that when the proportion of participants is either too small or too large, commonly-sought-after estimands and treatment effects are inherently group-specific, aiming for specific subpopulations, which implicitly move the goalposts. Furthermore,	 we demonstrated why and how ATE estimators can fail to identify logical treatment effect estimands and why using IPW trimming is not always a good idea.   Finally, we show that the equipoise estimators (matching, overlap, and matching weights) are more flexible  and have the ability to deliver robust results, when there is a violation of the positivity or important imbalance in treatment allocations. 
    We used a series of Monte-Carlo simulations to examine the performance of the different estimators and to highlight the shortcomings of the  estimators of ATE based on IPW (with and without trimming), when there is violations of the positivity assumption or in the presence of treatment allocation imbalance.  We also examine the performance of the corresponding augmented estimators, especially in the presence of underlying model misspecifications. Our simulations confirmed where ATE estimator via IPW  falls in the ATT--ATC spectrum compared to equipoise estimators (OW, MW, and EW), when there is an important imbalance in treatment allocations.%Furthermore, we briefly summarize some of our simulation results in the main paper and provide the detailed results in the Supplemental Material. We finally interpret these findings in the context of the current literature in causal inference.
    
	% Principled statistical practice prohibit us to conduct a number of statistical analyses (based on different estimators and methods), then cherry-pick the results we are comfortable with. By doing so, we might be comparing quantities that are not directly related and are not meant to be compared.  To better compare (and interpret causally) results from different methods (or weighting schemes) requires a good level of scrutiny since different methods call for different (or additional) sets of  assumptions. As such, we must be clear about the objectives of our study to help us embark on a good and expeditious data analysis. It is  incumbent on the researcher to be aware of these facts and make the necessary analytical choices, in accordance to what they want to accomplish.

    When there are violations of the positivity assumption or when the proportion of participant in the treatment group, $p$ , is small (or large), one should use ATE with caution as it may yield an estimate that is substantially different from what we have in mind, unless we assume that the treatment effect is constant. However, even under this assumption, the ATE estimate can still be biased due to the influence of extreme weights. Our findings demonstrate that the equipoise estimators (OW, MW, and EW) have better performance when extreme weights are present due to the lack of adequate positivity or important imbalance in treatment allocation, which aligns with prior studies  \citep{li2018balancing, li2018addressing, zhou2020propensity, matsouaka2020framework}. The advantages of the equipoise estimators is that they smoothly and gradually diminish the weights (i.e., the influence) of observations whose $e(x)(1-e(x))$ is closer to 0. They  do not require a user-specified threshold and target the subpopulation patients for whom there is equipoise. Nevertheless, in our study we found that IPW estimator of ATE without trimming and equipoise treatment effect estimators have comparable performance when the treatment allocation was nearly balanced ($p=45.92\%$). However, when $p=20.77\%$ and $p = 79.59\%$, we have noticed drastically different behaviors among all the IPW estimators of ATE (with or without trimming), those of ATT and ATC  and  the equipoise estimators (OW, MW, and EW). We concluded that the differences in estimating the treatment effects between the two methods were heightened by the presence of extreme weights as well as the tails of the treatment allocation proportions (i.e., small or large $p$ ). Although one may think of making their job easier by targeting ATT (or ATC), our study indicates that this might also be a bad idea. While choosing ATT (or ATC) results in relatively smaller absolute relative percentage bias, the  root mean squared errors and more importantly, the coverage probability will be greatly and negatively affected whenever $p \leq 20.30\%$ or $p\geq 79.59\%$. This conundrum is well illustrated in our data application, particularly in the standard errors of ATE and ATT for the White-Asian comparison.

%	\color{red}
	 In practice, heterogeneity of treatment effects is the norm rather than the exception; we should estimate our treatment effect(s) under these premises. Thus, it is important to have a better understanding of our data and the target population to whom your results will be applied to. ``What are we weighting for?'' is thus a call to seriously examine the reasons we use specific weighting schemes and to  make sure that (most of) the related assumptions are satisfied.  In this context, it is worth asking questions that can shed light on the factors contributing to the lack of adequate positivity or a small $p$. For intance, does the lack of positivity or reason for a small $p$ is just related to our specific sample? In other word, do we have a case of random violation of the positivity assumptions? Does our sample happen to have an allocation imbalance, but not target population? Imbalance allocations from perfectly well balance target population are often used in education research interventions. Most studies tend to enroll a smaller number of participants than can be found in the general population   \citep{terada_2020, bartlett2017schools}. Imbalance allocation can just be a characteristic of the target population (e.g., unusual occupational exposures or exposure to toxic chemicals, see  \cite{hall2020considering,checkoway2004research}) or a dynamic phenomenon in the study of a drug (e.g., early users of a newly released drug---with the promised of improved effectiveness or safety---versus broaden user base after a longer period of time as mentioned by \cite{schneeweiss2011assessing}) 

    Our work sheds light on some of the criticisms raised in the early days of PS methods, often caricatured through Basu's elephant analogy \citep{basu2011essay}, when applications of IPW led to extreme weights, unstable and often questionable causal estimates. These issues, as we now have demonstrated, go beyond the presence of extreme weights to also include the influence of treatment allocations imbalances and the mitigating role the ratio $r$ of the variances of PSs (between the treatment groups) can have. 
	In addition, from a statistical standpoint, the question can be framed as: when $p$ is extremely small (or large), what comparisons ATE gives rise to? The ATE estimand and the related IPW estimation method(s) should not be the default approach, as if we have a hammer and view everything as a nail. Because ATE is the average treatment effect in the entire population if all participants were treated vs. if all participants were untreated, we need to ask ourselves probing and fundamental questions as the one above. We should be clear about our estimand of interest before we start analyzing the data, based on subject matter knowledge and our scientific question. This allows us to be
		intentional and explicitly make meaningful and appropriate comparisons; do not compare ``apples'' to ``oranges''.
		
		Finally, each estimand that is estimated via any weighting method comes with a specific set of weights and targets a specific population. We must choose our estimand and corresponding weights wisely to recover the estimated or specific causal effects (or parameters) of interest that align with our scientific question(s). This will help better interpret the results we obtain. Similarly, we should be careful of using trimming blindly as it can be problematic. \textcolor{black}{Ignorance in this context can be costly: trimming leads to biased estimation of the ATE and calls for a different estimand  \citep{crump2006moving}, unless one uses bias-corrected IPW trimming methods \citep{chaudhuri2014heavy, sasaki2022estimation, ma2020robust}. We have indicated that the above biased-corrected ATE estimators are reliable only when  there is random violation of the positivity assumption. When such a  violation is  structural, there is no guarantee that these methods would lead to consistent estimators of ATE. In addition, their asymptotic behaviors are unknown when there is a structural violation or when there is an important imbalance in treatment allocation. }
  
  {\color{black}\subsection{Limitations and perspectives}
  The methods evaluated and compared in our study may have limitations, including but not limited to the following. 
  
  First, to be concise, we did not elaborate on the impact of the ratio $r$. A thorough and granular investigation needs to be conducted to complement the work presented in this paper. Second, the sandwich variance estimation method is not always without drawbacks. It is possible to face challenges when there is ``non-regularity'', whether due to chance or structural issues. Empirical studies, such as those discussed in \cite{matsouaka2023variance}, have demonstrated that when the sample size is limited and PS overlap between groups is poor, sandwich variance estimation may not be feasible. We also encountered this issue in our data analysis in Section \ref{sec:data}. In practice, when sandwich variance estimation fails, people may turn to bootstrap for help. However, bootstrap approaches have their own potential drawbacks. For instance, when dealing with big data, bootstrapping can become computationally time-consuming. There can also be random violations of positivity in some bootstrap replications. Moreover, in cases where the tilting function contains non-smooth and jumping points, such as ATE trimming, the sandwich variance is expected to fail, even though it is still used in practice for this cases (e.g., the implementation of PSweight package \citep{zhou2020psweight}). Therefore, there is a need for an in-depth assessment of different variance estimation methods. We should also consider the development of alternative approaches that are flexible, computationally efficient, and robust for estimating WATE estimators. This could involve exploring ideas from other nonparametric resampling techniques, such as the wild bootstrap using semiparametric efficient influence functions (see \cite{matsouaka2023variance}). 

  In addition, there is currently no doubly robust estimator proposed for WATE with a general tilting function. Double robustness is only applicable for the special cases of the tilting functions of ATE, ATT, and ATC. For the equipoise estimators (ATO, ATM, and ATEN), empirical studies suggest that the augmented estimators are still nearly double robust. However, the existing asymptotic analyses have not provide yet any theoretical justification of their double robustness. As part of our future research, we will explore the development of potentially doubly or multiply robust estimators while maintaining good efficiency, under certain regularity conditions. 

  Finally, although our study considered a number of useful alternatives for ATE when there is lack of positivity, we have not considered developing robust bias-corrected methods that still target ATE when positivity is violated, either randomly or structurally. Instead of just trimming observations or truncating their weights in regions of poor overlap, i.e., where $e(x)(1-e(x))\approx 0$ or use equipoise estimators, we may need to preserve an overall population level-inference \citep{nethery2019estimating}.  In this case, one can first estimate the treatment effect in regions with good overlap, then in a principled manner extrapolate  outside of these regions (if possible), and obtain an adequate estimate of a population level estimand, such as ATE, with less bias and better efficiency. \cite{chaudhuri2014heavy}, \cite{nethery2019estimating},  \cite{ma2020robust}, and  \cite{sasaki2022estimation} provided several strategies to better handle bias that may be induced by trimming. We can consider similar proposals, emulate their ideas and extend the methods to deal with both random and structural violations of the positivity assumption.  
  }
  
  {\color{black}\subsection{Conclusion} 
  Several papers have investigated issues related to violations of positivity assumptions or imbalances in treatment allocation when using the inverse probability weighting methods. When  using any weighting method, it is essential to ask yourself: what are you weighting for?
  You must be aware of what you ultimately get when using a specific weighting method and what estimand you are targeting. The ATE estimation via IPW, thanks to its many shortcomings, may not lead you where you  expect to land. However,  equipoise estimators (ATO, ATM, and ATEN) judiciously take you to the overlap/equipoise land, providing estimates of the treatment effect on the subgroup of participants for whom there is clinical equipoise  \citep{matsouaka2020framework}. 
  }
  %  \newpage
    \bibliography{OWWWF}
    \bibliographystyle{plainnat}
\newpage
    \appendix \label{appendix}
	\newcounter{Appendix}[section]
	\numberwithin{equation}{subsection}
	\renewcommand\theequation{\Alph{section}.\arabic{subsection}.\arabic{equation}}
	\numberwithin{table}{subsection}
	\numberwithin{figure}{subsection}
	\section{Appendix: Technical Proofs}\label{apx:proofs}
	\subsection{Sandwich variance estimation}
	In this section we calculate the variance estimation of the ATE, ATT, and ATC using M-theory (see \citet{stefanski2002calculus}). The estimator $\widehat\tau_{g}$ is derived using an estimating equation of the form $\displaystyle 0=\displaystyle\sum_{i=1}^{N} \Psi_\theta({X}_i, Z_i, Y_i)$ for which $\widehat\theta$ is solution to the equation and the estimator $\widehat\tau_{g}$  is a linear combination of the components of the component of $\theta$, for some matrix $ \Psi_\theta({X}_i, Z_i, Y_i).$ 
	
	Using $E[\Psi_\theta({X}_i, Z_i, Y_i)]=0$, we have $\widehat\theta\overset{p}{\longrightarrow}  \theta $ when  $N\longrightarrow \infty$ under some regularity conditions  \citep{stefanski2002calculus}, and then we can conclude that the estimator $ \widehat\tau_g$ is consistent by Slutsky's theorem. In addition,  $\sqrt{N}(\widehat{\theta}-{\theta})\overset{d}{\longrightarrow}N({0}, \Sigma({\theta})),$ with $ \Sigma({\theta})=A(\theta)^{-1}B(\theta)\{A(\theta)'\}^{-1}$. A consistent estimator of the asymptotic covariance matrix $\Sigma({\theta})$ is $ \widehat \Sigma({\widehat\theta})=A_N(\widehat\theta)^{-1}B_N(\widehat\theta)\{A_N(\widehat\theta)'\}^{-1}$, where  $A(\theta), B(\theta),$ $ A_N(\widehat\theta)$ and $B_N(\widehat\theta)$ are the following matrices: 
	\begin{align*}%\label{eq:matrices_appendix}
		A_N(\widehat{\theta})&=\frac{- 1}{N}\displaystyle\sum_{i=1}^{N}\frac{\partial \Psi_\theta({X}_i, Z_i, Y_i)}{\partial{\theta'}}\Big|_{\theta={\widehat\theta}}; ~~B_N(\widehat{\theta})= \frac{1}{N}\displaystyle\sum_{i=1}^{N}\Psi_\theta({X}_i, Z_i, Y_i)\Psi_\theta({X}_i, Z_i, Y_i)'\big|_{\theta=\widehat\theta}.
% 		\\
% 		A(\theta)&=-E_\theta\left[\dfrac{\partial\Psi_\theta(X_i,Y_i,Z_i)}{\partial\theta'}\right]~~\text{and}~~ 
% 		B(\theta)=E_\theta\left[ \Psi_\theta({X}_i, Z_i, Y_i)\Psi_\theta({X}_i, Z_i, Y_i)'\right] 
	\end{align*} 
	where $A_N(\widehat\theta)\overset{p}{\to} A(\theta)=-E_\theta\left[\dfrac{\partial\Psi_\theta(X_i,Y_i,Z_i)}{\partial\theta'}\right]$  and $B_N(\widehat\theta)\overset{p}{\to} B(\theta)=E_\theta\left[ \Psi_\theta({X}_i, Z_i, Y_i)\Psi_\theta({X}_i, Z_i, Y_i)'\right] $, as $N\to\infty$, by the weak law of large numbers and the consistency of $\widehat\theta$.
	
	 As an illustrative example, we provide the matrices $A_N(\widehat{\theta})$ and $B_N(\widehat{\theta})$ when the propensity score model $e({x_i};\widehat{\beta})=P(Z=1|{X_i=x_i}; \widehat{\beta}))$ and the regression models  $\widehat m_z(x_i)=m(x_i; \widehat\alpha_z)$ for $z=0$ and $z=1$ are estimated by maximum likelihood using, respectively, the logistic and linear regression models. We consider different combinations (or subsets) of covariates $X$ that are entered into the logistic and regression models, which we  denote them by $V$ and $W$, respectively.

	%%%%%%%%%%%%%%%%%%%%%%%%%%%%%%%%%%%%%%%%%%%%%%%%%%%%%%%%%%%%%%%%%%%%%%%%%%%%%%%%%%%%%%%%%%%%%%%%%%%%%%%%%%%%%%%%%%%%%%%%%%%%
	%%%%%%%%%%%%%%%%%%%%%%%%%%%%%%%%%%%%%%%%%%%%%%%%%%%%%%%%%%%%%%%%%%%%%%%%%%%%%%%%%%%%%%%%%%%%%%%%%%%%%%%%%%%%%%%%%%%%%%%%%%%%
	\subsubsection{Variance for the H\'ajek-type weighted estimator }\label{subapx:proofs-sand-wt}   
	For the weighted average treatment effect (WATE) estimator, we have 
	\begin{equation*}%\label{eq:li_est}
		\widehat\tau_{g}= \displaystyle\sum_{i=1}^{N}\left[ \frac{Z_i\widehat  w_1(x_i)}{N_{\widehat w_1}}-\frac{	(1-Z_i)\widehat w_{0}(x)}{N_{\widehat w_0}}\right] Y_i,~~~\text{with}~N_{\widehat w_k}=\displaystyle\sum_{i=1}^{N} Z_i^k(1-Z_i)^{1-k}\widehat w_k(x_i), ~k=0,1,
	\end{equation*} 
	where  $\widehat w_k(x) =\widehat g(x)\widehat e(x)^{-k}(1-\widehat e(x))^{k-1}.$ \\
	The estimated propensity score parameter vector $\widehat{\beta}$ and the estimator $(\widehat{\mu}_{1g}, ~\widehat{\mu}_{0g}$)  are derived as solutions to the estimating equation
	\begin{align*}
		\displaystyle 0=\displaystyle\sum_{i=1}^{N} \Psi_\theta({X}_i, Z_i, Y_i)
		= \displaystyle\sum_{i=1}^{N} 
		\begin{bmatrix}
			\psi_{\beta}({X}_i, Z_i)\\
			\psi_{\mu_{1g}}({X}_i, Z_i, Y_i) \\%[0.3em]
			\psi_{\mu_{0g}}({X}_i, Z_i, Y_i) \\%[0.3em]
		\end{bmatrix} 
		= \displaystyle\sum_{i=1}^{N} \begin{bmatrix}
			\psi_{\beta}({X}_i, Z_i)\\
			Z_iw_1({X}_i)(Y_i-\mu_{1g})\\%[0.3em]
			(1-Z_i)w_0({X}_i)(Y_i-\mu_{0g})\\%[0.3em]
		\end{bmatrix} 
	\end{align*}
	with respect to  ${\theta}=({\beta}',\mu_{1g}, \mu_{0g})'$ where $\widehat{\tau}_g=c_0'{\theta}=\widehat{\mu}_{1g}-\widehat{\mu}_{0g}$ and $c_0=(0,1,-1)'$. 
	
	The matrices $A(\theta)$, $A_N(\widehat\theta)$, $B(\theta)$ and $B_N(\widehat\theta)$ are 
	\begin{align*}
		A_N(\widehat{\theta})&=N^{-1}\displaystyle\sum_{i=1}^{N}\left[ - \frac{\partial}{\partial{\theta'}}\Psi_\theta({X}_i, Z_i, Y_i)\right]_{\theta={\widehat\theta}}=
		\begin{bmatrix} 
			\widehat A_{11} & 0 &0  \\
			\widehat A_{21} &\widehat A_{22} &0\\
			\widehat A_{31} & 0 &\widehat A_{33}
		\end{bmatrix}.
	\end{align*} 
	If we estimate the propensity scores via a logistic regression model $e(X_i)=[1+\exp(-V_{i}'\beta)]^{-1}$,   we have $\psi_{{\beta}}({X}_i, Z_i)=[Z_i-e({V_{i}};{\beta})]V_{i}$. The components of the matrix $A_N$ are given by 
	\allowdisplaybreaks\begin{align*}
		\widehat A_{11}&=N^{-1}\displaystyle\sum_{i=1}^{N} \widehat e_i(\mathrm{v})(1-\widehat e_i(\mathrm{v}))V_iV_i'\\
		\widehat A_{21}&=-N^{-1}\displaystyle\sum_{i=1}^{N} Z_i\left[\left[\frac{\partial g(V_i)}{\partial{\beta}}\right]_{\beta=\widehat\beta} - (1-\widehat e_i(\mathrm{v}))\widehat{g}(V_i)V_i'\right]\widehat e_i(\mathrm{v})^{-1} \left(Y_i-\widehat\mu_{1g}\right);\\
		\widehat A_{31}&=-N^{-1}\displaystyle\sum_{i=1}^{N} (1-Z_i)\left[\left[\frac{\partial g(V_i)}{\partial{\beta}}\right]_{\beta=\widehat\beta} + \widehat e_i(\mathrm{v})\widehat{g}(V_i)V_i'\right](1-\widehat e_i(\mathrm{v}))^{-1} \left( Y_i-\widehat\mu_{0g}\right); \\
		\widehat A_{22}&=N^{-1}\displaystyle\sum_{i=1}^{N} Z_i\widehat e_i(\mathrm{v})^{-1}\widehat{g}(V_i); ~~
		\widehat A_{33} =N^{-1}\displaystyle\sum_{i=1}^{N}(1- Z_i)(1-\widehat e_i(\mathrm{v}))^{-1}\widehat{g}(V_i).		
	\end{align*}  
	Therefore,  an estimator of the variance of $\widehat{\tau}_g$  is $\widehat{Var}({\widehat{\tau}_g})=N^{-1}c_0'\widehat\Sigma({\widehat\theta})c_0$.\\	
Note that $\displaystyle\left[\displaystyle\frac{\partial g(V_i)}{\partial{\beta}}\right]_{\beta=\widehat\beta}$ is equal to 0 for ATE, equal to $e(\mathrm{v})[1-e(\mathrm{v})]V_i'$ for ATT and  to  $-e(\mathrm{v})[1-e(\mathrm{v})]V_i'$ for ATC.

	%%%%%%%%%%%%%%%%%%%%%%%%%%%%%%%%%%%%%%%%%%%%%%%%%%%%%%%%%%%%%%%%%%%%%%%%%%%%%%%%%%%%%%%%%%%%%%%%%%%%%%%%%%%%%%%%%%%%%%%%%%%%
	%%%%%%%%%%%%%%%%%%%%%%%%%%%%%%%%%%%%%%%%%%%%%%%%%%%%%%%%%%%%%%%%%%%%%%%%%%%%%%%%%%%%%%%%%%%%%%%%%%%%%%%%%%%%%%%%%%%%%%%%%%%%
	\subsubsection{Variance for the doubly robust ATE, ATT, and ATC }\label{subapx:proofs-sand-DR}   
	%%%%%%%%%%%%%%%%%%%%%%%%%%%%%%%%%%%%%%%%%%%%%%%%%%%%%%%%%%%%%%%%%%%%%%%%%%%%%%%%%%%%%%%%%%%%%%%%%%%%%%%%%%%%%%%%%%%%%%%%%%%%
	The doubly robust estimators for ATE, ATT, and ATC are given, respectively, by 

	 \allowdisplaybreaks\begin{align}\label{tau_dr}
		\widehat \tau_{\text{ATE}}^{\text{dr}}&=\sum_{i=1}^{N}\frac{Z_i\widehat  e(x_i)^{-1}\left\lbrace Y_i- \widehat m_1(x_i)\right\rbrace}{\displaystyle\sum_{i=1}^{N} Z_i\widehat e(x_i)^{-1}} - \sum_{i=1}^{N}\frac{(1-Z_i)(1-\widehat e(x_i))^{-1}\left\lbrace Y_i- \widehat m_0(x_i)\right\rbrace}{\displaystyle\sum_{i=1}^{N} (1-Z_i)(1-\widehat e(x_i))^{-1}} \nonumber\\
		&+\frac{1}{N}\sum_{i=1}^{N} \left\lbrace  \widehat m_1(x_i) -  \widehat m_0(x_i)  \right\rbrace; \\
		\widehat \tau_{\text{ATT}}^{\text{dr}}&=\sum_{i=1}^{N}\frac{Z_i\left\lbrace Y_i- \widehat m_0(x_i)\right\rbrace}{\displaystyle\sum_{i=1}^{N} Z_i} - \sum_{i=1}^{N}\frac{(1-Z_i)\widehat  e(x_i)(1-\widehat e(x_i))^{-1}\left\lbrace Y_i- \widehat m_0(x_i)\right\rbrace}{\displaystyle\sum_{i=1}^{N} (1-Z_i)\widehat  e(x_i)(1-\widehat e(x_i))^{-1}}; \\
		\widehat \tau_{\text{ATC}}^{\text{dr}}&=\sum_{i=1}^{N}\frac{Z_i\widehat  e(x_i)^{-1}(1-\widehat e(x_i))\left\lbrace Y_i- \widehat m_1(x_i)\right\rbrace}{\displaystyle\sum_{i=1}^{N} Z_i\widehat e(x_i)^{-1}(1-\widehat e(x_i))} - \sum_{i=1}^{N}\frac{(1-Z_i)\left\lbrace Y_i- \widehat m_1(x_i)\right\rbrace}{\displaystyle\sum_{i=1}^{N} (1-Z_i)} 	\end{align}
	%%%%%%%%%%%%%%%%%%%%%%%%%%%%%%%%%%%%%%%%%%%%%%%%%%%% ATE %%%%%%%%%%%%%%%%%%%%%%%%%%%%%%%%%%%%%%%%%%%%%%%%%%%%%%%%%%%%%%%%%%%%%%%%
	In addition to the above function $\psi_{\beta}({X}_i, Z_i)$,   we also consider the estimating functions $\psi_{\alpha_z}(X)$ for the regression models $m_z(X)=m_z(X; \alpha_z),$ $z=0,1$. Let $c_1=(0,0,0,1,-1,1,-1)'$;  we can derive the estimator $\widehat\tau_{\text{ATE}}^{\text{dr}}=c_1'\widehat\theta_{ate}=\widehat\tau_{1g}^{m}-\widehat\tau_{0g}^{m}+\widehat\mu_{1g}-\widehat\mu_{0g}$ through $\widehat\theta_{ate}=(\widehat\beta', \widehat\alpha_1',  \widehat\alpha_0', \widehat\tau_{1g}^{m}, \widehat\tau_{0g}^{m},\widehat\mu_{1g}, \widehat\mu_{0g})'$, the  solution to the estimating equation 
	\allowdisplaybreaks \begin{align*}
	\displaystyle \displaystyle\sum_{i=1}^{N}	\Psi_{\theta_{ate}}({X}_i, Z_i, Y_i) & = 
	\displaystyle \displaystyle\sum_{i=1}^{N}	\begin{bmatrix}
			\psi_{\beta}({X}_i, Z_i)\\
			Z_i\psi_{\alpha_1}({X}_i, Y_i)\\
			(1-Z_i)\psi_{\alpha_0}({X}_i, Y_i)\\
			m_1(X_i)-\tau_{1}^m\\
			m_0(X_i)-\tau_{0}^m\\
			Z_ie(x_i)^{-1}(Y_i-m_1(X_i)-\mu_{1g})\\%[0.3em]
			(1-Z_i)(1-e(x_i))^{-1}(Y_i-m_0(X_i)-\mu_{0g})\\%[0.3em]
		\end{bmatrix} = 0
	\end{align*}
    with respect to $\theta_{ate}=(\beta',\alpha_1',  \alpha_0', \tau_{1g}^{m}, \tau_{0g}^{m},  \mu_{1g}, \mu_{0g})'.$ 
 
	If we estimate  $e(X)$ and  $m_z(X)$ using logistic and linear regression models  $e(V_i)=[1+\exp(-V_i'\beta)]^{-1}$ and $m_z(W_i)=W_i'\alpha_z$, $z=0,1$, then $\psi_{{\beta}}({X}_i, Z_i)=[Z_i-e({V_i};{\beta})]V_i$ and $\psi_{{\alpha}_z}({X}_i, Z_i)=W_i(Y_i-W_i'{\alpha}_z)$. 
	Assuming that the same covariates appear as predictors in the regression models $m_z(W)$, the non-zero components $\widehat A_{ij}$ of the matrix $ A_N$ are given by 
	\allowdisplaybreaks\begin{align*}
		\widehat A_{11}&=N^{-1}\displaystyle\sum_{i=1}^{N} \widehat e_i(\mathrm{v})(1-\widehat e_i(\mathrm{v}))V_iV_i'; ~
		\widehat A_{22}=N^{-1}\displaystyle\sum_{i=1}^{N} Z_iW_iW_i';~~ 
		\widehat A_{33}=N^{-1}\displaystyle\sum_{i=1}^{N}(1- Z_i)W_iW_i';\\ 
		\widehat A_{42}&=\widehat A_{53}=-N^{-1}\displaystyle\sum_{i=1}^{N} W_i'; ~\widehat A_{44}= \widehat A_{55}=1\\
		\widehat A_{61}&=N^{-1}\displaystyle\sum_{i=1}^{N} Z_i (1-\widehat e_i(\mathrm{v}))\widehat e_i(\mathrm{v})^{-1} \left( \widehat Y_i-\widehat m_1(W_i)-\widehat\mu_{1g}\right)V_i';\\
		\widehat A_{62}&=N^{-1}\displaystyle\sum_{i=1}^{N} Z_i\widehat e_i(\mathrm{v})^{-1}W_i'; ~~
		\widehat A_{66}=N^{-1}\displaystyle\sum_{i=1}^{N} Z_i\widehat e_i(\mathrm{v})^{-1};\\
		\widehat A_{71}&=- N^{-1}\displaystyle\sum_{i=1}^{N} (1-Z_i)\widehat e_i(\mathrm{v})(1-\widehat e_i(\mathrm{v}))^{-1} ( \widehat Y_i-\widehat m_0(W_i)-\widehat\mu_{0g})V_i'\\
		\widehat A_{73}&=N^{-1}\displaystyle\sum_{i=1}^{N} (1-Z_i)(1-\widehat e_i(\mathrm{v}))^{-1}W_i'; ~~ \widehat A_{77}=N^{-1}\displaystyle\sum_{i=1}^{N}(1- Z_i)(1-\widehat e_i(\mathrm{v}))^{-1}.
	\end{align*}   
	
	An estimator of  $\Sigma({\theta}_{ate})$ is then $ \widehat\Sigma(\widehat{\theta}_{ate})=A_N(\widehat\theta_{ate})^{-1}B(\widehat\theta_{ate})\{A(\widehat\theta_{ate})'\}^{-1},$ from which we can derive the variance of $\widehat\tau_{ATE}^{\text{dr}}$ as $\widehat{Var}(\widehat{\tau}_g^{ate})=N^{-1}c_1'\widehat\Sigma({\widehat{\theta}_{ate}})c_1.$
	
	%%%%%%%%%%%%%%%%%%%%%%%%%%%%%%%%%%%%%%%%%%%%%%%%%%%% ATT %%%%%%%%%%%%%%%%%%%%%%%%%%%%%%%%%%%%%%%%%%%%%%%%%%%%%%%%%%%%%%%%%%%%%%%%
	For the ATT estimator $\widehat\tau_{ATT}^{\text{dr}}$, we can use the  solution to the estimating equation 
	\allowdisplaybreaks \begin{align*}
			\displaystyle \displaystyle\sum_{i=1}^{N} \Psi_{\theta_{att}}({X}_i, Z_i, Y_i) & = 
		\displaystyle \displaystyle\sum_{i=1}^{N}	\begin{bmatrix}
			\psi_{\beta}({X}_i, Z_i)\\
			%Z_i\psi_{\alpha_1}({X}_i, Y_i)\\
			(1-Z_i)\psi_{\alpha_0}({X}_i, Y_i)\\
%			g(X_i)\{m_1(X_i)-\tau_{1g}^m\}\\
%			g(X_i)\{m_0(X_i)-\tau_{0g}^m\}\\
			Z_i(Y_i-m_0(X_i)-\mu_{1g})\\%[0.3em]
			(1-Z_i)e(X_i)(1-e(X_i))^{-1}(Y_i-m_0(X_i)-\mu_{0g})\\%[0.3em]
		\end{bmatrix} =0
	\end{align*}
      with respect to $\theta_{att}=(\beta',\alpha_0',  \mu_{1g}, \mu_{0g})',$ to calculate       
      $\widehat\tau_{ATT}^{\text{dr}}=c_2'\widehat\theta_{att}=\widehat\mu_{1g}-\widehat\mu_{0g}$ where  $c_2=(0,0,1, -1)$ and $\widehat\theta_{att}=(\widehat\beta', \widehat\alpha_0', \widehat\mu_{1g}, \widehat\mu_{0g})'$.
     
	When  $e(X)$ and  $m_z(X)$ are estimated via maximum likelihood based on logistic and linear regression models  $e(V_i)=[1+\exp(-V_i'\beta)]^{-1}$ and $m_z(W_i)=W_i'\alpha_z$, $z=0,1$, then $\psi_{{\beta}}({X}_i, Z_i)=[Z_i-e({V_i};{\beta})]V_i$ and $\psi_{{\alpha}_z}({X}_i, Z_i)=W_i(Y_i-W_i'{\alpha}_z)$. 
	Assuming that the same covariates appear as predictors in the regression models $m_z(W)$, the non-zero components $\widehat A_{ij}$ of the matrix $ A_N$ are given by 
	% \allowdisplaybreaks\begin{align*}
	% 	\widehat A_{11}&=N^{-1}\displaystyle\sum_{i=1}^{N} \widehat e_i(\mathrm{v})(1-\widehat e_i(\mathrm{v}))V_iV_i'; ~
	% 	\widehat A_{22}=N^{-1}\displaystyle\sum_{i=1}^{N} Z_iW_iW_i';\\ 
	% 	\widehat A_{33}&=N^{-1}\displaystyle\sum_{i=1}^{N}(1- Z_i)W_iW_i';~~ 
	% 	\widehat A_{43}=N^{-1}\displaystyle\sum_{i=1}^{N} Z_iW_i'; ~\widehat A_{44}= N^{-1}\displaystyle\sum_{i=1}^{N} Z_i\\
	% 	\widehat A_{51}&=-N^{-1}\displaystyle\sum_{i=1}^{N} (1-Z_i)\frac{\widehat e_i(\mathrm{v})}{(1-\widehat e_i(\mathrm{v}))}(Y_i-\widehat m_0(W_i)-\widehat \mu_{0g})V_i'; \\
	% 	\widehat A_{53}&=N^{-1}\displaystyle\sum_{i=1}^{N}(1- Z_i) \frac{\widehat e_i(\mathrm{v})W_i'}{(1-\widehat e_i(\mathrm{v}))};~~
	% 	\widehat A_{55}=N^{-1}\displaystyle\sum_{i=1}^{N}(1- Z_i) \frac{\widehat e_i(\mathrm{v})}{(1-\widehat e_i(\mathrm{v}))}.
	% \end{align*}   
        \allowdisplaybreaks\begin{align*}
		\widehat A_{11}&=N^{-1}\displaystyle\sum_{i=1}^{N} \widehat e_i(\mathrm{v})(1-\widehat e_i(\mathrm{v}))V_iV_i'; ~
		\widehat A_{22}=N^{-1}\displaystyle\sum_{i=1}^{N} Z_iW_iW_i';\\ 
		\widehat A_{32}&=N^{-1}\displaystyle\sum_{i=1}^{N} Z_iW_i'; ~\widehat A_{44}= N^{-1}\displaystyle\sum_{i=1}^{N} Z_i\\
		\widehat A_{41}&=-N^{-1}\displaystyle\sum_{i=1}^{N} (1-Z_i)\frac{\widehat e_i(\mathrm{v})}{(1-\widehat e_i(\mathrm{v}))}(Y_i-\widehat m_0(W_i)-\widehat \mu_{0g})V_i'; \\
		\widehat A_{42}&=N^{-1}\displaystyle\sum_{i=1}^{N}(1- Z_i) \frac{\widehat e_i(\mathrm{v})W_i'}{(1-\widehat e_i(\mathrm{v}))};~~
		\widehat A_{44}=N^{-1}\displaystyle\sum_{i=1}^{N}(1- Z_i) \frac{\widehat e_i(\mathrm{v})}{(1-\widehat e_i(\mathrm{v}))}.
	\end{align*} 
	The variance of  $\widehat\tau_{ATE}^{\text{dr}}$ is then estimated via $\widehat{Var}(\widehat{\tau}_g^{att})=N^{-1}c_2'\widehat\Sigma({\widehat{\theta}_{att}})c_2.$
	
	%%%%%%%%%%%%%%%%%%%%%%%%%%%%%%%%%%%%%%%%%%%%%%%%%%%%%%%%%%% ATC %%%%%%%%%%%%%%%%%%%%%%%%%%%%%%%%%%%%%%%%%%%%%%%%%%%%%%%%%%%%%%%%%%%%%%%%%%%%
	Finally, for the ATC, the estimators $\widehat\tau_{ATC}^{\text{dr}}$ can be derived by using the solution   $\theta_{atc}= (\beta',\alpha_1', \mu_{1g}, \mu_{0g})',$ to the estimating equation
		\allowdisplaybreaks \begin{align*}
		\displaystyle \displaystyle\sum_{i=1}^{N} \Psi_{\theta_{atc}}({X}_i, Z_i, Y_i)&= 
		\displaystyle \displaystyle\sum_{i=1}^{N}		\begin{bmatrix}
			\psi_{\beta}({X}_i, Z_i)\\
			Z_i\psi_{\alpha_1}({X}_i, Y_i)\\
			%(1-Z_i)\psi_{\alpha_0}({X}_i, Y_i)\\
			Z_ie(X_i)^{-1}(1-e(X_i))(Y_i-m_1(X_i)-\mu_{1g})\\%[0.3em]
			(1-Z_i)(Y_i-m_1(X_i)-\mu_{0g})\\%[0.3em]
		\end{bmatrix} =0
	\end{align*}
	where        
	$\widehat\tau_{ATC}^{\text{dr}}=c_2'\widehat\theta_{atc}=\widehat\mu_{1g}-\widehat\mu_{0g}$, with $\widehat\theta_{atc}=(\widehat\beta', \widehat\alpha_1',  \widehat\alpha_0', \widehat\mu_{1g}, \widehat\mu_{0g})'$.
	
	In this case, the non-zero components of the matrix $A_N$ are then 
	% \allowdisplaybreaks\begin{align*}
	% 	\widehat A_{11}&=N^{-1}\displaystyle\sum_{i=1}^{N} \widehat e_i(\mathrm{v})(1-\widehat e_i(\mathrm{v}))V_iV_i'; ~
	% 	\widehat A_{22}=N^{-1}\displaystyle\sum_{i=1}^{N} Z_iW_iW_i';\\ 
	% 	\widehat A_{33}&=N^{-1}\displaystyle\sum_{i=1}^{N}(1- Z_i)W_iW_i';~~
	% 	\widehat A_{41}=N^{-1}\displaystyle\sum_{i=1}^{N} \frac{Z_i\widehat e_i(\mathrm{v})}{(1-\widehat e_i(\mathrm{v}))}(Y_i-\widehat m_0(W_i)-\widehat \mu_{0g})V_i'; \\	
	% 	\widehat A_{42}&=N^{-1}\displaystyle\sum_{i=1}^{N} \frac{Z_i\widehat e_i(\mathrm{v})}{(1-\widehat e_i(\mathrm{v}))}W_i; ~~
	% 	\widehat A_{44}= N^{-1}\displaystyle\sum_{i=1}^{N} \frac{Z_i\widehat e_i(\mathrm{v})}{(1-\widehat e_i(\mathrm{v}))}\\
	% 	\widehat A_{52}&=N^{-1}\displaystyle\sum_{i=1}^{N} (1-Z)W_i';~~
	% 	\widehat A_{55}=N^{-1}\displaystyle\sum_{i=1}^{N}(1- Z_i).
	% \end{align*}    
	\allowdisplaybreaks\begin{align*}
		\widehat A_{11}&=N^{-1}\displaystyle\sum_{i=1}^{N} \widehat e_i(\mathrm{v})(1-\widehat e_i(\mathrm{v}))V_iV_i'; ~
		\widehat A_{22}=N^{-1}\displaystyle\sum_{i=1}^{N} Z_iW_iW_i';\\ 
		\widehat A_{31}&=N^{-1}\displaystyle\sum_{i=1}^{N} \frac{Z_i\widehat e_i(\mathrm{v})}{(1-\widehat e_i(\mathrm{v}))}(Y_i-\widehat m_0(W_i)-\widehat \mu_{0g})V_i'; \\	
		\widehat A_{32}&=N^{-1}\displaystyle\sum_{i=1}^{N} \frac{Z_i\widehat e_i(\mathrm{v})}{(1-\widehat e_i(\mathrm{v}))}W_i; ~~
		\widehat A_{33}= N^{-1}\displaystyle\sum_{i=1}^{N} \frac{Z_i\widehat e_i(\mathrm{v})}{(1-\widehat e_i(\mathrm{v}))}\\
		\widehat A_{42}&=N^{-1}\displaystyle\sum_{i=1}^{N} (1-Z)W_i';~~
		\widehat A_{44}=N^{-1}\displaystyle\sum_{i=1}^{N}(1- Z_i).
	\end{align*}   
	%%%%%%%%%%%%%%%%%%%%%%%%%%%%%%%%%%%%%%%%%%%%%%%% Augmented WATE %%%%%%%%%%%%%%%%%%%%%%%%%%%%%%%%%%%%%%%%%%%%%%%%%%%%%%%%%%%%%%%%%%%%%%%%%%%%
	
	\subsubsection{Sandwich variance for the augmented estimator}\label{subapx:proofs-sand-aug}  
	For the estimator $\widehat\tau_{g}^{\text{aug}}$, we also consider  $c=(0,0,0,1,-1,1,-1)'$ such that $\widehat\tau_{g}^{\text{aug}}=c'\widehat\theta_{\text{aug}}=\widehat\tau_{1g}^{m}-\widehat\tau_{0g}^{m}+\widehat\mu_{1g}-\widehat\mu_{0g},$ where $\widehat\theta_{\text{aug}}=(\widehat\beta', \widehat\alpha_1',  \widehat\alpha_0', \widehat\tau_{1g}^{m}, \widehat\tau_{0g}^{m},\widehat\mu_{1g}, \widehat\mu_{0g})'$ is the  solution to the estimating equation 
    \allowdisplaybreaks \begin{align*}%\label{eq:est.eqaug}
    \displaystyle \displaystyle\sum_{i=1}^{N} \Psi_{\theta_{\text{aug}}}({X}_i, Z_i, Y_i)&= 
	\displaystyle \displaystyle\sum_{i=1}^{N} \begin{bmatrix}
		\psi_{\beta}({X}_i, Z_i)\\
		Z_i\psi_{\alpha_1}({X}_i, Y_i)\\
		(1-Z_i)\psi_{\alpha_0}({X}_i, Y_i)\\
		g(X_i)\{m_1(X_i)-\tau_{1g}^m\}\\
		g(X_i)\{m_0(X_i)-\tau_{0g}^m\}\\
		Z_iw_1({X}_i)(Y_i-m_1(X_i)-\mu_{1g})\\
		(1-Z_i)w_0({X}_i)(Y_i-m_0(X_i)-\mu_{0g})\\
	\end{bmatrix} =0
	\end{align*}
	with respect to $\theta_{\text{aug}}=(\beta',\alpha_1',  \alpha_0', \tau_{1g}^{m}, \tau_{0g}^{m},  \mu_{1g}, \mu_{0g})'.$
	
	When we estimate the propensity score $e(X)$ and the regression models $m_z(X)$ using, respectively, a logistic regression model and a linear regression model, i.e., $e(V_i)=[1+\exp(-V_i'\beta)]^{-1}$ and the regression models $m_z(W_i)=W_i'\alpha_z$, $z=0,1$. Hence, $\psi_{{\beta}}({X}_i, Z_i)=[Z_i-e({V_i};{\beta})]V_i$ and $\psi_{{\alpha}_z}({X}_i, Z_i)=W_i(Y_i-W_i'{\alpha}_z)$. 
	Assuming that the same covariates appear as predictors in the regression models $m_z(W)$, the non-zero components $\widehat A_{ij}$ of the matrix $ A_N$ are given by 
	\allowdisplaybreaks\begin{align*}
		\widehat A_{11}&=N^{-1}\displaystyle\sum_{i=1}^{N} \widehat e_i(\mathrm{v})(1-\widehat e_i(\mathrm{v}))V_iV_i'; ~
		\widehat A_{22}=N^{-1}\displaystyle\sum_{i=1}^{N} Z_iW_iW_i';~~ 
		\widehat A_{33}=N^{-1}\displaystyle\sum_{i=1}^{N}(1- Z_i)W_iW_i';\\ 
		\widehat A_{41}&=-N^{-1}\displaystyle\sum_{i=1}^{N} \left[\frac{\partial g(V_i)}{\partial{\beta}}\right]_{\beta=\widehat\beta}\{\widehat m_1(W_i)- \widehat \tau_{1g}^m\}; ~~
		\widehat A_{42}=\widehat A_{53}=-N^{-1}\displaystyle\sum_{i=1}^{N} \widehat g(V_i)W_i'; \\
		\widehat A_{44}&= \widehat A_{55}=N^{-1}\displaystyle\sum_{i=1}^{N} \widehat g(V_i);~~
		\widehat A_{51}=-N^{-1}\displaystyle\sum_{i=1}^{N} \left[\frac{\partial g(V_i)}{\partial{\beta}}\right]_{\beta=\widehat\beta}\{\widehat m_0(W_i)- \widehat \tau_{0g}^m\}; \\
		\widehat A_{61}&=-N^{-1}\displaystyle\sum_{i=1}^{N} Z_i\left[\left[\frac{\partial g(V_i)}{\partial{\beta}}\right]_{\beta=\widehat\beta} - (1-\widehat e_i(\mathrm{v}))\widehat{g}(V_i)V_i'\right]\widehat e_i(\mathrm{v})^{-1} \left( \widehat Y_i-\widehat m_1(W_i)-\widehat\mu_{1g}\right);\\
		\widehat A_{62}&=N^{-1}\displaystyle\sum_{i=1}^{N} Z_i\widehat w_1(\mathrm{v})W_i'; ~~
		\widehat A_{66}=N^{-1}\displaystyle\sum_{i=1}^{N} Z_i\widehat w_1(\mathrm{v});\\
		\widehat A_{71}&=- N^{-1}\displaystyle\sum_{i=1}^{N} (1-Z_i)\left[\left[\frac{\partial g(V_i)}{\partial{\beta}}\right]_{\beta=\widehat\beta} + \widehat e_i(\mathrm{v})\widehat{g}(V_i)V_i'\right](1-\widehat e_i(\mathrm{v}))^{-1} ( \widehat Y_i-\widehat m_0(W_i)-\widehat\mu_{0g});\\
		\widehat A_{73}&=N^{-1}\displaystyle\sum_{i=1}^{N} (1-Z_i)\widehat w_0(\mathrm{v})W_i'; ~~ \widehat A_{77}=N^{-1}\displaystyle\sum_{i=1}^{N}(1- Z_i)\widehat w_0(\mathrm{v}).
	\end{align*}   
	
	An estimator of  $\Sigma({\theta}_{\text{aug}})$ is then $ \widehat\Sigma(\widehat{\theta}_{\text{aug}})=A_N(\widehat\theta_{\text{aug}})^{-1}B(\widehat\theta_{\text{aug}})\{A(\widehat\theta_{\text{aug}})'\}^{-1},$ from which we can derive the variance of $\widehat\tau_{g}^{\text{aug}}=c'\theta_{\text{aug}}$ as $\widehat{Var}(\widehat{\tau}_g^{\text{aug}})=N^{-1}c'\widehat\Sigma({\widehat{\theta}_{\text{aug}}})c.$
	
	%%%%%%%%%%%%%%%%%%%%%%%%%%%%%%%%%%%% %%%%%%%%%%%%%%%%%%%%%%%%%%%%%%%%%%%%%%%%%%%%%%%%%%%%%%%%%% %%%%%%%%%%%%%%%%%%%%%%%%%%%%%%%%%%%%%%%%%%%%%%%%%%%%%%%%%%
	
	\section{Appendix: Additional Simulation Details}\label{apx:simulation}
	%%%%%%%%%%%%%%%%%%%%%%%%%%%%%%%%%%%%%%%%%%%%%%%%%%%%%%%%%%%%%%%%%%%%%%%%%%%%%%%%%%%%%%%%%%%%%%%%%%%%%%%%%%%%%%%%%%%%%%%%%%%%%%%%
	%%%%%%%%%%%%%%%%%%%%%%%%%%%%%%%%%%%%%%%%%%%%%%%%%%%%%%%%%%%%%%%%%%%%%%%%%%%%%%%%%%%%%%%%%%%%%%%%%%%%%%%%%%%%%%%%%%%%%%%%%%%%%%%%
	\subsection{Propensity score analysis}\label{subapx:sim-ps-analysis}
	
	We considered 6 propensity score (PS) models for different proportions ($p=P(Z=1)$) and ratio of variances ($r$). For each correctly specified PS model, the coefficients ($\beta_0$ to $\beta_7$) of the PS model via logistic regression, proportion $p$ as well as the ratio ($r$) of estimated variance of propensity score of treatment group to control group are provided in table \ref{tab:coef}. 	
	\begin{table}[htbp]
		\centering
		\begin{threeparttable}
			\caption {Propensity score model parameters specified by our DGP}\label{tab:coef}
			\begin{tabular}{cccccccccccccc}
				\toprule
				Model & $\beta_0$ & $\beta_1$ & $\beta_2$ & $\beta_3$ & $\beta_4$ & $\beta_5$ & $\beta_6$ & $\beta_7$ & $p$ & $r$ \\ 
				\midrule
				1 & -3.07 & 0.3 & 0.4 & 0.4 & 0.4 & -0.1 & -0.1 & 0.1 & 10.05\% & 2.54 \\
				2 & -1.82 & -0.25 & 0.45 & -0.3 & 0.65 & -0.03 & -0.03 & 0.07 & 20.77\% & 1.80 \\ 
				3 & -0.37 & -0.25 & 0.45 & -0.3 & 0.65 & -0.03 & -0.03 & 0.07 & 49.72\% & 1.13 \\
				4 & 0.98 & 0.3 & 0.4 & 0.4 & 0.4 & -0.1 & -0.1 & 0.1 & 79.22\% & 0.42 \\ 
				5 & 1.86 & 0.3 & 0.4 & 0.4 & 0.4 & -0.1 & -0.1 & 0.1 & 89.18\% & 0.26 \\ 
				6 & 1.12 & -0.25 & 0.45 & -0.3 & 0.65 & -0.03 & -0.03 & 0.07 & 79.59\% & 0.75 \\
				\bottomrule
			\end{tabular}
			\begin{tablenotes}
				\tiny
				\item  $p$: proportion of treated participants; $r$: ratio of variances of propensity scores (treatment vs. control group)
			\end{tablenotes}
		\end{threeparttable}
	\end{table}
    
\begin{figure}[htbp]
		\centering
		\includegraphics[trim=5 15 0 5, clip, width=0.9\textwidth]{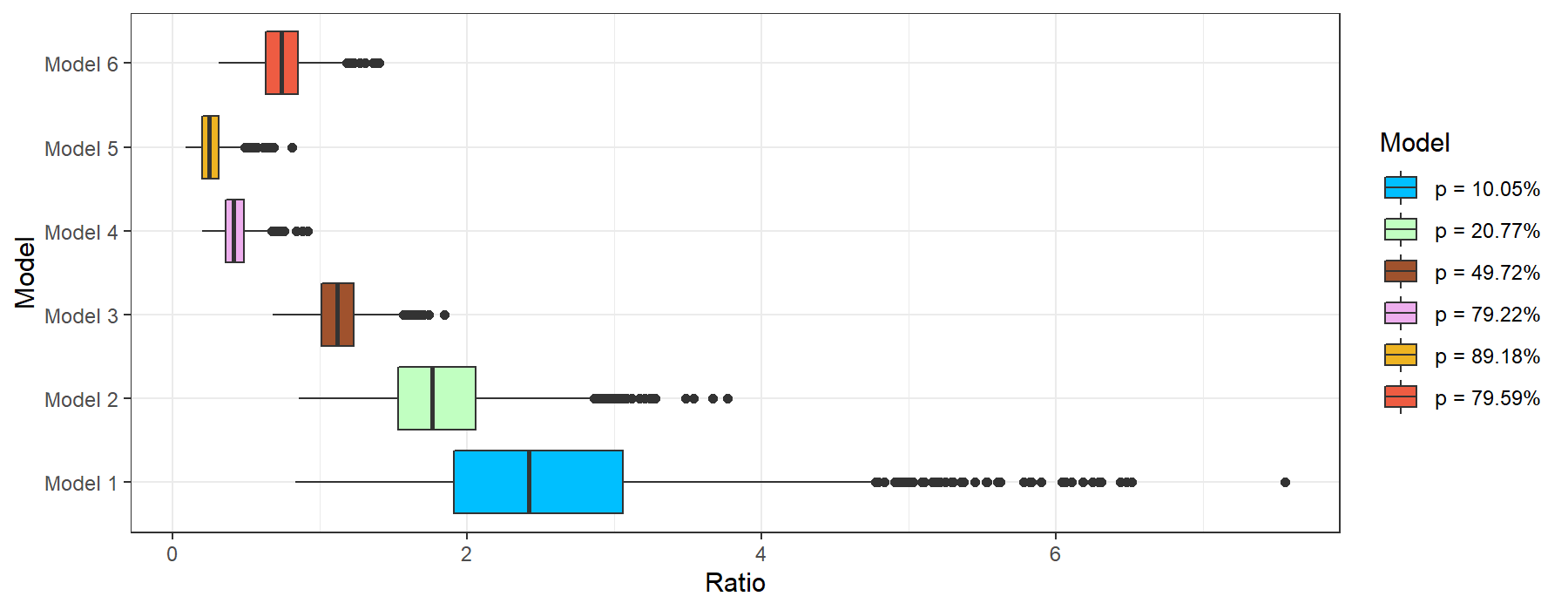}
		\caption{Ratio of estimated variance of propensity score in treatment group to control group}\label{fig:apx-varratio}
	\end{figure}

    Figure     
	\ref{fig:apx-varratio} shows the boxplot of $r$'s of the 6 models over the 2000  simulation replicates.
 
\begin{figure}[htbp]
    	\centering
    	\includegraphics[trim=5 10 0 12, clip, width=1\textwidth]{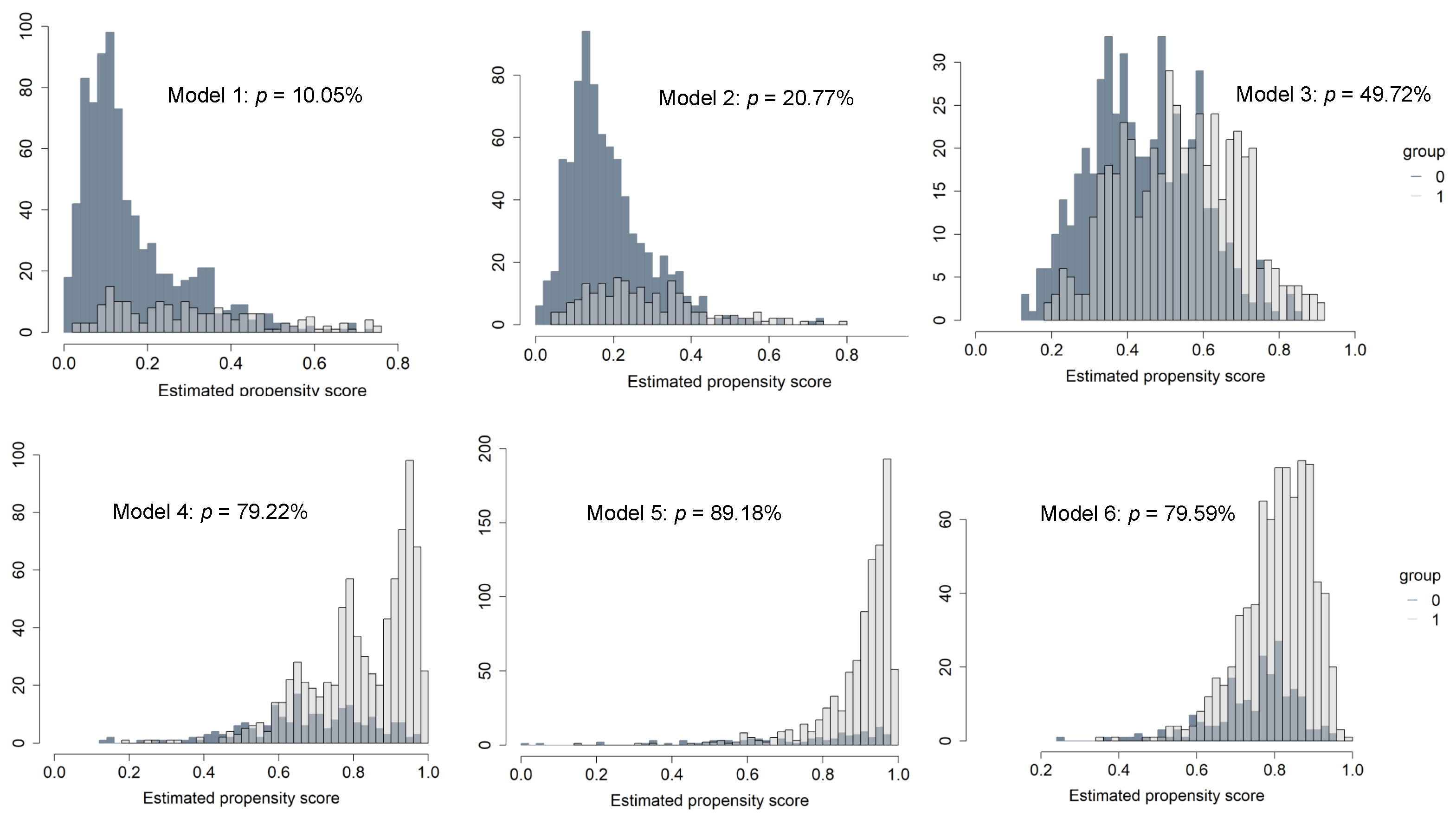}
    	\caption{Estimated propensity score histograms by treatment group of models 1--6}	\label{fig:apx-ps-sim}
    \end{figure}

	We also provide the estimated propensity score (PS) distributions of all 6 models in Figure \ref{fig:apx-ps-sim}. To do so, we randomly pick a sample out of the 2000 simulation replicates and plot the estimated PS histograms using the correctly specified PS model. Table \ref{tab:ess} shows the effective sample sizes (ESS) by group of the 6 models, which reflects the performance of different PS weights. 
	
	Finally, we give the true values of the heterogeneous treatment effects of all models in table \ref{tab:truth-all}.
	
	\begin{table}[htbp]
	\small
	\centering
	\begin{threeparttable}
		\caption {True heterogeneous treatment effects by different weighted methods}\label{tab:truth-all}
		\begin{tabular}{cccccccccccc}
			\toprule
			Model & $p$ & $r$ & ATE & ATE (0.05)& ATE (0.1)& ATE (015)& ATO & ATM & ATEN & ATC & ATT \\ 
			\midrule
			1 & 10.05\% & 2.54 & 17.22 & 16.61 & 22.79 & 30.46 & 20.53 & 22.28 & 19.26 & 16.62 & 22.59 \\
	 		2 & 20.77\% & 1.80 & 17.22 & 17.15 & 16.90 & 17.08 & 17.72 & 18.33 & 17.55 & 16.81 & 18.76 \\ 
	 		3 & 49.72\% & 1.13 & 17.22 & 17.20 & 17.12 & 16.91 & 16.69 & 16.30 & 16.81 & 16.61 & 17.83 \\
			4 & 79.22\% & 0.42 & 17.22 & 15.28 & 13.48 & 13.87 & 15.39 & 15.81 & 15.55 & 18.58 & 16.86 \\ 
			5 & 89.18\% & 0.26 & 17.22 & 13.67 & 16.59 & 22.78 & 17.36 & 18.84 & 16.78 & 20.79 & 16.78 \\ 
			6 & 79.59\% & 0.75 & 17.22 & 16.75 & 15.78 & 15.77 & 16.56 & 16.63 & 16.63 & 16.70 & 17.35\\
			\bottomrule
		\end{tabular}
		\begin{tablenotes}
			\tiny
	\item  $p$: percentage of treated participants; $r$: ratio of variances of propensity scores (treatment vs. control group); 
   \item ATE: average treatment effect; ATE ($\alpha$): ATE by trimming PS outside of $\alpha< e(X)<1-\alpha$, $\alpha=0.05, 0.1, 0.15$; 
   \item ATO (resp. ATM, ATEN, ATC, ATT): average treatment effect on overlap (resp. matching, entropy, controls, and treated)
		\end{tablenotes}
	\end{threeparttable}
	\end{table}
	
    \begin{table}[htbp]\footnotesize
    	\centering
    \begin{threeparttable}
    	\caption {Effective sample sizes (ESS) for all propensity score weights (Simulation models 1--6)}
    	\label{tab:ess}
    	\begin{tabular}{crcccccccccccccccccccccccccccccccccccccc}
    		\toprule
 Group & IPW & \makecell{ATE (0.05)} & \makecell{ATE  (0.1)} & \makecell{ATE   (0.15)} & OW & MW & EW & ATC & ATT \\ 
    		\midrule
    		\multicolumn{10}{c}{\textbf{Model 1}, $p=10.05\%$, $r=2.54$} \\\cmidrule(lr){2-10}
    Control & 886.28 & 580.09 & 297.93 & 159.48 & 530.09 & 437.23 & 620.19 & 898.41 & 410.27 \\ 
    Treated & 52.35 & 69.12 & 60.32 & 45.73 & 99.25 & 101.14 & 94.15 & 47.15 & 101.59 \\ 
    		\addlinespace
    		\multicolumn{10}{c}{\textbf{Model 2}, $p=20.77\%$, $r=1.80$} \\\cmidrule(lr){2-10}
    Control & 767.65 & 748.33 & 635.68 & 458.58 & 617.22 & 528.77 & 661.71 & 792.37 & 471.82 \\ 
    Treated & 152.23 & 158.77 & 164.76 & 151.44 & 200.46 & 205.42 & 194.03 & 132.09 & 207.63 \\
    		\addlinespace
    		\multicolumn{10}{c}{\textbf{Model 3}, $p=49.72\%$, $r=1.13$} \\\cmidrule(lr){2-10}
    Control & 441.66 & 438.16 & 432.71 & 421.32 & 444.86 & 429.45 & 447.79 & 541.46 & 265.54 \\ 
    Treated & 333.81 & 350.74 & 367.48 & 372.51 & 395.51 & 392.80 & 390.76 & 207.89 & 458.54 \\ 
    		\addlinespace
    		\multicolumn{10}{c}{\textbf{Model 4}, $p=79.22\%$, $r=0.42$} \\\cmidrule(lr){2-10}
    Control & 121.82 & 140.07 & 150.83 & 143.26 & 192.69 & 198.93 & 183.87 & 207.96 & 98.52 \\ 
    Treated & 732.93 & 672.33 & 517.12 & 381.13 & 539.84 & 451.14 & 590.98 & 309.95 & 792.04 \\ 
    		\addlinespace
    		\multicolumn{10}{c}{\textbf{Model 5}, $p=89.18\%$, $r=0.26$} \\\cmidrule(lr){2-10}
    Control & 55.09 & 70.94 & 61.35 & 46.07 & 102.73 & 105.52 & 96.91 & 107.92 & 49.13  \\ 
    Treated & 867.74 & 570.55 & 294.78 & 144.05 & 500.88 & 392.97 & 595.16 & 308.56 & 892.08 \\
    		\addlinespace
    		\multicolumn{10}{c}{\textbf{Model 6}, $p = 79.59\%$, $r=0.75$} \\\cmidrule(lr){2-10}
    Control & 148.63 & 158.37 & 165.38 & 153.22 & 199.45 & 203.66 & 193.31 & 204.42 & 129.12 \\ 
    Treated  & 778.19 & 750.15 & 640.26 & 473.07 & 628.90 & 547.46 & 672.03 & 525.07 & 795.58 \\ 
    		\bottomrule 
    	\end{tabular}
    	\begin{tablenotes}
    		\tiny
    		\item ATE : average treatment effect; ATE ($\alpha$): ATE  by trimming outside of $\alpha < e(\boldsymbol{x})<1-\alpha$, $\alpha=0.05, 0.1, 0.15$;
            \item OW: overlap weight; MW: matching weight; EW: entropy weight; IPWC: IPW on the controls; IPWT: IPW on the treated; $p$: proportion of treated participants; $r$: ratio of variances of propensity scores (treatment vs. control group);
    	\end{tablenotes}
    \end{threeparttable}
    \end{table}

\subsection{Full simulation results}\label{subapx:sim-allres}

We provide the full set of simulation results of all the models we considered in Tables \ref{tab:md1-all}-- \ref{tab:md6-all}. In each model, the results by H\'ajek-type (weighted) estimator and the augmented estimator of all weighted average treatment effects (WATE), including ATE with and without trimming (at 0.05, 0.1, 0.15), ATO, ATM, ATEN, ATT and ATC, are reported. For augmented estimators, we considered 4 cases of model specifications: (i) both propensity score (PS) and outcome (OR) models are correctly specified; (ii) only PS model is correctly specified; (iii) only OR model is correctly specified; (iv) both PS and OR models are misspecified. 

%%%%%%%%%%%%%%%%%%%%%%%%%%%%%%%%%%%%%%%%%%%%%%%%%%%%%%%%%%%%%%%%%%%%%%%%%%%%%%%%%%%%%%%%%%%%
%%%%%%%%%%%%%%%%%%%%%%%%%%%%%%%%%%%% figure results %%%%%%%%%%%%%%%%%%%%%%%%%%%%%%%%%%%%%%%%
%%%%%%%%%%%%%%%%%%%%%%%%%%%%%%%%%%%%%%%%%%%%%%%%%%%%%%%%%%%%%%%%%%%%%%%%%%%%%%%%%%%%%%%%%%%%

%%%% Point estimates %%%%
\begin{figure}[htbp]
	\begin{center}
	\includegraphics[trim=10 15 0 5, clip, width=1.05\textwidth]{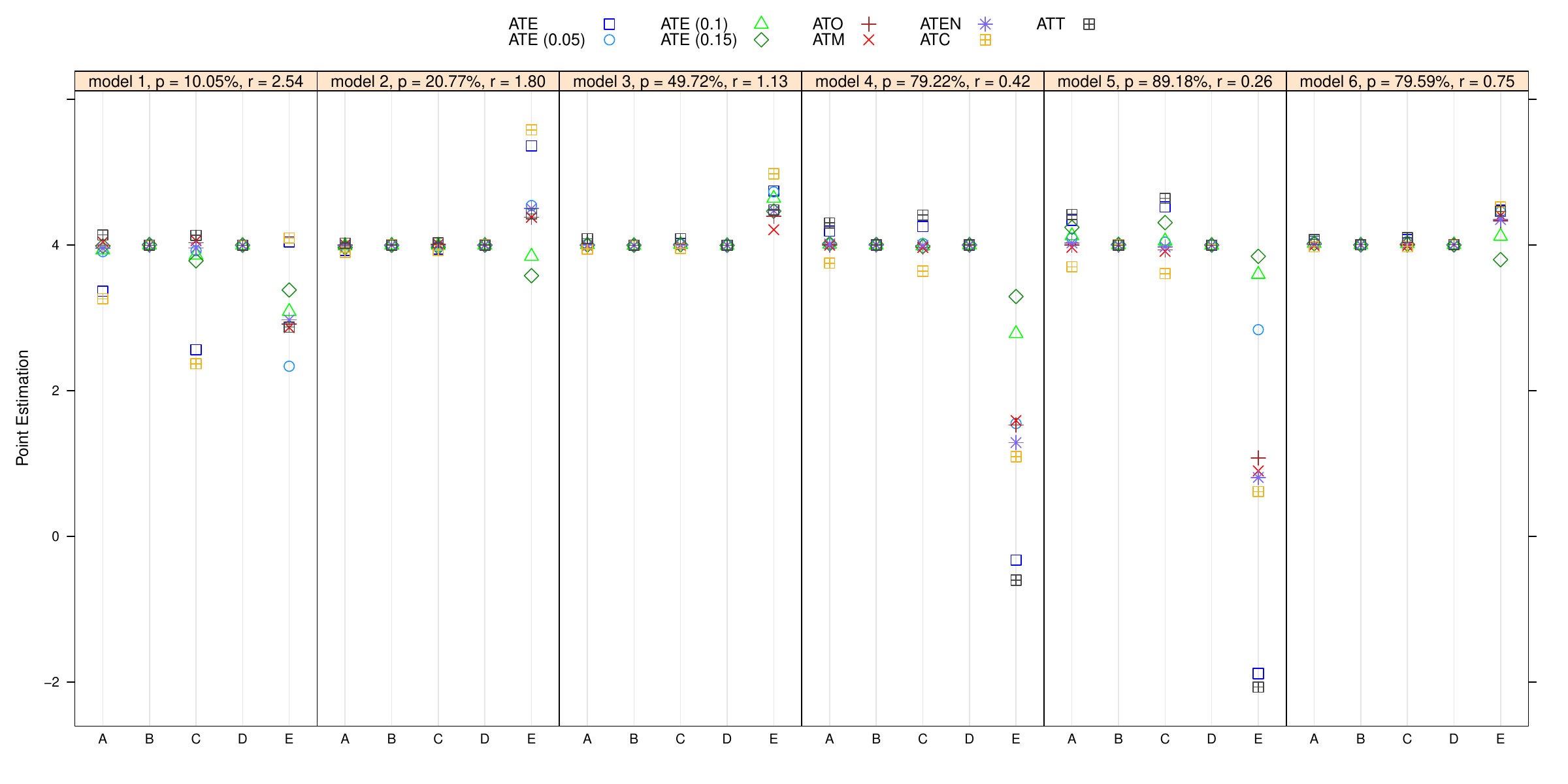}
	
% 	{\footnotesize A: H\'ajek-type (weighted) estimator; B (resp. C, D, and E): augmented estimator, with both models correctly specified (resp. PS model correctly specified, OR model correctly specified, both PS and OR models misspecified)}
% 	\caption{Point estimates under constant treatment effects of models 1--6}\label{fig:PE-cons-xy-all}
% \end{figure}  
\begin{threeparttable}
% \begin{figure}[htbp]
% 	\centering
% \captionsetup{format = hang}

\includegraphics[trim=10 15 0 35, clip, width=1.05\textwidth]{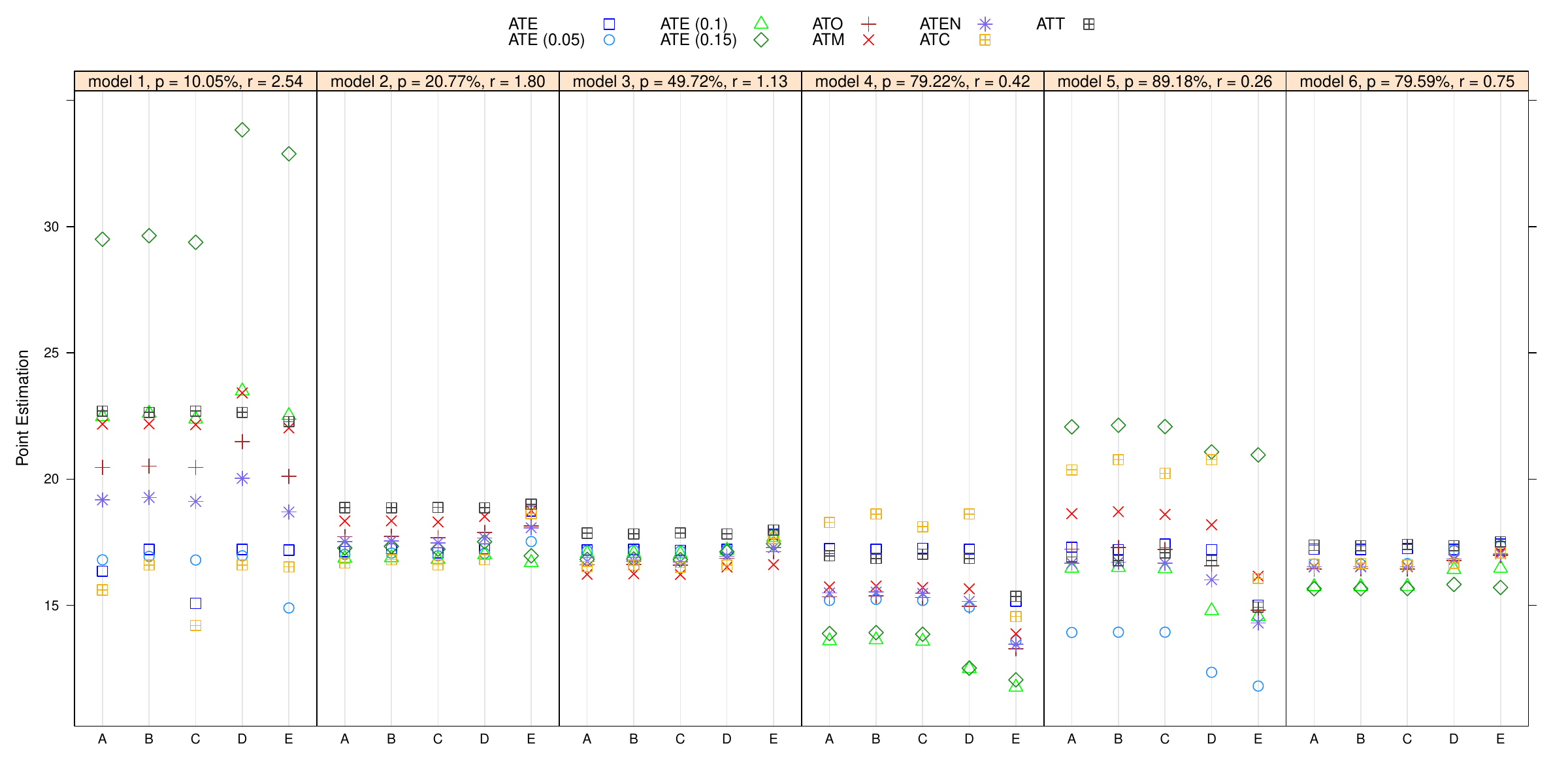}
{ 

\begin{tablenotes} \tiny 
\item A: H\'ajek-type (weighted) estimator; B (resp. C, D, and E): augmented estimator, with both models correctly specified  (resp. PS model correctly specified, OR model correctly specified, both PS and OR models misspecified)\end{tablenotes}}
 \end{threeparttable}
 \end{center} 
 \caption{Point estimates under constant (top) and heterogeneous (bottom) treatment effects}\label{fig:PE-hete-xy-all}
\end{figure} 

%%%% Bias boxplots %%%%
\begin{figure}[htbp]
	\centering
	\includegraphics[width=1.05\textwidth]{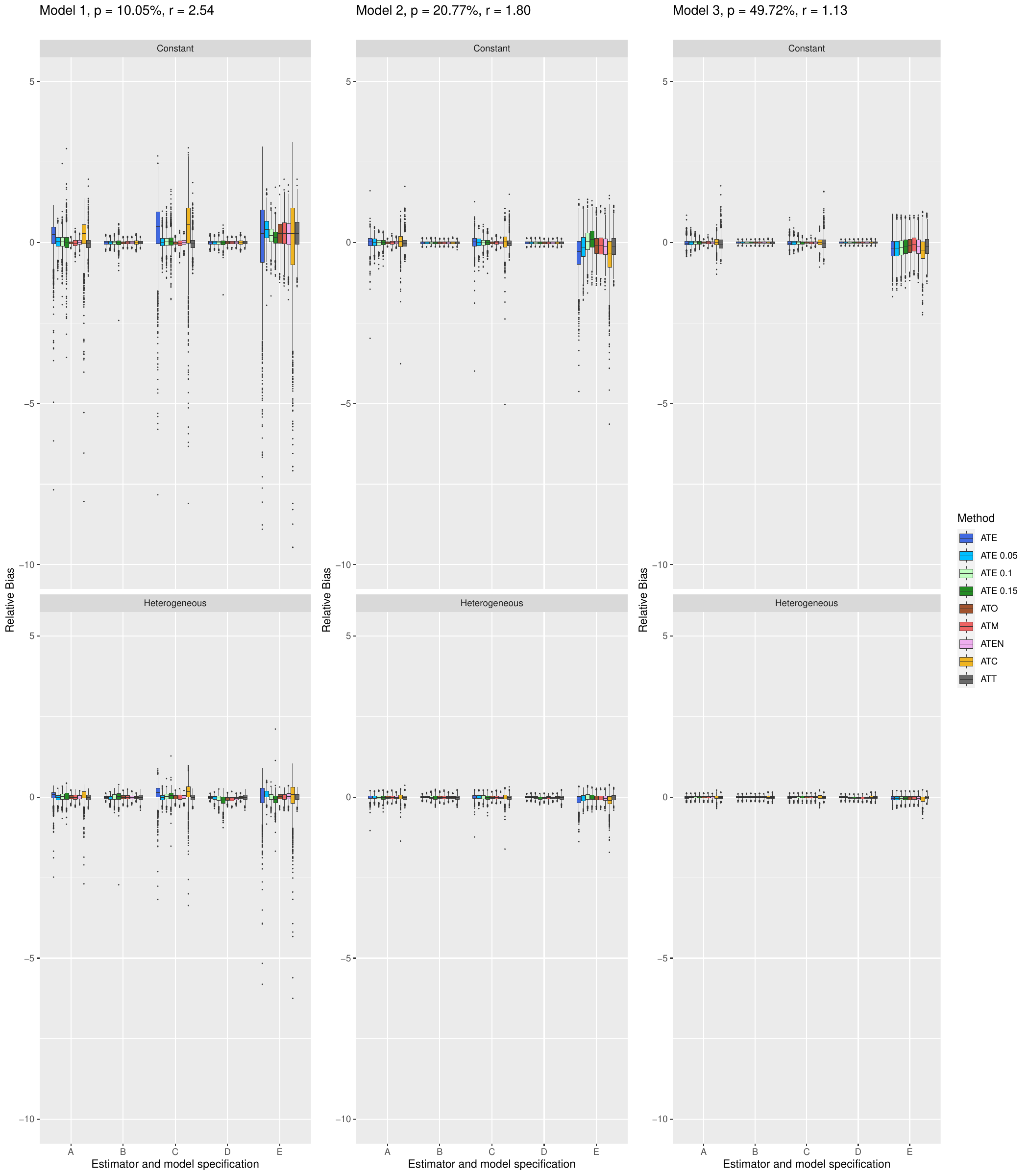}
	{\begin{tablenotes} \tiny 
 \item  A: H\'ajek-type (weighted) estimator; B (resp. C, D, and E): augmented estimator, with both models correctly specified (resp. PS model correctly specified, OR model correctly specified, both PS and OR models misspecified); the ``Heterogeneous'' (resp. ``Constant'') in the green boxes means the window is under heterogeneous (resp. constant) treatment effect.\end{tablenotes}}
	\caption{Relative bias boxplots of point estimates (for models 1--3)}\label{fig:RB-md123}
\end{figure} 

\begin{figure}[htbp]
	\centering
	\includegraphics[width=1.05\textwidth]{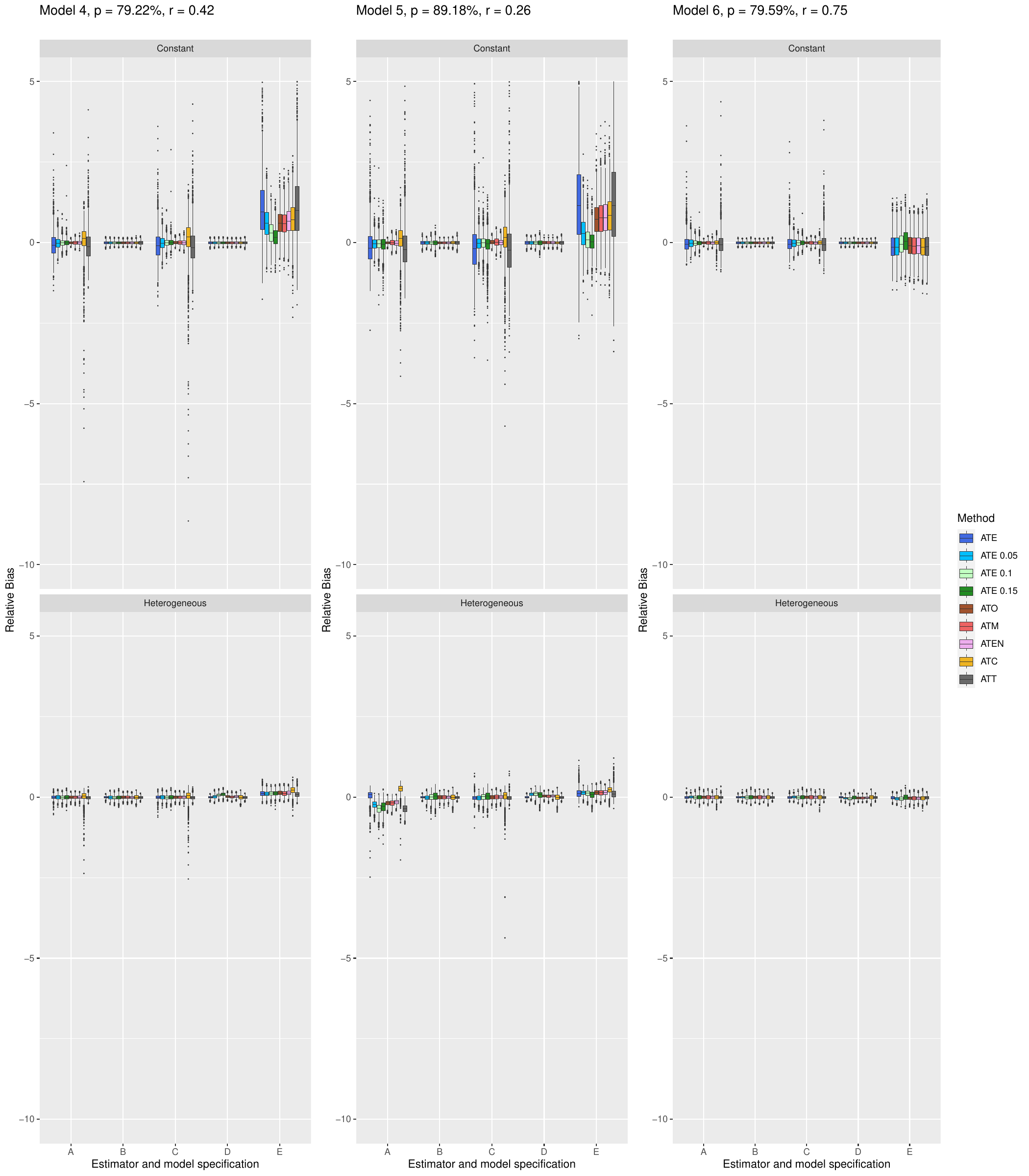}
	{\begin{tablenotes} 
 \tiny 
 \item  A: H\'ajek-type (weighted) estimator; B (resp. C, D, and E): augmented estimator, with both models correctly specified (resp. PS model correctly specified, OR model correctly specified, both PS and OR models misspecified); the ``Heterogeneous'' (resp. ``Constant'') in the green boxes means the window is under heterogeneous (resp. constant) treatment effect.\end{tablenotes}
 }
	\caption{Relative bias boxplots of point estimates (for models 4--6)}\label{fig:RB-md456}
\end{figure} 

%%%% Relative efficiency %%%%
\begin{figure}[htbp]
\begin{threeparttable}
	\centering
	\includegraphics[trim=10 15 10 5, clip, width=1.05\textwidth]{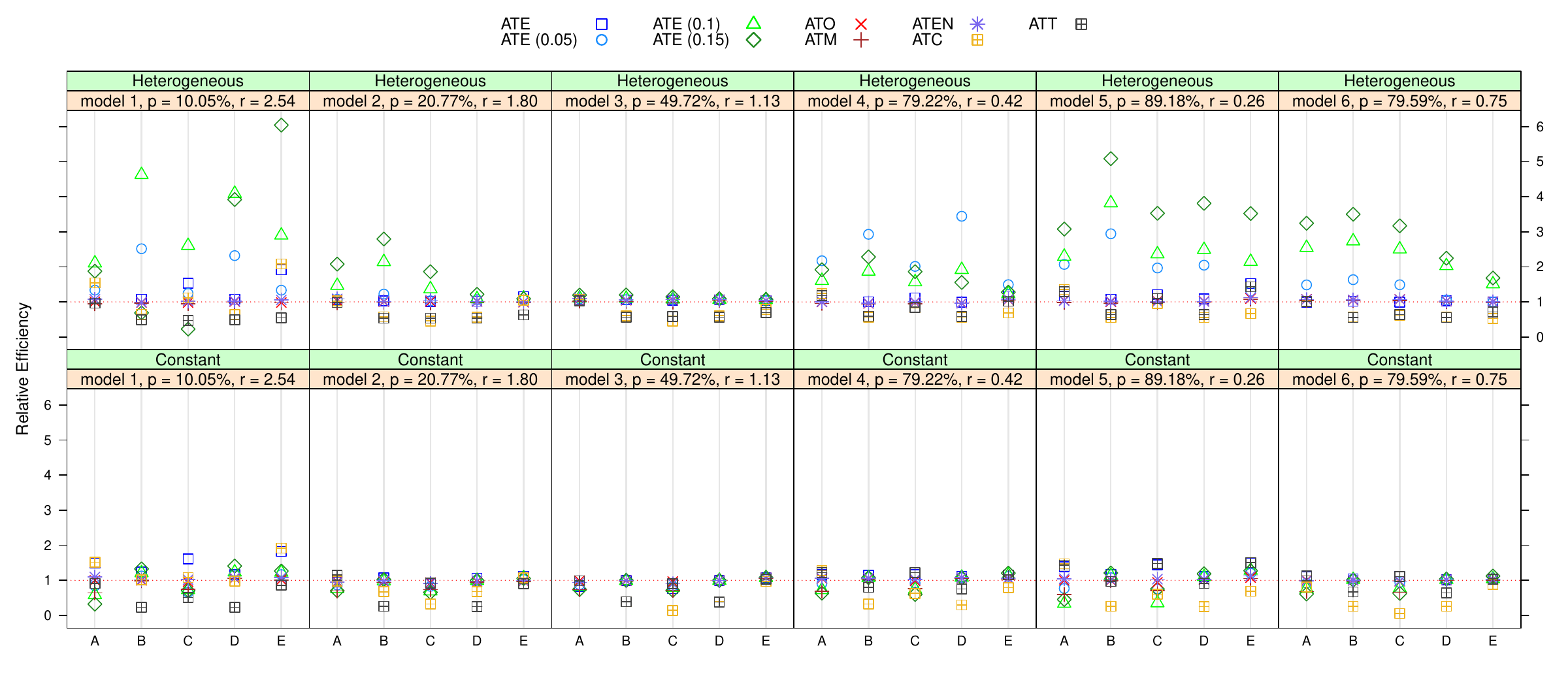}
	{\begin{tablenotes} \tiny 
 \item   The red dotted line indicates the relative efficiency of 1; A: H\'ajek-type (weighted) estimator; B (resp. C, D, and E): augmented estimator, with both models correctly specified (resp. PS model correctly specified, OR model correctly specified, both PS and OR models misspecified); the ``Heterogeneous'' (resp. ``Constant'') in the green boxes means the window is under heterogeneous (resp. constant) treatment effect.\end{tablenotes}}
	\caption{Relative efficiencies of sandwich variance estimations of all WATE estimators (models 1--6)}\label{fig:RE-xy-all}
 \end{threeparttable}
\end{figure}  

%%%% Coverage probability %%%%
\begin{figure}[htbp]
\begin{threeparttable}
	\centering
	\includegraphics[trim=10 15 10 5, clip, width=1.05\textwidth]{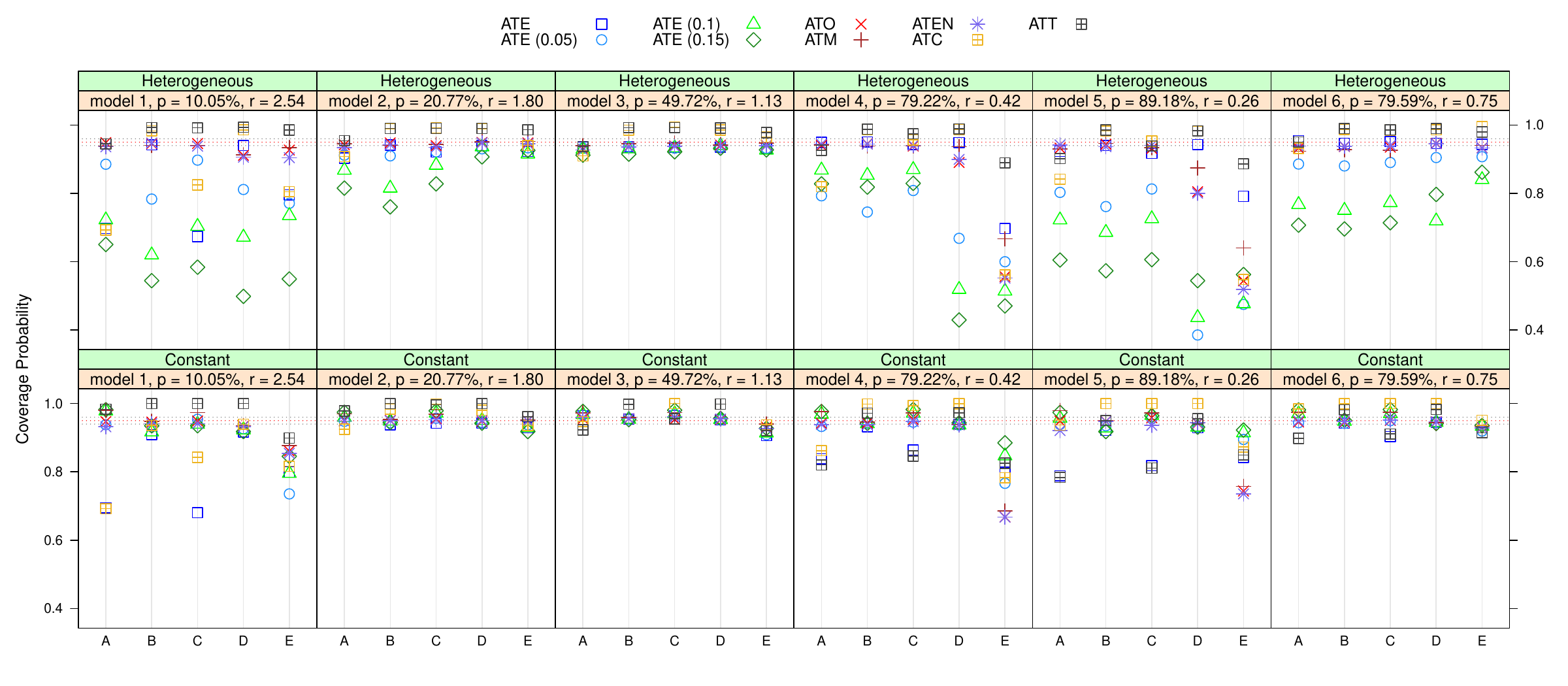}	
	{\begin{tablenotes} \tiny 
 \item    The red dotted line indicates the nominal 0.95 coverage level, the two gray dotted lines indicate 0.94 and 0.96 which cover a range that the coverage probability does not significantly different from the 0.95 nominal level; A: H\'ajek-type (weighted) estimator; B (resp. C, D, and E): augmented estimator, with both models correctly specified (resp. PS model correctly specified, OR model correctly specified, both PS and OR models misspecified); the ``Heterogeneous'' (resp. ``Constant'') in the green boxes means the window is under heterogeneous (resp. constant) treatment effect.\end{tablenotes}}
	\caption{Coverage probabilities of confidence intervals for the estimands (models 1--6)}\label{fig:CP-xy-all}
 \end{threeparttable}
\end{figure}  

%%%%%%%%%%%%%%%%%%%%%%%%%%%%%%%%%%%%%%%%%%%%%%%%%%%%%%%%%%%%%%%%%%%%%%%%%%%%%%%%%%%%%%%%%%%%
%%%%%%%%%%%%%%%%%%%%%%%%%%%%%%%%%%%% model 1 result %%%%%%%%%%%%%%%%%%%%%%%%%%%%%%%%%%%%%%%%
%%%%%%%%%%%%%%%%%%%%%%%%%%%%%%%%%%%%%%%%%%%%%%%%%%%%%%%%%%%%%%%%%%%%%%%%%%%%%%%%%%%%%%%%%%%%

\begin{table}[htbp]\tiny
	\centering
	\begin{threeparttable}
		\caption{Model 1, with $p = 10.05\%$, $r=2.54$}
		\label{tab:md1-all}
		\begin{tabular}{llcccccccccccccccccc}
			\toprule	
			 \multicolumn{16}{c}{\textbf{Constant treatment effect}} \\ \cmidrule(lr){2-16}
			& \multicolumn{5}{c}{H\'ajek-type estimator} & \multicolumn{5}{c}{Aug: both models correctly specified} & \multicolumn{5}{c}{Aug: PS model correctly specified} \\ \cmidrule(lr){2-6}\cmidrule(lr){7-11}\cmidrule(lr){12-16}
			Estimand & PE & Bias & RMSE & RE & CP & PE & Bias & RMSE & RE & CP & PE & Bias & RMSE & RE & CP \\ 
			\midrule
			ATE & 3.37 & 15.84 & 2.33 & 1.49 & 0.69 & 4.00 & 0.09 & 0.31 & 1.23 & 0.91 & 2.56 & 35.89 & 4.53 & 1.62 & 0.68 \\ 
			ATE (0.05) & 3.91 & 2.26 & 0.92 & 0.84 & 0.94 & 4.00 & 0.12 & 0.26 & 1.12 & 0.93 & 3.92 & 1.91 & 0.82 & 0.65 & 0.95 \\ 
			ATE (0.1) & 3.93 & 1.63 & 1.01 & 0.59 & 0.98 & 4.00 & 0.04 & 0.32 & 1.21 & 0.92 & 3.86 & 3.43 & 0.81 & 0.73 & 0.94 \\ 
			ATE (0.15) & 3.99 & 0.35 & 1.50 & 0.32 & 0.98 & 4.00 & 0.12 & 0.48 & 1.33 & 0.94 & 3.78 & 5.39 & 1.05 & 0.71 & 0.94 \\ 
			ATO & 4.00 & 0.04 & 0.22 & 1.00 & 0.95 & 4.00 & 0.04 & 0.22 & 1.00 & 0.95 & 4.03 & 0.80 & 0.26 & 0.96 & 0.95 \\ 
			ATM & 4.04 & 1.10 & 0.49 & 0.64 & 0.98 & 4.00 & 0.04 & 0.22 & 0.99 & 0.95 & 4.06 & 1.52 & 0.47 & 0.75 & 0.97 \\ 
			ATEN & 3.96 & 1.12 & 0.37 & 1.09 & 0.93 & 4.00 & 0.05 & 0.22 & 1.01 & 0.94 & 3.96 & 0.95 & 0.40 & 1.02 & 0.94 \\ 
			ATC & 3.26 & 18.46 & 2.61 & 1.52 & 0.69 & 4.00 & 0.10 & 0.33 & 1.01 & 0.94 & 2.37 & 40.71 & 4.96 & 1.08 & 0.84 \\ 
			ATT & 4.14 & 3.48 & 1.00 & 0.91 & 0.98 & 4.00 & 0.04 & 0.22 & 0.23 & 1.00 & 4.13 & 3.35 & 0.91 & 0.51 & 1.00 \\ 
			\addlinespace 
			& \multicolumn{5}{c}{Aug: OR model correctly specified} & \multicolumn{5}{c}{Aug: both models misspecified} & \multicolumn{5}{c}{} \\ \cmidrule(lr){2-6}\cmidrule(lr){7-11}\cmidrule(lr){12-16}
			Estimand & PE & Bias & RMSE & RE & CP & PE & Bias & RMSE & RE & CP & \multicolumn{5}{c}{} \\
			\midrule
			ATE & 4.00 & 0.10 & 0.30 & 1.16 & 0.92 & 4.04 & 1.06 & 6.65 & 1.83 & 0.80 & \multicolumn{5}{c}{}\\ 
			ATE (0.05) & 4.00 & 0.12 & 0.28 & 1.15 & 0.93 & 2.34 & 41.58 & 2.29 & 1.16 & 0.74 & \multicolumn{5}{c}{}\\ 
			ATE (0.1) & 4.00 & 0.05 & 0.32 & 1.23 & 0.92 & 3.09 & 22.77 & 1.50 & 1.23 & 0.80 & \multicolumn{5}{c}{}\\ 
			ATE (0.15) & 4.00 & 0.03 & 0.43 & 1.41 & 0.92 & 3.38 & 15.41 & 1.28 & 1.27 & 0.85 & \multicolumn{5}{c}{}\\ 
			ATO & 4.00 & 0.11 & 0.23 & 1.06 & 0.93 & 2.92 & 27.11 & 2.07 & 1.02 & 0.86 & \multicolumn{5}{c}{}\\ 
			ATM & 4.00 & 0.11 & 0.23 & 1.05 & 0.94 & 2.87 & 28.29 & 2.24 & 1.01 & 0.88 & \multicolumn{5}{c}{}\\ 
			ATEN & 4.00 & 0.10 & 0.24 & 1.07 & 0.93 & 2.97 & 25.68 & 2.19 & 1.08 & 0.85 & \multicolumn{5}{c}{}\\ 
			ATC & 4.00 & 0.11 & 0.32 & 0.98 & 0.94 & 4.10 & 2.46 & 7.11 & 1.92 & 0.82 & \multicolumn{5}{c}{}\\ 
			ATT & 4.00 & 0.04 & 0.22 & 0.23 & 1.00 & 2.88 & 28.11 & 2.30 & 0.86 & 0.90 & \multicolumn{5}{c}{}\\ 
			\addlinespace
			\multicolumn{16}{c}{\textbf{Heterogeneous treatment effect}} \\ \cmidrule(lr){2-16}
			& \multicolumn{5}{c}{H\'ajek-type estimator} & \multicolumn{5}{c}{Aug: both models correctly specified} & \multicolumn{5}{c}{Aug: PS model correctly specified} \\ \cmidrule(lr){2-6}\cmidrule(lr){7-11}\cmidrule(lr){12-16}
			Estimand & PE & Bias & RMSE & RE & CP & PE & Bias & RMSE & RE & CP & PE & Bias & RMSE & RE & CP \\ 
			\midrule
			ATE & 16.35 & 5.04 & 3.22 & 1.54 & 0.70 & 17.21 & 0.04 & 0.68 & 1.08 & 0.94 & 15.08 & 12.43 & 5.37 & 1.53 & 0.67 \\ 
			ATE (0.05) & 16.80 & 1.14 & 1.72 & 1.33 & 0.89 & 16.94 & 1.96 & 1.25 & 2.52 & 0.78 & 16.79 & 1.10 & 1.66 & 1.24 & 0.90 \\ 
			ATE (0.1) & 22.47 & 1.39 & 2.81 & 2.11 & 0.72 & 22.61 & 0.79 & 2.45 & 4.63 & 0.62 & 22.37 & 1.82 & 2.67 & 2.61 & 0.70 \\ 
			ATE (0.15) & 29.50 & 3.16 & 4.34 & 1.87 & 0.65 & 29.64 & 2.70 & 4.41 & 0.69 & 0.54 & 29.38 & 3.55 & 8.68 & 0.22 & 0.58 \\ 
			ATO & 20.46 & 0.34 & 1.55 & 1.00 & 0.95 & 20.51 & 0.10 & 1.52 & 0.98 & 0.95 & 20.46 & 0.35 & 1.60 & 0.98 & 0.95 \\ 
			ATM & 22.18 & 0.45 & 1.98 & 0.94 & 0.94 & 22.18 & 0.42 & 1.96 & 0.97 & 0.94 & 22.16 & 0.55 & 2.01 & 0.95 & 0.94 \\ 
			ATEN & 19.17 & 0.45 & 1.36 & 1.07 & 0.94 & 19.27 & 0.06 & 1.23 & 0.99 & 0.95 & 19.12 & 0.75 & 1.43 & 1.02 & 0.94 \\ 
			ATC & 15.61 & 6.08 & 3.58 & 1.54 & 0.69 & 16.60 & 0.13 & 0.70 & 0.65 & 0.98 & 14.20 & 14.54 & 5.90 & 1.12 & 0.82 \\ 
			ATT & 22.69 & 0.45 & 2.31 & 0.96 & 0.94 & 22.64 & 0.23 & 2.26 & 0.49 & 0.99 & 22.69 & 0.46 & 2.30 & 0.47 & 0.99 \\ 
			\hline
			\addlinespace 
			& \multicolumn{5}{c}{Aug: OR model correctly specified} & \multicolumn{5}{c}{Aug: both models misspecified} & \multicolumn{5}{c}{} \\ \cmidrule(lr){2-6}\cmidrule(lr){7-11}\cmidrule(lr){12-16}
			Estimand & PE & Bias & RMSE & RE & CP & PE & Bias & RMSE & RE & CP & \multicolumn{5}{c}{} \\
			\midrule
			ATE & 17.21 & 0.03 & 0.68 & 1.07 & 0.94 & 17.19 & 0.17 & 8.10 & 1.91 & 0.80  & \multicolumn{5}{c}{} \\ 
			ATE (0.05) & 16.96 & 2.10 & 1.20 & 2.32 & 0.81 & 14.90 & 10.33 & 2.77 & 1.33 & 0.77  & \multicolumn{5}{c}{} \\ 
			ATE (0.1) & 23.49 & 3.08 & 2.44 & 4.09 & 0.67 & 22.51 & 1.20 & 2.80 & 2.90 & 0.73  & \multicolumn{5}{c}{} \\ 
			ATE (0.15) & 33.83 & 11.07 & 19.40 & 3.92 & 0.50 & 32.88 & 7.96 & 11.20 & 6.04 & 0.55  & \multicolumn{5}{c}{} \\ 
			ATO & 21.48 & 4.61 & 1.64 & 1.01 & 0.91 & 20.11 & 2.09 & 2.23 & 0.98 & 0.93  & \multicolumn{5}{c}{} \\ 
			ATM & 23.41 & 5.09 & 2.12 & 1.00 & 0.91 & 22.02 & 1.13 & 2.51 & 0.97 & 0.93  & \multicolumn{5}{c}{} \\ 
			ATEN & 20.03 & 3.99 & 1.32 & 1.02 & 0.91 & 18.70 & 2.93 & 2.39 & 1.06 & 0.90  & \multicolumn{5}{c}{} \\ 
			ATC & 16.60 & 0.12 & 0.70 & 0.64 & 0.99 & 16.52 & 0.57 & 8.68 & 2.08 & 0.80  & \multicolumn{5}{c}{} \\ 
			ATT & 22.64 & 0.23 & 2.26 & 0.49 & 0.99 & 22.28 & 1.39 & 2.65 & 0.54 & 0.99  & \multicolumn{5}{c}{} \\ 
			\bottomrule
		\end{tabular}
		\begin{tablenotes}
			\tiny
			\item PE: point estimation; Bias: absolute relative bias$\times 100$; RMSE: root mean squared error; RE: relative efficiency; CP: coverage probability; Aug: augmented estimator; PS: propensity score; OR: outcome regression; ATE: average treatment effect; ATE ($\alpha$): ATE by trimming PS $>1-\alpha$ or PS $<\alpha$, $\alpha=0.05, 0.1, 0.15$; ATO (resp. ATM, ATEN, ATC, and ATT): average treatment effect on overlap (resp. matching, entropy, controls, and treated)
		\end{tablenotes}
	\end{threeparttable}
\end{table}	 

%%%%%%%%%%%%%%%%%%%%%%%%%%%%%%%%%%%%%%%%%%%%%%%%%%%%%%%%%%%%%%%%%%%%%%%%%%%%%%%%%%%%%%%%%%%%
%%%%%%%%%%%%%%%%%%%%%%%%%%%%%%%%%%%% model 2 result %%%%%%%%%%%%%%%%%%%%%%%%%%%%%%%%%%%%%%%%
%%%%%%%%%%%%%%%%%%%%%%%%%%%%%%%%%%%%%%%%%%%%%%%%%%%%%%%%%%%%%%%%%%%%%%%%%%%%%%%%%%%%%%%%%%%%
\begin{table}[htbp]
	\centering
	\begin{threeparttable}\tiny
		\caption{Model 2, with $p = 20.77\%$, $r=1.80$}
		\label{tab:md2-all}
		\begin{tabular}{llcccccccccccccccccc}
			\toprule	
			\multicolumn{16}{c}{\textbf{Constant treatment effect}} \\ \cmidrule(lr){2-16}
			& \multicolumn{5}{c}{H\'ajek-type estimator} & \multicolumn{5}{c}{Aug: both models correctly specified} & \multicolumn{5}{c}{Aug: PS model correctly specified} \\ \cmidrule(lr){2-6}\cmidrule(lr){7-11}\cmidrule(lr){12-16}
			Estimand & PE & Bias & RMSE & RE & CP & PE & Bias & RMSE & RE & CP & PE & Bias & RMSE & RE & CP \\ 
			\midrule
			ATE & 3.93 & 1.82 & 0.83 & 0.95 & 0.93 & 4.00 & 0.06 & 0.18 & 1.07 & 0.94 & 3.95 & 1.33 & 0.90 & 0.90 & 0.94 \\ 
			ATE (0.05) & 3.97 & 0.76 & 0.61 & 0.86 & 0.95 & 4.00 & 0.05 & 0.17 & 1.04 & 0.94 & 3.98 & 0.55 & 0.65 & 0.83 & 0.96 \\ 
			ATE (0.1) & 3.99 & 0.22 & 0.49 & 0.76 & 0.96 & 4.00 & 0.01 & 0.17 & 0.98 & 0.95 & 4.00 & 0.09 & 0.49 & 0.69 & 0.97 \\ 
			ATE (0.15) & 3.97 & 0.86 & 0.43 & 0.70 & 0.97 & 4.00 & 0.07 & 0.19 & 1.02 & 0.94 & 3.97 & 0.74 & 0.44 & 0.66 & 0.98 \\ 
			ATO & 4.00 & 0.01 & 0.16 & 0.96 & 0.95 & 4.00 & 0.02 & 0.16 & 0.96 & 0.95 & 4.01 & 0.20 & 0.17 & 0.93 & 0.96 \\ 
			ATM & 4.01 & 0.19 & 0.31 & 0.69 & 0.97 & 4.00 & 0.00 & 0.16 & 0.94 & 0.95 & 4.01 & 0.32 & 0.29 & 0.73 & 0.97 \\ 
			ATEN & 3.99 & 0.25 & 0.21 & 0.95 & 0.95 & 4.00 & 0.03 & 0.16 & 0.98 & 0.95 & 4.00 & 0.04 & 0.21 & 0.91 & 0.95 \\ 
			ATC & 3.90 & 2.45 & 1.07 & 0.97 & 0.92 & 4.00 & 0.08 & 0.19 & 0.67 & 0.98 & 3.93 & 1.85 & 1.16 & 0.32 & 1.00 \\ 
			ATT & 4.03 & 0.67 & 0.92 & 1.15 & 0.98 & 4.00 & 0.01 & 0.16 & 0.26 & 1.00 & 4.03 & 0.74 & 0.82 & 0.92 & 1.00 \\
			\addlinespace 
			& \multicolumn{5}{c}{Aug: OR model correctly specified} & \multicolumn{5}{c}{Aug: both models misspecified} & \multicolumn{5}{c}{} \\ \cmidrule(lr){2-6}\cmidrule(lr){7-11}\cmidrule(lr){12-16}
			Estimand & PE & Bias & RMSE & RE & CP & PE & Bias & RMSE & RE & CP & \multicolumn{5}{c}{} \\
			\midrule
			ATE & 4.00 & 0.06 & 0.18 & 1.06 & 0.94 & 5.36 & 34.05 & 2.76 & 1.12 & 0.93 & \multicolumn{5}{c}{} \\ 
			ATE (0.05) & 4.00 & 0.04 & 0.17 & 1.01 & 0.94 & 4.54 & 13.56 & 1.86 & 1.02 & 0.94 & \multicolumn{5}{c}{} \\ 
			ATE (0.1) & 4.00 & 0.00 & 0.17 & 0.99 & 0.95 & 3.85 & 3.83 & 1.57 & 1.07 & 0.93 & \multicolumn{5}{c}{} \\ 
			ATE (0.15) & 4.00 & 0.03 & 0.18 & 0.99 & 0.94 & 3.58 & 10.51 & 1.53 & 1.05 & 0.92 & \multicolumn{5}{c}{} \\ 
			ATO & 4.00 & 0.01 & 0.16 & 0.97 & 0.95 & 4.38 & 9.54 & 1.47 & 0.98 & 0.95 & \multicolumn{5}{c}{} \\ 
			ATM & 4.00 & 0.00 & 0.16 & 0.95 & 0.95 & 4.38 & 9.56 & 1.50 & 0.97 & 0.95 & \multicolumn{5}{c}{} \\ 
			ATEN & 4.00 & 0.02 & 0.16 & 0.98 & 0.95 & 4.50 & 12.55 & 1.56 & 0.99 & 0.95 & \multicolumn{5}{c}{} \\ 
			ATC & 4.00 & 0.08 & 0.19 & 0.68 & 0.98 & 5.58 & 39.55 & 3.18 & 1.07 & 0.94 & \multicolumn{5}{c}{} \\ 
			ATT & 4.00 & 0.00 & 0.16 & 0.25 & 1.00 & 4.43 & 10.72 & 1.54 & 0.91 & 0.96 & \multicolumn{5}{c}{} \\  
			\addlinespace
			\multicolumn{16}{c}{\textbf{Heterogeneous treatment effect}} \\ \cmidrule(lr){2-16}
			& \multicolumn{5}{c}{H\'ajek-type estimator} & \multicolumn{5}{c}{Aug: both models correctly specified} & \multicolumn{5}{c}{Aug: PS model correctly specified} \\ \cmidrule(lr){2-6}\cmidrule(lr){7-11}\cmidrule(lr){12-16}
			Estimand & PE & Bias & RMSE & RE & CP & PE & Bias & RMSE & RE & CP & PE & Bias & RMSE & RE & CP \\ 
			\midrule
			ATE & 17.13 & 0.50 & 1.29 & 1.06 & 0.90 & 17.24 & 0.11 & 0.63 & 1.02 & 0.94 & 17.08 & 0.82 & 1.40 & 1.01 & 0.92 \\ 
			ATE (0.05) & 17.02 & 0.77 & 1.09 & 1.07 & 0.91 & 17.07 & 0.48 & 0.69 & 1.22 & 0.91 & 16.96 & 1.11 & 1.19 & 1.03 & 0.92 \\ 
			ATE (0.1) & 16.86 & 0.24 & 1.17 & 1.47 & 0.87 & 16.89 & 0.10 & 0.96 & 2.14 & 0.82 & 16.82 & 0.50 & 1.23 & 1.37 & 0.88 \\ 
			ATE (0.15) & 17.27 & 1.13 & 1.44 & 2.08 & 0.82 & 17.33 & 1.50 & 1.33 & 2.79 & 0.76 & 17.22 & 0.84 & 1.46 & 1.86 & 0.83 \\ 
			ATO & 17.72 & 0.03 & 0.99 & 1.04 & 0.94 & 17.73 & 0.05 & 0.96 & 1.01 & 0.95 & 17.68 & 0.25 & 1.04 & 1.01 & 0.94 \\ 
			ATM & 18.34 & 0.03 & 1.21 & 0.97 & 0.95 & 18.35 & 0.06 & 1.20 & 0.99 & 0.94 & 18.30 & 0.19 & 1.25 & 0.97 & 0.94 \\ 
			ATEN & 17.52 & 0.17 & 0.92 & 1.09 & 0.93 & 17.55 & 0.02 & 0.85 & 1.01 & 0.94 & 17.48 & 0.41 & 0.99 & 1.05 & 0.94 \\ 
			ATC & 16.67 & 0.83 & 1.58 & 1.02 & 0.91 & 16.81 & 0.02 & 0.68 & 0.57 & 0.99 & 16.60 & 1.25 & 1.71 & 0.46 & 0.99 \\ 
			ATT & 18.87 & 0.60 & 1.47 & 0.98 & 0.95 & 18.86 & 0.56 & 1.43 & 0.54 & 0.99 & 18.88 & 0.63 & 1.46 & 0.53 & 0.99 \\ 
			\addlinespace 
			& \multicolumn{5}{c}{Aug: OR model correctly specified} & \multicolumn{5}{c}{Aug: both models misspecified} & \multicolumn{5}{c}{} \\ \cmidrule(lr){2-6}\cmidrule(lr){7-11}\cmidrule(lr){12-16}
			Estimand & PE & Bias & RMSE & RE & CP & PE & Bias & RMSE & RE & CP & \multicolumn{5}{c}{} \\
			\midrule
			ATE & 17.24 & 0.10 & 0.63 & 1.02 & 0.94 & 18.73 & 8.80 & 3.27 & 1.15 & 0.92 & \multicolumn{5}{c}{}\\ 
			ATE (0.05) & 17.06 & 0.53 & 0.65 & 1.06 & 0.93 & 17.53 & 2.19 & 2.14 & 1.06 & 0.93 & \multicolumn{5}{c}{}\\ 
			ATE (0.1) & 17.00 & 0.58 & 0.70 & 1.09 & 0.94 & 16.70 & 1.22 & 1.88 & 1.10 & 0.92 & \multicolumn{5}{c}{}\\ 
			ATE (0.15) & 17.52 & 2.56 & 0.95 & 1.22 & 0.91 & 16.96 & 0.69 & 1.76 & 1.09 & 0.93 & \multicolumn{5}{c}{}\\ 
			ATO & 17.88 & 0.91 & 0.77 & 1.01 & 0.95 & 18.14 & 2.34 & 1.70 & 0.99 & 0.95 & \multicolumn{5}{c}{}\\ 
			ATM & 18.52 & 1.02 & 0.95 & 1.00 & 0.95 & 18.75 & 2.25 & 1.75 & 0.98 & 0.95 & \multicolumn{5}{c}{}\\ 
			ATEN & 17.67 & 0.65 & 0.71 & 1.01 & 0.95 & 18.07 & 2.98 & 1.80 & 1.00 & 0.94 & \multicolumn{5}{c}{}\\ 
			ATC & 16.81 & 0.03 & 0.68 & 0.57 & 0.99 & 18.63 & 10.80 & 3.79 & 1.05 & 0.94 & \multicolumn{5}{c}{}\\ 
			ATT & 18.86 & 0.56 & 1.43 & 0.54 & 0.99 & 19.00 & 1.31 & 1.80 & 0.63 & 0.99 & \multicolumn{5}{c}{}\\  
			\bottomrule
		\end{tabular}
		\begin{tablenotes}
			\tiny
			\item  PE: point estimation; Bias: absolute relative bias$\times 100$; RMSE: root mean squared error; RE: relative efficiency; CP: coverage probability; Aug: augmented estimator; PS: propensity score; OR: outcome regression; ATE: average treatment effect; ATE ($\alpha$): ATE by trimming PS $>1-\alpha$ or PS $<\alpha$, $\alpha=0.05, 0.1, 0.15$; ATO (resp. ATM, ATEN, ATC, and ATT): average treatment effect on overlap (resp. matching, entropy, controls, and treated)
		\end{tablenotes}
	\end{threeparttable}
\end{table}	 

%%%%%%%%%%%%%%%%%%%%%%%%%%%%%%%%%%%%%%%%%%%%%%%%%%%%%%%%%%%%%%%%%%%%%%%%%%%%%%%%%%%%%%%%%%%%
%%%%%%%%%%%%%%%%%%%%%%%%%%%%%%%%%%%% model 3 result %%%%%%%%%%%%%%%%%%%%%%%%%%%%%%%%%%%%%%%%
%%%%%%%%%%%%%%%%%%%%%%%%%%%%%%%%%%%%%%%%%%%%%%%%%%%%%%%%%%%%%%%%%%%%%%%%%%%%%%%%%%%%%%%%%%%%
\begin{table}[htbp]
	\centering
	\begin{threeparttable}\tiny
		\caption{Model 3, with $p = 49.72\%$, $r=1.13$}
		\label{tab:md3-all}
		\begin{tabular}{llcccccccccccccccccc}
			\toprule	
			\multicolumn{16}{c}{\textbf{Constant treatment effect}} \\ \cmidrule(lr){2-16}
			& \multicolumn{5}{c}{H\'ajek-type estimator} & \multicolumn{5}{c}{Aug: both models correctly specified} & \multicolumn{5}{c}{Aug: PS model correctly specified} \\ \cmidrule(lr){2-6}\cmidrule(lr){7-11}\cmidrule(lr){12-16}
			Estimand & PE & Bias & RMSE & RE & CP & PE & Bias & RMSE & RE & CP & PE & Bias & RMSE & RE & CP \\ 
			\midrule
			ATE & 4.02 & 0.41 & 0.41 & 0.83 & 0.96 & 4.00 & 0.08 & 0.13 & 1.00 & 0.95 & 4.02 & 0.50 & 0.37 & 0.76 & 0.96 \\ 
			ATE (0.05) & 4.01 & 0.24 & 0.37 & 0.87 & 0.97 & 4.00 & 0.08 & 0.13 & 1.00 & 0.95 & 4.01 & 0.32 & 0.34 & 0.83 & 0.97 \\ 
			ATE (0.1) & 4.01 & 0.18 & 0.30 & 0.83 & 0.97 & 4.00 & 0.08 & 0.13 & 0.99 & 0.95 & 4.01 & 0.24 & 0.29 & 0.79 & 0.97 \\ 
			ATE (0.15) & 4.00 & 0.00 & 0.24 & 0.74 & 0.98 & 4.00 & 0.07 & 0.13 & 1.00 & 0.95 & 4.00 & 0.10 & 0.23 & 0.71 & 0.98 \\ 
			ATO & 4.00 & 0.08 & 0.13 & 0.99 & 0.96 & 4.00 & 0.09 & 0.13 & 0.99 & 0.96 & 4.00 & 0.01 & 0.14 & 0.96 & 0.95 \\ 
			ATM & 3.99 & 0.19 & 0.21 & 0.73 & 0.97 & 4.00 & 0.09 & 0.13 & 0.99 & 0.96 & 4.00 & 0.12 & 0.20 & 0.71 & 0.97 \\ 
			ATEN & 4.00 & 0.01 & 0.15 & 0.94 & 0.96 & 4.00 & 0.09 & 0.13 & 0.99 & 0.96 & 4.00 & 0.06 & 0.15 & 0.90 & 0.96 \\ 
			ATC & 3.95 & 1.32 & 0.57 & 0.98 & 0.95 & 4.00 & 0.09 & 0.14 & 0.39 & 1.00 & 3.95 & 1.13 & 0.54 & 0.14 & 1.00 \\ 
			ATT & 4.09 & 2.18 & 0.87 & 0.99 & 0.92 & 4.00 & 0.07 & 0.14 & 0.39 & 1.00 & 4.09 & 2.16 & 0.80 & 0.90 & 0.96 \\  
			\addlinespace
			& \multicolumn{5}{c}{Aug: OR model correctly specified} & \multicolumn{5}{c}{Aug: both models misspecified} & \multicolumn{5}{c}{} \\ \cmidrule(lr){2-6}\cmidrule(lr){7-11}\cmidrule(lr){12-16}
			Estimand & PE & Bias & RMSE & RE & CP & PE & Bias & RMSE & RE & CP & \multicolumn{5}{c}{} \\
			\midrule
			ATE & 4.00 & 0.08 & 0.13 & 1.00 & 0.95 & 4.74 & 18.55 & 1.56 & 1.07 & 0.91 & \multicolumn{5}{c}{} \\ 
			ATE (0.05) & 4.00 & 0.08 & 0.13 & 1.00 & 0.95 & 4.73 & 18.25 & 1.54 & 1.06 & 0.91 & \multicolumn{5}{c}{}\\ 
			ATE (0.1) & 4.00 & 0.09 & 0.13 & 1.00 & 0.95 & 4.64 & 16.03 & 1.49 & 1.08 & 0.91 & \multicolumn{5}{c}{}\\ 
			ATE (0.15) & 4.00 & 0.08 & 0.13 & 1.00 & 0.96 & 4.47 & 11.65 & 1.39 & 1.08 & 0.93 & \multicolumn{5}{c}{}\\ 
			ATO & 4.00 & 0.09 & 0.13 & 1.00 & 0.95 & 4.40 & 9.89 & 1.29 & 1.07 & 0.93 & \multicolumn{5}{c}{}\\ 
			ATM & 4.00 & 0.09 & 0.13 & 1.00 & 0.96 & 4.21 & 5.27 & 1.20 & 1.07 & 0.94 & \multicolumn{5}{c}{}\\ 
			ATEN & 4.00 & 0.09 & 0.13 & 1.00 & 0.95 & 4.47 & 11.77 & 1.34 & 1.07 & 0.93 & \multicolumn{5}{c}{}\\ 
			ATC & 4.00 & 0.08 & 0.14 & 0.39 & 1.00 & 4.98 & 24.46 & 1.90 & 0.96 & 0.94 & \multicolumn{5}{c}{}\\ 
			ATT & 4.00 & 0.07 & 0.14 & 0.38 & 1.00 & 4.48 & 12.09 & 1.41 & 1.03 & 0.92 & \multicolumn{5}{c}{}\\  
			\addlinespace
			\multicolumn{16}{c}{\textbf{Heterogeneous treatment effect}} \\ \cmidrule(lr){2-16}
			& \multicolumn{5}{c}{H\'ajek-type estimator} & \multicolumn{5}{c}{Aug: both models correctly specified} & \multicolumn{5}{c}{Aug: PS model correctly specified} \\ \cmidrule(lr){2-6}\cmidrule(lr){7-11}\cmidrule(lr){12-16}
			Estimand & PE & Bias & RMSE & RE & CP & PE & Bias & RMSE & RE & CP & PE & Bias & RMSE & RE & CP \\ 
			\midrule
			ATE & 17.19 & 0.14 & 0.71 & 1.07 & 0.94 & 17.21 & 0.04 & 0.63 & 1.06 & 0.94 & 17.18 & 0.22 & 0.74 & 1.05 & 0.94 \\ 
			ATE (0.05) & 17.16 & 0.25 & 0.71 & 1.08 & 0.93 & 17.18 & 0.16 & 0.63 & 1.07 & 0.94 & 17.15 & 0.33 & 0.74 & 1.06 & 0.93 \\ 
			ATE (0.1) & 17.05 & 0.41 & 0.70 & 1.11 & 0.93 & 17.06 & 0.33 & 0.63 & 1.10 & 0.93 & 17.03 & 0.48 & 0.73 & 1.07 & 0.93 \\ 
			ATE (0.15) & 16.82 & 0.56 & 0.71 & 1.19 & 0.91 & 16.83 & 0.47 & 0.66 & 1.20 & 0.92 & 16.81 & 0.63 & 0.74 & 1.14 & 0.92 \\ 
			ATO & 16.61 & 0.49 & 0.67 & 1.09 & 0.93 & 16.62 & 0.38 & 0.63 & 1.05 & 0.94 & 16.60 & 0.53 & 0.69 & 1.05 & 0.94 \\ 
			ATM & 16.23 & 0.44 & 0.71 & 1.02 & 0.94 & 16.25 & 0.32 & 0.66 & 1.01 & 0.94 & 16.22 & 0.47 & 0.72 & 0.99 & 0.95 \\ 
			ATEN & 16.73 & 0.43 & 0.66 & 1.10 & 0.93 & 16.75 & 0.32 & 0.62 & 1.05 & 0.94 & 16.73 & 0.48 & 0.69 & 1.06 & 0.94 \\ 
			ATC & 16.53 & 0.44 & 1.12 & 1.08 & 0.91 & 16.60 & 0.05 & 0.86 & 0.60 & 0.98 & 16.50 & 0.63 & 1.16 & 0.45 & 0.99 \\ 
			ATT & 17.86 & 0.15 & 0.96 & 1.03 & 0.94 & 17.83 & 0.03 & 0.90 & 0.56 & 0.99 & 17.87 & 0.17 & 0.94 & 0.58 & 0.99 \\  
			\addlinespace
			& \multicolumn{5}{c}{Aug: OR model correctly specified} & \multicolumn{5}{c}{Aug: both models misspecified} & \multicolumn{5}{c}{} \\ \cmidrule(lr){2-6}\cmidrule(lr){7-11}\cmidrule(lr){12-16}
			Estimand & PE & Bias & RMSE & RE & CP & PE & Bias & RMSE & RE & CP & \multicolumn{5}{c}{} \\
			\midrule
			ATE & 17.21 & 0.04 & 0.62 & 1.06 & 0.94 & 17.83 & 3.56 & 1.48 & 1.04 & 0.93 & \multicolumn{5}{c}{}\\ 
			ATE (0.05) & 17.21 & 0.03 & 0.63 & 1.06 & 0.94 & 17.81 & 3.55 & 1.46 & 1.03 & 0.93 & \multicolumn{5}{c}{}\\ 
			ATE (0.1) & 17.19 & 0.41 & 0.63 & 1.06 & 0.94 & 17.69 & 3.35 & 1.42 & 1.06 & 0.93 & \multicolumn{5}{c}{}\\ 
			ATE (0.15) & 17.12 & 1.21 & 0.67 & 1.09 & 0.93 & 17.44 & 3.12 & 1.37 & 1.07 & 0.93 & \multicolumn{5}{c}{}\\ 
			ATO & 16.86 & 1.04 & 0.66 & 1.06 & 0.94 & 17.12 & 2.62 & 1.26 & 1.04 & 0.94 & \multicolumn{5}{c}{}\\ 
			ATM & 16.52 & 1.34 & 0.70 & 1.05 & 0.94 & 16.61 & 1.93 & 1.21 & 1.02 & 0.95 & \multicolumn{5}{c}{}\\ 
			ATEN & 16.95 & 0.83 & 0.64 & 1.06 & 0.94 & 17.29 & 2.84 & 1.29 & 1.04 & 0.94 & \multicolumn{5}{c}{}\\ 
			ATC & 16.60 & 0.05 & 0.86 & 0.60 & 0.99 & 17.65 & 6.31 & 2.04 & 0.78 & 0.96 & \multicolumn{5}{c}{}\\ 
			ATT & 17.83 & 0.03 & 0.90 & 0.56 & 0.99 & 17.99 & 0.85 & 1.20 & 0.70 & 0.98 & \multicolumn{5}{c}{}\\  
			\bottomrule
		\end{tabular}
		\begin{tablenotes}
			\tiny
			\item PE: point estimation; Bias: absolute relative bias$\times 100$; RMSE: root mean squared error; RE: relative efficiency; CP: coverage probability; Aug: augmented estimator; PS: propensity score; OR: outcome regression; ATE: average treatment effect; ATE ($\alpha$): ATE by trimming PS $>1-\alpha$ or PS $<\alpha$, $\alpha=0.05, 0.1, 0.15$; ATO (resp. ATM, ATEN, ATC, and ATT): average treatment effect on overlap (resp. matching, entropy, controls, and treated)
		\end{tablenotes}
	\end{threeparttable}
\end{table}	 

%%%%%%%%%%%%%%%%%%%%%%%%%%%%%%%%%%%%%%%%%%%%%%%%%%%%%%%%%%%%%%%%%%%%%%%%%%%%%%%%%%%%%%%%%%%%
%%%%%%%%%%%%%%%%%%%%%%%%%%%%%%%%%%%% model 4 result %%%%%%%%%%%%%%%%%%%%%%%%%%%%%%%%%%%%%%%%%
%%%%%%%%%%%%%%%%%%%%%%%%%%%%%%%%%%%%%%%%%%%%%%%%%%%%%%%%%%%%%%%%%%%%%%%%%%%%%%%%%%%%%%%%%%%%
\begin{table}[htbp]
	\centering
	\begin{threeparttable}\tiny
		\caption{Model 4, with $p = 79.22\%$, $r=0.42$}
		\label{tab:md4-all}
		\begin{tabular}{llcccccccccccccccccc}
			\toprule	
			\multicolumn{16}{c}{\textbf{Constant treatment effect}} \\ \cmidrule(lr){2-16}
			& \multicolumn{5}{c}{H\'ajek-type estimator} & \multicolumn{5}{c}{Aug: both models correctly specified} & \multicolumn{5}{c}{Aug: PS model correctly specified} \\ \cmidrule(lr){2-6}\cmidrule(lr){7-11}\cmidrule(lr){12-16}
			Estimand & PE & Bias & RMSE & RE & CP & PE & Bias & RMSE & RE & CP & PE & Bias & RMSE & RE & CP \\ 
			\midrule
			ATE & 4.19 & 4.76 & 1.65 & 1.16 & 0.84 & 4.00 & 0.03 & 0.21 & 1.15 & 0.93 & 4.25 & 6.33 & 1.99 & 1.17 & 0.86 \\ 
			ATE (0.05) & 4.02 & 0.59 & 0.76 & 0.91 & 0.93 & 4.00 & 0.05 & 0.19 & 1.09 & 0.94 & 4.02 & 0.51 & 0.90 & 0.92 & 0.95 \\ 
			ATE (0.1) & 4.01 & 0.31 & 0.47 & 0.73 & 0.97 & 4.00 & 0.01 & 0.19 & 1.07 & 0.94 & 3.99 & 0.28 & 0.54 & 0.74 & 0.97 \\ 
			ATE (0.15) & 4.02 & 0.46 & 0.47 & 0.63 & 0.98 & 4.01 & 0.18 & 0.20 & 1.07 & 0.94 & 3.98 & 0.46 & 0.53 & 0.60 & 0.98 \\ 
			ATO & 4.00 & 0.05 & 0.17 & 1.05 & 0.94 & 4.00 & 0.05 & 0.17 & 1.05 & 0.94 & 3.97 & 0.72 & 0.24 & 0.99 & 0.95 \\ 
			ATM & 4.00 & 0.06 & 0.29 & 0.69 & 0.98 & 4.00 & 0.06 & 0.17 & 1.04 & 0.95 & 3.97 & 0.82 & 0.32 & 0.76 & 0.97 \\ 
			ATEN & 4.02 & 0.41 & 0.27 & 1.06 & 0.94 & 4.00 & 0.04 & 0.17 & 1.07 & 0.94 & 4.00 & 0.12 & 0.35 & 1.03 & 0.95 \\ 
			ATC & 3.75 & 6.21 & 2.30 & 1.27 & 0.86 & 4.00 & 0.01 & 0.19 & 0.32 & 1.00 & 3.64 & 8.92 & 2.98 & 0.63 & 0.99 \\ 
			ATT & 4.30 & 7.61 & 2.09 & 1.22 & 0.82 & 4.00 & 0.04 & 0.23 & 0.80 & 0.97 & 4.41 & 10.29 & 2.44 & 1.22 & 0.85 \\  
			\addlinespace
			& \multicolumn{5}{c}{Aug: OR model correctly specified} & \multicolumn{5}{c}{Aug: both models misspecified} & \multicolumn{5}{c}{} \\ \cmidrule(lr){2-6}\cmidrule(lr){7-11}\cmidrule(lr){12-16}
			Estimand & PE & Bias & RMSE & RE & CP & PE & Bias & RMSE & RE & CP & \multicolumn{5}{c}{} \\
			\midrule
			ATE & 4.00 & 0.04 & 0.20 & 1.11 & 0.94 & -0.32 & 108.05 & 5.85 & 1.15 & 0.80 & \multicolumn{5}{c}{}\\ 
			ATE (0.05) & 4.00 & 0.03 & 0.19 & 1.08 & 0.94 & 1.55 & 61.23 & 3.20 & 1.11 & 0.77 & \multicolumn{5}{c}{}\\ 
			ATE (0.1) & 4.00 & 0.05 & 0.20 & 1.09 & 0.94 & 2.79 & 30.37 & 1.90 & 1.15 & 0.85 & \multicolumn{5}{c}{}\\ 
			ATE (0.15) & 4.00 & 0.05 & 0.20 & 1.07 & 0.94 & 3.29 & 17.65 & 1.45 & 1.20 & 0.88 & \multicolumn{5}{c}{}\\ 
			ATO & 4.00 & 0.04 & 0.18 & 1.07 & 0.94 & 1.53 & 61.71 & 2.96 & 1.05 & 0.67 & \multicolumn{5}{c}{}\\ 
			ATM & 4.00 & 0.04 & 0.18 & 1.06 & 0.94 & 1.59 & 60.20 & 2.92 & 1.04 & 0.69 & \multicolumn{5}{c}{}\\ 
			ATEN & 4.00 & 0.03 & 0.18 & 1.07 & 0.94 & 1.29 & 67.76 & 3.26 & 1.06 & 0.67 & \multicolumn{5}{c}{}\\ 
			ATC & 4.00 & 0.05 & 0.18 & 0.30 & 1.00 & 1.10 & 72.62 & 3.71 & 0.79 & 0.78 & \multicolumn{5}{c}{}\\ 
			ATT & 4.00 & 0.04 & 0.22 & 0.75 & 0.97 & -0.60 & 114.96 & 6.46 & 1.15 & 0.83 & \multicolumn{5}{c}{}\\ 
			\addlinespace
			\multicolumn{16}{c}{\textbf{Heterogeneous treatment effect}} \\ \cmidrule(lr){2-16}
			& \multicolumn{5}{c}{H\'ajek-type estimator} & \multicolumn{5}{c}{Aug: both models correctly specified} & \multicolumn{5}{c}{Aug: PS model correctly specified} \\ \cmidrule(lr){2-6}\cmidrule(lr){7-11}\cmidrule(lr){12-16}
			Estimand & PE & Bias & RMSE & RE & CP & PE & Bias & RMSE & RE & CP & PE & Bias & RMSE & RE & CP \\ 
			\midrule
			ATE & 17.24 & 0.14 & 1.07 & 1.09 & 0.95 & 17.22 & 0.01 & 0.63 & 1.00 & 0.95 & 17.25 & 0.21 & 1.17 & 1.11 & 0.94 \\ 
			ATE (0.05) & 15.20 & 0.58 & 1.09 & 2.17 & 0.79 & 15.24 & 0.31 & 0.98 & 2.93 & 0.75 & 15.19 & 0.59 & 1.12 & 2.01 & 0.81 \\ 
			ATE (0.1) & 13.60 & 0.89 & 0.92 & 1.61 & 0.87 & 13.63 & 1.16 & 0.83 & 1.87 & 0.85 & 13.58 & 0.77 & 0.94 & 1.57 & 0.87 \\ 
			ATE (0.15) & 13.88 & 0.11 & 1.09 & 1.92 & 0.83 & 13.92 & 0.36 & 1.02 & 2.28 & 0.82 & 13.86 & 0.09 & 1.10 & 1.86 & 0.83 \\ 
			ATO & 15.33 & 0.41 & 0.86 & 1.00 & 0.94 & 15.37 & 0.15 & 0.81 & 0.96 & 0.95 & 15.31 & 0.53 & 0.89 & 0.99 & 0.94 \\ 
			ATM & 15.73 & 0.55 & 0.98 & 0.96 & 0.94 & 15.76 & 0.32 & 0.91 & 0.94 & 0.94 & 15.71 & 0.68 & 1.01 & 0.95 & 0.94 \\ 
			ATEN & 15.49 & 0.39 & 0.82 & 1.01 & 0.94 & 15.53 & 0.16 & 0.76 & 0.97 & 0.94 & 15.48 & 0.49 & 0.85 & 1.00 & 0.94 \\ 
			ATC & 18.28 & 1.61 & 3.36 & 1.24 & 0.82 & 18.62 & 0.19 & 1.45 & 0.56 & 0.99 & 18.11 & 2.52 & 3.87 & 0.83 & 0.96 \\ 
			ATT & 16.96 & 0.59 & 1.06 & 1.18 & 0.93 & 16.85 & 0.04 & 0.70 & 0.59 & 0.99 & 17.02 & 0.95 & 1.09 & 0.85 & 0.97 \\ 
			\addlinespace
			& \multicolumn{5}{c}{Aug: OR model correctly specified} & \multicolumn{5}{c}{Aug: both models misspecified} & \multicolumn{5}{c}{} \\ \cmidrule(lr){2-6}\cmidrule(lr){7-11}\cmidrule(lr){12-16}
			Estimand & PE & Bias & RMSE & RE & CP & PE & Bias & RMSE & RE & CP & \multicolumn{5}{c}{} \\
			\midrule
			ATE & 17.22 & 0.02 & 0.62 & 1.00 & 0.95 & 15.16 & 11.96 & 2.63 & 1.17 & 0.70 & \multicolumn{5}{c}{}\\ 
			ATE (0.05) & 14.93 & 2.31 & 1.04 & 3.44 & 0.67 & 13.60 & 11.02 & 2.14 & 1.49 & 0.60 & \multicolumn{5}{c}{}\\ 
			ATE (0.1) & 12.49 & 7.29 & 1.23 & 1.92 & 0.52 & 11.76 & 12.73 & 2.01 & 1.23 & 0.51 & \multicolumn{5}{c}{}\\ 
			ATE (0.15) & 12.51 & 9.78 & 1.57 & 1.55 & 0.43 & 12.05 & 13.12 & 2.13 & 1.28 & 0.47 & \multicolumn{5}{c}{}\\ 
			ATO & 14.96 & 2.80 & 0.78 & 0.96 & 0.89 & 13.28 & 13.73 & 2.45 & 1.06 & 0.56 & \multicolumn{5}{c}{}\\ 
			ATM & 15.65 & 1.02 & 0.78 & 0.96 & 0.94 & 13.88 & 12.26 & 2.39 & 1.04 & 0.67 & \multicolumn{5}{c}{}\\ 
			ATEN & 15.16 & 2.52 & 0.73 & 0.96 & 0.90 & 13.45 & 13.48 & 2.43 & 1.07 & 0.55 & \multicolumn{5}{c}{}\\ 
			ATC & 18.62 & 0.19 & 1.45 & 0.56 & 0.99 & 14.56 & 21.66 & 4.58 & 0.68 & 0.56 & \multicolumn{5}{c}{}\\ 
			ATT & 16.85 & 0.03 & 0.69 & 0.58 & 0.99 & 15.36 & 8.91 & 2.29 & 1.02 & 0.89 & \multicolumn{5}{c}{}\\ 
			\bottomrule
		\end{tabular}
		\begin{tablenotes}
			\tiny
			\item PE: point estimation; Bias: absolute relative bias$\times 100$; RMSE: root mean squared error; RE: relative efficiency; CP: coverage probability; Aug: augmented estimator; PS: propensity score; OR: outcome regression; ATE: average treatment effect; ATE ($\alpha$): ATE by trimming PS $>1-\alpha$ or PS $<\alpha$, $\alpha=0.05, 0.1, 0.15$; ATO (resp. ATM, ATEN, ATC, and ATT): average treatment effect on overlap (resp. matching, entropy, controls, and treated)
		\end{tablenotes}
	\end{threeparttable}
\end{table}

%%%%%%%%%%%%%%%%%%%%%%%%%%%%%%%%%%%%%%%%%%%%%%%%%%%%%%%%%%%%%%%%%%%%%%%%%%%%%%%%%%%%%%%%%%%%
%%%%%%%%%%%%%%%%%%%%%%%%%%%%%%%%%%%% model 5 result %%%%%%%%%%%%%%%%%%%%%%%%%%%%%%%%%%%%%%%%
%%%%%%%%%%%%%%%%%%%%%%%%%%%%%%%%%%%%%%%%%%%%%%%%%%%%%%%%%%%%%%%%%%%%%%%%%%%%%%%%%%%%%%%%%%%%
\begin{table}[htbp]
	\centering
	\begin{threeparttable}\tiny
		\caption{Model 5, with $p = 89.18\%$, $r=0.26$}
		\label{tab:md5-all}
		\begin{tabular}{llcccccccccccccccccc}
			\toprule	
			\multicolumn{16}{c}{\textbf{Constant treatment effect}} \\ \cmidrule(lr){2-16}
			& \multicolumn{5}{c}{H\'ajek-type estimator} & \multicolumn{5}{c}{Aug: both models correctly specified} & \multicolumn{5}{c}{Aug: PS model correctly specified} \\ \cmidrule(lr){2-6}\cmidrule(lr){7-11}\cmidrule(lr){12-16}
			Estimand & PE & Bias & RMSE & RE & CP & PE & Bias & RMSE & RE & CP & PE & Bias & RMSE & RE & CP \\ 
			\midrule
			ATE & 4.34 & 8.48 & 2.84 & 1.39 & 0.79 & 4.00 & 0.05 & 0.29 & 1.16 & 0.92 & 4.52 & 13.11 & 3.69 & 1.44 & 0.82 \\ 
			ATE (0.05) & 4.10 & 2.38 & 0.90 & 0.77 & 0.94 & 4.00 & 0.08 & 0.25 & 1.08 & 0.94 & 4.04 & 1.01 & 1.06 & 0.76 & 0.96 \\ 
			ATE (0.1) & 4.13 & 3.20 & 0.90 & 0.34 & 0.96 & 4.00 & 0.08 & 0.29 & 1.14 & 0.93 & 4.06 & 1.40 & 1.07 & 0.35 & 0.97 \\ 
			ATE (0.15) & 4.24 & 6.08 & 1.13 & 0.45 & 0.97 & 4.01 & 0.18 & 0.39 & 1.21 & 0.92 & 4.31 & 7.71 & 1.53 & 0.75 & 0.96 \\ 
			ATO & 4.00 & 0.00 & 0.21 & 1.00 & 0.95 & 4.00 & 0.01 & 0.21 & 1.00 & 0.95 & 3.93 & 1.69 & 0.32 & 0.94 & 0.96 \\ 
			ATM & 3.97 & 0.77 & 0.43 & 0.60 & 0.98 & 4.00 & 0.02 & 0.22 & 0.99 & 0.95 & 3.91 & 2.15 & 0.55 & 0.70 & 0.97 \\ 
			ATEN & 4.03 & 0.73 & 0.39 & 1.05 & 0.92 & 4.00 & 0.00 & 0.22 & 1.01 & 0.95 & 3.97 & 0.63 & 0.47 & 1.03 & 0.94 \\ 
			ATC & 3.70 & 7.45 & 2.82 & 1.47 & 0.94 & 4.00 & 0.05 & 0.23 & 0.26 & 1.00 & 3.61 & 9.80 & 3.51 & 0.60 & 1.00 \\ 
			ATT & 4.42 & 10.51 & 3.18 & 1.44 & 0.78 & 4.00 & 0.06 & 0.31 & 0.97 & 0.95 & 4.64 & 16.02 & 4.08 & 1.48 & 0.81 \\ 
			\addlinespace
			& \multicolumn{5}{c}{Aug: OR model correctly specified} & \multicolumn{5}{c}{Aug: both models misspecified} & \multicolumn{5}{c}{} \\ \cmidrule(lr){2-6}\cmidrule(lr){7-11}\cmidrule(lr){12-16}
			Estimand & PE & Bias & RMSE & RE & CP & PE & Bias & RMSE & RE & CP & \multicolumn{5}{c}{} \\
			\midrule
			ATE & 4.00 & 0.06 & 0.28 & 1.11 & 0.93 & -1.88 & 147.03 & 9.27 & 1.49 & 0.84 & \multicolumn{5}{c}{}\\ 
			ATE (0.05) & 4.00 & 0.01 & 0.27 & 1.11 & 0.93 & 2.84 & 29.03 & 2.36 & 1.20 & 0.90 & \multicolumn{5}{c}{}\\ 
			ATE (0.1) & 4.00 & 0.04 & 0.30 & 1.10 & 0.93 & 3.60 & 10.04 & 1.58 & 1.28 & 0.91 & \multicolumn{5}{c}{}\\ 
			ATE (0.15) & 4.00 & 0.01 & 0.37 & 1.19 & 0.93 & 3.85 & 3.86 & 1.38 & 1.27 & 0.92 & \multicolumn{5}{c}{}\\ 
			ATO & 4.00 & 0.03 & 0.23 & 1.02 & 0.94 & 1.08 & 73.04 & 3.68 & 1.08 & 0.74 & \multicolumn{5}{c}{}\\ 
			ATM & 4.00 & 0.03 & 0.23 & 1.02 & 0.94 & 0.90 & 77.49 & 3.96 & 1.05 & 0.76 & \multicolumn{5}{c}{}\\ 
			ATEN & 4.00 & 0.04 & 0.23 & 1.03 & 0.94 & 0.81 & 79.74 & 4.04 & 1.13 & 0.74 & \multicolumn{5}{c}{}\\ 
			ATC & 4.00 & 0.01 & 0.22 & 0.25 & 1.00 & 0.62 & 84.59 & 4.37 & 0.69 & 0.87 & \multicolumn{5}{c}{}\\ 
			ATT & 4.00 & 0.07 & 0.30 & 0.91 & 0.96 & -2.07 & 151.68 & 9.81 & 1.51 & 0.85 & \multicolumn{5}{c}{}\\ 
			\addlinespace
			\multicolumn{16}{c}{\textbf{Heterogeneous treatment effect}} \\ \cmidrule(lr){2-16}
			& \multicolumn{5}{c}{H\'ajek-type estimator} & \multicolumn{5}{c}{Aug: both models correctly specified} & \multicolumn{5}{c}{Aug: PS model correctly specified} \\ \cmidrule(lr){2-6}\cmidrule(lr){7-11}\cmidrule(lr){12-16}
			Estimand & PE & Bias & RMSE & RE & CP & PE & Bias & RMSE & RE & CP & PE & Bias & RMSE & RE & CP \\ 
			\midrule
			ATE & 17.30 & 0.46 & 1.31 & 1.17 & 0.92 & 17.20 & 0.08 & 0.68 & 1.08 & 0.94 & 17.42 & 1.20 & 1.45 & 1.20 & 0.92 \\ 
			ATE (0.05) & 13.92 & 1.83 & 1.24 & 2.07 & 0.80 & 13.94 & 1.91 & 1.17 & 2.94 & 0.76 & 13.94 & 1.91 & 1.27 & 1.97 & 0.81 \\ 
			ATE (0.1) & 16.47 & 0.69 & 1.97 & 2.30 & 0.72 & 16.50 & 0.52 & 1.89 & 3.82 & 0.69 & 16.45 & 0.83 & 1.95 & 2.36 & 0.73 \\ 
			ATE (0.15) & 22.07 & 3.14 & 3.44 & 3.08 & 0.60 & 22.13 & 2.86 & 3.34 & 5.09 & 0.57 & 22.08 & 3.09 & 3.43 & 3.53 & 0.61 \\ 
			ATO & 17.23 & 0.74 & 1.29 & 1.04 & 0.93 & 17.29 & 0.40 & 1.23 & 1.00 & 0.94 & 17.21 & 0.87 & 1.33 & 1.02 & 0.93 \\ 
			ATM & 18.63 & 1.12 & 1.54 & 0.98 & 0.93 & 18.71 & 0.69 & 1.42 & 0.97 & 0.95 & 18.60 & 1.27 & 1.61 & 0.98 & 0.93 \\ 
			ATEN & 16.67 & 0.64 & 1.12 & 1.03 & 0.94 & 16.71 & 0.39 & 1.08 & 1.02 & 0.94 & 16.66 & 0.67 & 1.15 & 1.00 & 0.94 \\ 
			ATC & 20.37 & 2.03 & 4.23 & 1.35 & 0.84 & 20.77 & 0.08 & 2.22 & 0.56 & 0.98 & 20.23 & 2.69 & 4.67 & 0.96 & 0.95 \\ 
			ATT & 16.92 & 0.80 & 1.36 & 1.29 & 0.90 & 16.77 & 0.08 & 0.70 & 0.64 & 0.99 & 17.08 & 1.75 & 1.49 & 1.10 & 0.94 \\ 
			\addlinespace
			& \multicolumn{5}{c}{Aug: OR model correctly specified} & \multicolumn{5}{c}{Aug: both models misspecified} & \multicolumn{5}{c}{} \\ \cmidrule(lr){2-6}\cmidrule(lr){7-11}\cmidrule(lr){12-16}
			Estimand & PE & Bias & RMSE & RE & CP & PE & Bias & RMSE & RE & CP & \multicolumn{5}{c}{} \\
			\midrule
			ATE & 17.20 & 0.08 & 0.68 & 1.08 & 0.94 & 15.01 & 12.82 & 3.30 & 1.52 & 0.79 & \multicolumn{5}{c}{}\\ 
			ATE (0.05) & 12.35 & 9.71 & 1.57 & 2.05 & 0.39 & 11.80 & 13.70 & 2.19 & 1.30 & 0.47 & \multicolumn{5}{c}{}\\ 
			ATE (0.1) & 14.79 & 10.85 & 2.27 & 2.49 & 0.44 & 14.56 & 12.23 & 2.60 & 2.15 & 0.48 & \multicolumn{5}{c}{}\\ 
			ATE (0.15) & 21.08 & 7.49 & 3.14 & 3.81 & 0.54 & 20.96 & 8.01 & 3.35 & 3.52 & 0.56 & \multicolumn{5}{c}{}\\ 
			ATO & 16.56 & 4.59 & 1.19 & 1.01 & 0.80 & 14.81 & 14.69 & 2.99 & 1.10 & 0.54 & \multicolumn{5}{c}{}\\ 
			ATM & 18.19 & 3.43 & 1.36 & 1.00 & 0.87 & 16.15 & 14.25 & 3.35 & 1.08 & 0.64 & \multicolumn{5}{c}{}\\ 
			ATEN & 16.01 & 4.55 & 1.07 & 1.01 & 0.80 & 14.30 & 14.75 & 2.86 & 1.12 & 0.52 & \multicolumn{5}{c}{}\\ 
			ATC & 20.77 & 0.08 & 2.22 & 0.56 & 0.98 & 16.06 & 22.72 & 5.31 & 0.66 & 0.55 & \multicolumn{5}{c}{}\\ 
			ATT & 16.77 & 0.08 & 0.70 & 0.64 & 0.98 & 14.93 & 11.07 & 3.19 & 1.43 & 0.89 & \multicolumn{5}{c}{}\\ 
			\bottomrule
		\end{tabular}
		\begin{tablenotes}
			\tiny
			\item PE: point estimation; Bias: absolute relative bias$\times 100$; RMSE: root mean squared error; RE: relative efficiency; CP: coverage probability; Aug: augmented estimator; PS: propensity score; OR: outcome regression; ATE: average treatment effect; ATE ($\alpha$): ATE by trimming PS $>1-\alpha$ or PS $<\alpha$, $\alpha=0.05, 0.1, 0.15$; ATO (resp. ATM, ATEN, ATC, and ATT): average treatment effect on overlap (resp. matching, entropy, controls, and treated)
		\end{tablenotes}
	\end{threeparttable}
\end{table}	 

%%%%%%%%%%%%%%%%%%%%%%%%%%%%%%%%%%%%%%%%%%%%%%%%%%%%%%%%%%%%%%%%%%%%%%%%%%%%%%%%%%%%%%%%%%%%
%%%%%%%%%%%%%%%%%%%%%%%%%%%%%%%%%%%% model 6 result %%%%%%%%%%%%%%%%%%%%%%%%%%%%%%%%%%%%%%%%
%%%%%%%%%%%%%%%%%%%%%%%%%%%%%%%%%%%%%%%%%%%%%%%%%%%%%%%%%%%%%%%%%%%%%%%%%%%%%%%%%%%%%%%%%%%%
\begin{table}[htbp]
	\centering
	\begin{threeparttable}\tiny
		\caption{Model 6, with $p = 79.59\%$, $r=0.75$}
		\label{tab:md6-all}
		\begin{tabular}{llcccccccccccccccccc}
			\toprule	
			\multicolumn{16}{c}{\textbf{Constant treatment effect}} \\ \cmidrule(lr){2-16}
			& \multicolumn{5}{c}{H\'ajek-type estimator} & \multicolumn{5}{c}{Aug: both models correctly specified} & \multicolumn{5}{c}{Aug: PS model correctly specified} \\ \cmidrule(lr){2-6}\cmidrule(lr){7-11}\cmidrule(lr){12-16}
			Estimand & PE & Bias & RMSE & RE & CP & PE & Bias & RMSE & RE & CP & PE & Bias & RMSE & RE & CP \\ 
			\midrule
			ATE & 4.05 & 1.33 & 1.21 & 1.11 & 0.90 & 4.00 & 0.12 & 0.18 & 1.04 & 0.94 & 4.07 & 1.86 & 1.11 & 1.11 & 0.90 \\ 
			ATE (0.05) & 4.04 & 0.97 & 0.67 & 0.92 & 0.94 & 4.01 & 0.13 & 0.18 & 1.04 & 0.94 & 4.05 & 1.20 & 0.65 & 0.94 & 0.95 \\ 
			ATE (0.1) & 4.02 & 0.57 & 0.45 & 0.76 & 0.97 & 4.00 & 0.12 & 0.17 & 1.02 & 0.95 & 4.02 & 0.55 & 0.44 & 0.77 & 0.97 \\ 
			ATE (0.15) & 4.02 & 0.50 & 0.39 & 0.61 & 0.98 & 4.00 & 0.05 & 0.19 & 1.01 & 0.95 & 4.01 & 0.29 & 0.39 & 0.63 & 0.98 \\ 
			ATO & 4.00 & 0.11 & 0.16 & 1.00 & 0.95 & 4.00 & 0.11 & 0.16 & 1.00 & 0.95 & 4.00 & 0.04 & 0.17 & 0.98 & 0.95 \\ 
			ATM & 3.99 & 0.15 & 0.28 & 0.68 & 0.97 & 4.00 & 0.10 & 0.16 & 0.99 & 0.95 & 3.99 & 0.25 & 0.27 & 0.66 & 0.97 \\ 
			ATEN & 4.01 & 0.30 & 0.23 & 0.98 & 0.95 & 4.00 & 0.11 & 0.16 & 1.00 & 0.95 & 4.01 & 0.32 & 0.23 & 0.98 & 0.95 \\ 
			ATC & 3.98 & 0.56 & 0.42 & 0.81 & 0.98 & 4.00 & 0.10 & 0.17 & 0.26 & 1.00 & 3.98 & 0.53 & 0.39 & 0.06 & 1.00 \\ 
			ATT & 4.08 & 1.90 & 1.53 & 1.12 & 0.90 & 4.01 & 0.13 & 0.19 & 0.67 & 0.98 & 4.10 & 2.54 & 1.41 & 1.10 & 0.91 \\
			\addlinespace
			& \multicolumn{5}{c}{Aug: OR model correctly specified} & \multicolumn{5}{c}{Aug: both models misspecified} & \multicolumn{5}{c}{} \\ \cmidrule(lr){2-6}\cmidrule(lr){7-11}\cmidrule(lr){12-16}
			Estimand & PE & Bias & RMSE & RE & CP & PE & Bias & RMSE & RE & CP & \multicolumn{5}{c}{} \\
			\midrule
			ATE & 4.01 & 0.14 & 0.18 & 1.02 & 0.95 & 4.48 & 11.94 & 1.68 & 1.04 & 0.92 & \multicolumn{5}{c}{}\\ 
			ATE (0.05) & 4.01 & 0.16 & 0.18 & 1.02 & 0.94 & 4.46 & 11.38 & 1.66 & 1.04 & 0.92 & \multicolumn{5}{c}{}\\ 
			ATE (0.1) & 4.01 & 0.13 & 0.17 & 1.02 & 0.94 & 4.12 & 3.02 & 1.52 & 1.04 & 0.93 & \multicolumn{5}{c}{}\\ 
			ATE (0.15) & 4.00 & 0.04 & 0.19 & 1.04 & 0.94 & 3.80 & 4.99 & 1.67 & 1.12 & 0.94 & \multicolumn{5}{c}{}\\ 
			ATO & 4.00 & 0.12 & 0.16 & 1.01 & 0.95 & 4.33 & 8.36 & 1.52 & 1.02 & 0.93 & \multicolumn{5}{c}{}\\ 
			ATM & 4.00 & 0.12 & 0.17 & 1.01 & 0.94 & 4.41 & 10.14 & 1.58 & 1.00 & 0.93 & \multicolumn{5}{c}{}\\ 
			ATEN & 4.01 & 0.13 & 0.16 & 1.01 & 0.95 & 4.35 & 8.78 & 1.53 & 1.03 & 0.93 & \multicolumn{5}{c}{}\\ 
			ATC & 4.00 & 0.11 & 0.17 & 0.26 & 1.00 & 4.52 & 13.09 & 1.66 & 0.89 & 0.95 & \multicolumn{5}{c}{}\\ 
			ATT & 4.01 & 0.15 & 0.19 & 0.64 & 0.98 & 4.46 & 11.57 & 1.75 & 1.04 & 0.91 & \multicolumn{5}{c}{}\\ 
			\addlinespace
			\multicolumn{16}{c}{\textbf{Heterogeneous treatment effect}} \\ \cmidrule(lr){2-16}
			& \multicolumn{5}{c}{H\'ajek-type estimator} & \multicolumn{5}{c}{Aug: both models correctly specified} & \multicolumn{5}{c}{Aug: PS model correctly specified} \\ \cmidrule(lr){2-6}\cmidrule(lr){7-11}\cmidrule(lr){12-16}
			Estimand & PE & Bias & RMSE & RE & CP & PE & Bias & RMSE & RE & CP & PE & Bias & RMSE & RE & CP \\ 
			\midrule
			ATE & 17.23 & 0.05 & 0.74 & 0.98 & 0.95 & 17.22 & 0.02 & 0.63 & 1.02 & 0.95 & 17.25 & 0.16 & 0.72 & 0.99 & 0.95 \\ 
			ATE (0.05) & 16.64 & 0.67 & 0.80 & 1.48 & 0.89 & 16.63 & 0.71 & 0.78 & 1.63 & 0.88 & 16.65 & 0.60 & 0.80 & 1.49 & 0.89 \\ 
			ATE (0.1) & 15.74 & 0.23 & 1.02 & 2.55 & 0.77 & 15.74 & 0.23 & 1.01 & 2.73 & 0.75 & 15.75 & 0.16 & 1.03 & 2.50 & 0.77 \\ 
			ATE (0.15) & 15.66 & 0.70 & 1.31 & 3.24 & 0.71 & 15.66 & 0.70 & 1.30 & 3.50 & 0.70 & 15.67 & 0.66 & 1.31 & 3.17 & 0.71 \\ 
			ATO & 16.44 & 0.75 & 1.01 & 1.07 & 0.93 & 16.45 & 0.69 & 1.00 & 1.05 & 0.94 & 16.44 & 0.73 & 1.02 & 1.07 & 0.94 \\ 
			ATM & 16.52 & 0.63 & 1.25 & 1.04 & 0.92 & 16.54 & 0.53 & 1.21 & 1.04 & 0.93 & 16.52 & 0.62 & 1.26 & 1.04 & 0.93 \\ 
			ATEN & 16.52 & 0.64 & 0.89 & 1.07 & 0.94 & 16.53 & 0.60 & 0.88 & 1.06 & 0.94 & 16.53 & 0.60 & 0.90 & 1.07 & 0.94 \\ 
			ATC & 16.63 & 0.41 & 1.38 & 1.02 & 0.93 & 16.66 & 0.21 & 1.30 & 0.56 & 0.99 & 16.60 & 0.58 & 1.45 & 0.62 & 0.99 \\ 
			ATT & 17.38 & 0.17 & 0.88 & 1.01 & 0.95 & 17.36 & 0.03 & 0.69 & 0.56 & 0.99 & 17.41 & 0.35 & 0.84 & 0.64 & 0.99 \\
			\addlinespace
			& \multicolumn{5}{c}{Aug: OR model correctly specified} & \multicolumn{5}{c}{Aug: both models misspecified} & \multicolumn{5}{c}{} \\ \cmidrule(lr){2-6}\cmidrule(lr){7-11}\cmidrule(lr){12-16}
			Estimand & PE & Bias & RMSE & RE & CP & PE & Bias & RMSE & RE & CP & \multicolumn{5}{c}{} \\
			\midrule
			ATE & 17.22 & 0.02 & 0.63 & 1.02 & 0.95 & 17.43 & 1.23 & 1.03 & 0.99 & 0.94 & \multicolumn{5}{c}{}\\ 
			ATE (0.05) & 17.17 & 2.48 & 0.76 & 1.05 & 0.90 & 17.37 & 3.69 & 1.18 & 1.00 & 0.91 & \multicolumn{5}{c}{}\\ 
			ATE (0.1) & 16.42 & 4.08 & 1.09 & 2.03 & 0.72 & 16.47 & 4.38 & 1.41 & 1.51 & 0.84 & \multicolumn{5}{c}{}\\ 
			ATE (0.15) & 15.83 & 0.37 & 1.05 & 2.25 & 0.80 & 15.71 & 0.39 & 1.49 & 1.68 & 0.86 & \multicolumn{5}{c}{}\\ 
			ATO & 16.77 & 1.28 & 0.72 & 1.02 & 0.95 & 16.97 & 2.50 & 1.18 & 1.01 & 0.93 & \multicolumn{5}{c}{}\\ 
			ATM & 16.82 & 1.15 & 0.74 & 1.00 & 0.95 & 17.09 & 2.81 & 1.28 & 0.98 & 0.94 & \multicolumn{5}{c}{}\\ 
			ATEN & 16.84 & 1.29 & 0.70 & 1.02 & 0.95 & 17.04 & 2.47 & 1.14 & 1.00 & 0.93 & \multicolumn{5}{c}{}\\ 
			ATC & 16.66 & 0.21 & 1.30 & 0.56 & 0.99 & 17.07 & 2.26 & 1.39 & 0.52 & 1.00 & \multicolumn{5}{c}{}\\ 
			ATT & 17.36 & 0.03 & 0.69 & 0.56 & 0.99 & 17.52 & 0.96 & 1.02 & 0.71 & 0.98 & \multicolumn{5}{c}{}\\  
			\bottomrule
		\end{tabular}
		\begin{tablenotes}
			\tiny
			\item PE: point estimation; Bias: absolute relative bias$\times 100$; RMSE: root mean squared error; RE: relative efficiency; CP: coverage probability; Aug: augmented estimator; PS: propensity score; OR: outcome regression; ATE: average treatment effect; ATE ($\alpha$): ATE by trimming PS $>1-\alpha$ or PS $<\alpha$, $\alpha=0.05, 0.1, 0.15$; ATO (resp. ATM, ATEN, ATC, and ATT): average treatment effect on overlap (resp. matching, entropy, controls, and treated)
		\end{tablenotes}
	\end{threeparttable}
\end{table}	 

\newpage 
\section{Appendix: Additional Data Analysis Results}\label{apx:data}

\begin{figure}[h]
 \begin{threeparttable}
		\begin{center}
			{\includegraphics[trim=5 10 5 8, clip, width=1\textwidth]{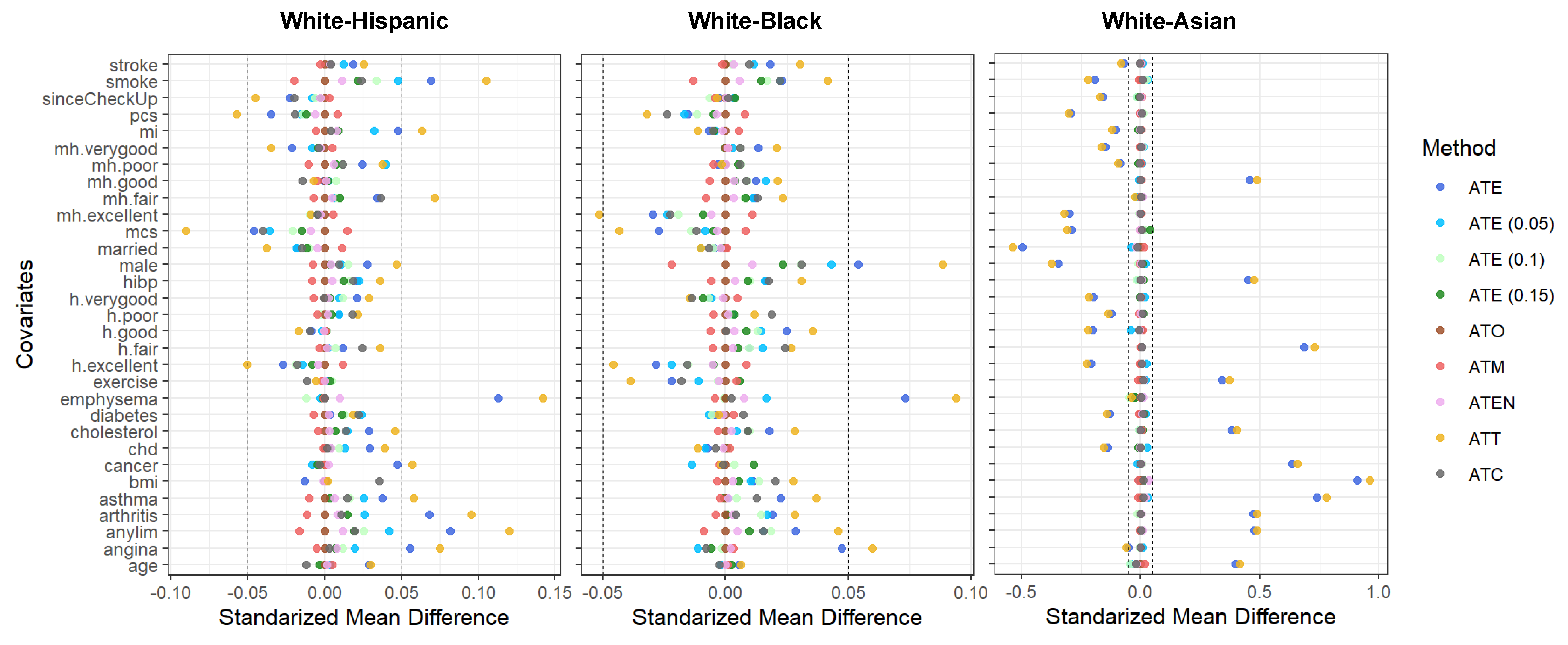}}
		\end{center}
			\begin{tablenotes}
			\tiny
			\item  ATE: average treatment effect; ATE ($\alpha$): ATE by trimming PS $>1-\alpha$ or PS $<\alpha$, $\alpha=0.05, 0.1, 0.15$; ATO (resp. ATM, ATEN, ATC, and ATT): average treatment effect on overlap (resp. matching, entropy, controls, and treated)	
   \end{tablenotes}
 \end{threeparttable}	
 \caption{MEPS Data: covariates balance by the three sub-population of race comparisons}\label{fig:asd-meps}
	\end{figure}

In this section, we show additional details of the MEPS data analysis in Section \ref{sec:data}. The love plots in Figure \ref{fig:asd-meps} display the standardized mean differences (SMD) of covariates between the 2-by-2 groups. Most SMDs in the first two sub-population are within the 0.1 threshold, but those in White-Asian group by ATT and ATC exceed 0.1 a lot. At the same time, in general the equipoise estimands (ATO, ATM and ATEN) balance the covariates the best. 

	\begin{figure}[h]
% \begin{threeparttable}
		\begin{center}
			{\includegraphics[trim=5 10 0 5, clip, width=1\textwidth]{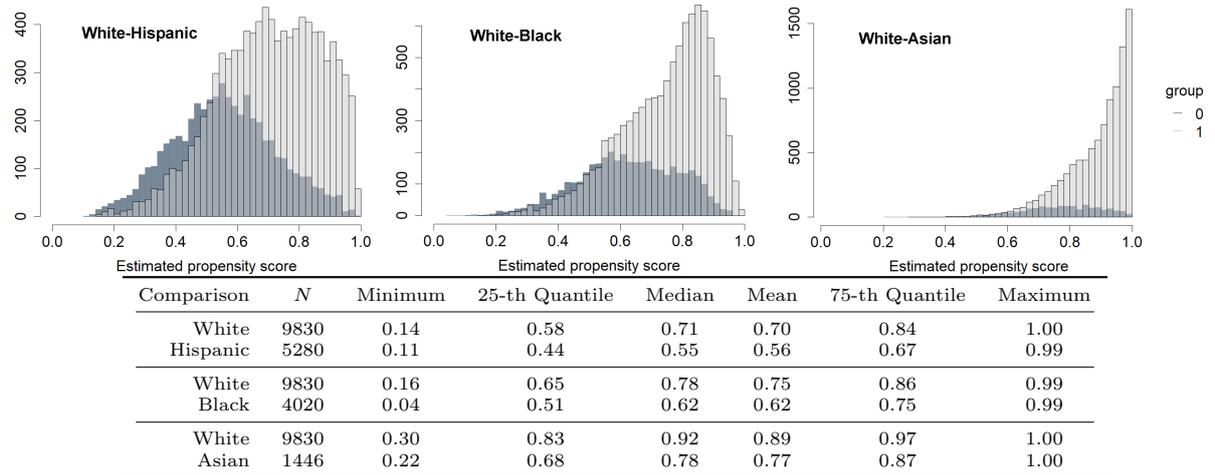}}
			{\scriptsize
			\begin{tabular}{rccccccc}
            \toprule
            Comparison & $N$ & Minimum & 25-th Quantile & Median & Mean & 75-th Quantile & Maximum\\ 
            
            \cmidrule(lr){1-8}
            White & 9830 & 0.14 & 0.58 & 0.71 & 0.70 & 0.84 & 1.00  \\ 
            Hispanic & 5280 & 0.11 & 0.44 & 0.55 & 0.56 & 0.67 & 0.99 \\ 
            
            \cmidrule(lr){1-8}
            White & 9830 & 0.16 & 0.65 & 0.78 & 0.75 & 0.86 & 0.99 \\ 
            Black & 4020 & 0.04 & 0.51 & 0.62 & 0.62 & 0.75 & 0.99 \\ 
            
            \cmidrule(lr){1-8}
            White & 9830 & 0.30 & 0.83 & 0.92 & 0.89 & 0.97 & 1.00 \\ 
            Asian & 1446 & 0.22 & 0.68 & 0.78 & 0.77 & 0.87 & 1.00 \\
            \bottomrule
            \end{tabular}}
		\end{center}
 %  \end{threeparttable}	
		\caption{Propensity score distributions and summary statistics of the three comparison groups of the medical expenditure data}\label{fig:ps-meps}
	\end{figure}
	
	\begin{table}[htbp]\footnotesize
    	\centering
    \begin{threeparttable}
    	\caption {Effective sample sizes (ESS) for the medical expenditure data}
    	\label{tab:ess_MEPS}
    	\begin{tabular}{ccrcccccccccccccccccccccccccccccccccccccc}
    		\toprule
    Group & N & ATE & \makecell{ ATE  (0.05)} & \makecell{ ATE  (0.1)} & \makecell{ ATE   (0.15)} & OW & MW & EW & ATT & ATC \\ 
    		\midrule
    		\multicolumn{10}{c}{\textbf{White-Hispanic: $p = 65.06\%, r = 1.00$}} \\
Hispanic & 5280 & 2018.91 & 3584.18 & 4078.39 & 4284.12 & 4843.23 & 4987.47 & 4636.77 & 1124.28 & 5280 \\ 
  White & 9830 & 8802.92 & 8420.37 & 7707.37 & 6952.06 & 7481.73 & 6773.92 & 7787.66 & 9830 & 5029.29 \\  
    		\addlinespace
    		\multicolumn{10}{c}{\textbf{White-Black: $p = 70.97\%, r = 0.81$}} \\
    Black & 4020 & 2311.46 & 2897.44 & 3095.89 & 3143.82 & 3767.29 & 3903.97 & 3624.32 & 1627.97 & 4020 \\ 
  White & 9830 & 9132.53 & 8909.64 & 8048.16 & 6643.04 & 7414.88 & 6428.39 & 7856.98 & 9830 & 5177.50 \\ 
    		\addlinespace
    		\multicolumn{10}{c}{\textbf{White-Asian: $p = 87.18\%, r = 0.60$}} \\
 Asian & 1446 & 9.07 & 947.79 & 1006.26 & 899.32 & 1404.83 & 1442.12 & 1298.08 & 7.72 & 1446 \\ 
 White & 9830 & 9623.01 & 6251.80 & 4172.28 & 2854.51 & 5410.03 & 4453.29 & 6244.44 & 9830 & 4229.11 \\ 
    		
    		\bottomrule 
    	\end{tabular}
    	\begin{tablenotes}
    		\tiny
    		\item  ATE : average treatment effect; ATE ($\alpha$):  ATE with trimming outside of $\alpha < e(\boldsymbol{x})<1-\alpha$, $\alpha=0.05, 0.1, 0.15$; OW: overlap weight; MW: matching weight; EW: entropy weight; IPWC: inverse probability weight on controls; IPWT: inverse probability weight on treated; $p$: proportion of participants in the treatment group; $r$: ratio of variances of propensity scores in treatment group to control group;
    	\end{tablenotes}
    \end{threeparttable}
    \end{table}
    
    \begin{table}[htbp]\small
		\centering
		\caption{Disparities in the health care expenditure of the three racial comparison groups of the MEPS data}
		\label{tab:meps-WATE}
	\begin{threeparttable}
	\begin{tabular}{lccrccrcccc}
		\toprule
		& \multicolumn{3}{c}{\textbf{H\'ajek-type Estimator}} & \multicolumn{3}{c}{\textbf{Augmented Estimator}} \\
		\cmidrule(lr){2-4}\cmidrule(lr){5-7}
		Estimand & Estimation & Standard error & p-value & Estimation & Standard error & p-value \\ 
		\midrule
		&\multicolumn{6}{c}{\textbf{White-Hispanic: $p = 65.06\%, r = 1.00$}} \\
		\cmidrule(lr){2-7}
		ATE & 699.12 & 304.77 & 0.022 & 1154.44 & 274.31 & $<$0.001 \\
      ATE (0.05) & 1326.22 & 198.59 & $<$0.001 & 1448.01 & 189.55 & $<$0.001 \\ 
      ATE (0.1) & 1170.73 & 198.52 & $<$0.001 & 1234.85 & 188.67 & $<$0.001 \\ 
      ATE (0.15) & 1240.09 & 188.38 & $<$0.001 & 1284.77 & 181.44 & $<$0.001 \\ 
      ATO & 1264.21 & 165.61 & $<$0.001 & 1282.59 & 166.30 & $<$0.001 \\ 
      ATM & 1306.46 & 158.35 & $<$0.001 & 1285.72 & 161.91 & $<$0.001 \\ 
      ATEN & 1202.48 & 173.40 & $<$0.001 & 1260.32 & 170.76 & $<$0.001 \\ 
      ATT & 345.26 & 419.58 & 0.411 & 1080.89 & 374.99 & 0.004 \\ 
      ATC & 1426.19 & 171.75 & $<$0.001 & 1289.21 & 170.10 & $<$0.001 \\ 
	\addlinespace
	&\multicolumn{6}{c}{\textbf{White-Black: $p = 70.97\%, r = 0.81$}} \\
	\cmidrule(lr){2-7}
	ATE & 850.82 & 234.56 & $<$0.001 & 992.82 & 235.79 & $<$0.001 \\ 
      ATE (0.05) & 738.62 & 238.16 & 0.002 & 764.25 & 235.05 & 0.001 \\ 
      ATE (0.1) & 802.82 & 223.56 & $<$0.001 & 836.40 & 219.82 & $<$0.001 \\ 
      ATE (0.15) & 846.31 & 228.58 & $<$0.001 & 868.74 & 223.59 & $<$0.001 \\ 
      ATO & 818.97 & 210.55 & $<$0.001 & 834.23 & 210.79 & $<$0.001 \\ 
      ATM & 824.31 & 213.90 & $<$0.001 & 814.78 & 215.29 & $<$0.001 \\ 
      ATEN & 823.11 & 212.55 & $<$0.001 & 841.77 & 212.56 & $<$0.001 \\ 
      ATT & 850.92 & 261.23 & 0.001 & 1088.15 & 264.84 & $<$0.001 \\ 
      ATC & 850.42 & 244.03 & $<$0.001 & 760.23 & 239.60 & 0.002 \\ 
		\addlinespace
		&\multicolumn{6}{c}{\textbf{White-Asian: $p = 87.18\%, r = 0.60$}} \\
		\cmidrule(lr){2-7}
		ATE & 2253.00 & 653.03 & $<$0.001 & 4712.69 & 2143.97 & 0.028 \\ 
      ATE (0.05) & 1248.64 & 256.82 & $<$0.001 & 1244.45 & 252.67 & $<$0.001 \\ 
      ATE (0.1) & 1293.97 & 216.63 & $<$0.001 & 1279.91 & NA  & NA \\ 
      ATE (0.15) & 1456.62 & 241.66 & $<$0.001 & 1442.50 & NA & NA  \\ 
      ATO & 1273.73 & 224.80 & $<$0.001 & 1303.53 & 227.28 & $<$0.001 \\ 
      ATM & 1391.96 & 219.19 & $<$0.001 & 1400.46 & 223.57 & $<$0.001 \\ 
      ATEN & 1231.66 & 243.22 & $<$0.001 & 1229.33 & 245.26 & $<$0.001 \\ 
      ATT & 2399.32 & 711.52 & $<$0.001 & 4960.53 & 2224.62 & 0.026 \\ 
      ATC & 1392.45 & 220.43 & $<$0.001 & 1388.95 & 224.85 & $<$0.001 \\
		\bottomrule
	\end{tabular}
	\begin{tablenotes}
		\tiny
		\item  ATE: average treatment effect; ATE ($\alpha$): ATE by trimming those with PS $>1-\alpha$ or PS $<\alpha$, $\alpha=0.05, 0.1, 0.15$; ATO (resp. ATM, ATEN, ATC, and ATT): average treatment effect on overlap (resp. matching, entropy, controls, and treated)
		\item $p$: proportion of White people in the sub-population; $r$: ratio of variance of propensity scores in White group to variance of propensity scores in minority groups (Hispanic, Black and Asian)
		\item Augmented ATE, ATE ($\alpha$), $\alpha = 0.05, 0.1, 0.15$, ATT and ATC are actually doubly-robust
	\end{tablenotes}
	\end{threeparttable}
	\end{table}
	
	Table \ref{tab:meps-WATE} shows all the estimated causal effects from the estimators we considered to evaluate racial disparities of the health care expenditure between the Whites participants and each of three comparison groups in the MEPS data. 	
	We need to point out the reason that the NA's shown in the standard errors and p-values of augmented ATE $(0.1)$ and ATE $(0.15)$ in the White-Asian comparison group. This is potentially due to the trimming. As can be seen in the Figure \ref{fig:ps-meps}, the propensity scores of White group in White-Asian figure has many values close to 1, which means when we trim those who have extreme propensity score, there is a large loss in sample size. In fact, after trimming by $0.1$, the sample size reduces to only $5458~ (48.40\%)$, and after trimming by $0.15$, only $3905~(34.63\%)$ remains in the whole White-Asian group $(N = 11276)$. When we lose these many participants (which implies a substantial information loss), the matrix $A_N(\widehat\theta)$ (estimated information matrix of the parameter vector, see Section \ref{subapx:proofs-sand-wt} for the details) used in the sandwich variance estimator becomes singular. This makes it difficult to invert the matrix and thus  disrupts the calculation of the sandwich variance estimate.
	
\end{document}